\documentclass[acmtog,nonacm]{acmart}
\usepackage{colortbl}
\usepackage{tikz}
\usepackage{multirow}
\usepackage{varwidth} 
\usepackage{pifont} 
\usepackage{boldline}
\usepackage{amsmath}
\usepackage{xcolor}
\usepackage[ruled]{algorithm2e} 
\usepackage{wrapfig}
\usepackage{bm}
\usepackage{makecell}
\usepackage{arydshln}
\usepackage[utf8x]{inputenc}
\usepackage{siunitx}
\usepackage{afterpage}
\usepackage{graphicx}  
\newcolumntype{P}[1]{>{\centering\arraybackslash}p{#1}}
\AtBeginDocument{%
  }

\graphicspath{ {../} }
\setcopyright{acmlicensed}
\acmJournal{TOG}
\acmYear{2025} \acmVolume{44} \acmNumber{4} \acmArticle{} \acmMonth{8}\acmDOI{10.1145/3731163}

\acmJournal{TOG}
\acmVolume{44}
\acmNumber{4}
\acmArticle{111}
\acmMonth{8}

\acmSubmissionID{525}

\begin{document}

\title{Gaussian Wave Splatting for Computer-Generated Holography}


\author{Suyeon Choi}
\email{suyeon@stanford.edu}
\authornote{denotes equal contribution.}
\orcid{0000-0001-9030-0960}
\affiliation{
  \institution{Stanford University}
  \country{USA}
}
\author{Brian Chao}
\email{brianchc@stanford.edu}
\orcid{0000-0002-4581-6850}
\authornotemark[1]
\affiliation{
  \institution{Stanford University}
  \country{USA}
}
\author{Jacqueline Yang}
\email{jyang01@stanford.edu}
\orcid{0009-0002-3101-3026}
\affiliation{
  \institution{Stanford University}
  \country{USA}
}
\author{Manu Gopakumar}
\email{manugopa@stanford.edu}
\orcid{0000-0001-9017-4968}
\affiliation{
  \institution{Stanford University}
  \country{USA}
}
\author{Gordon Wetzstein}
\email{gordon.wetzstein@stanford.edu}
\orcid{0000-0002-9243-6885}
\affiliation{
  \institution{Stanford University}
  \country{USA}
}


\renewcommand{\shortauthors}{Choi, Chao et al.}

\begin{abstract}
State-of-the-art neural rendering methods optimize Gaussian scene representations from a few photographs for novel-view synthesis. Building on these representations, we develop an efficient algorithm, dubbed Gaussian Wave Splatting, to turn these Gaussians into holograms. Unlike existing computer-generated holography (CGH) algorithms, Gaussian Wave Splatting supports accurate occlusions and view-dependent effects for photorealistic scenes by leveraging recent advances in neural rendering.  
Specifically, we derive a closed-form solution for a 2D Gaussian-to-hologram transform that supports occlusions and alpha blending. Inspired by classic computer graphics techniques, we also derive an efficient approximation of the aforementioned process in the Fourier domain that is easily parallelizable and implement it using custom CUDA kernels. 
By integrating emerging neural rendering pipelines with holographic display technology, our Gaussian-based CGH framework paves the way for next-generation holographic displays.
%
\end{abstract}




\received{23 January 2025}
\received[revised]{12 March 2025}
\received[accepted]{5 June 2025}

\begin{teaserfigure}
  \centering
	\includegraphics[width=\columnwidth]{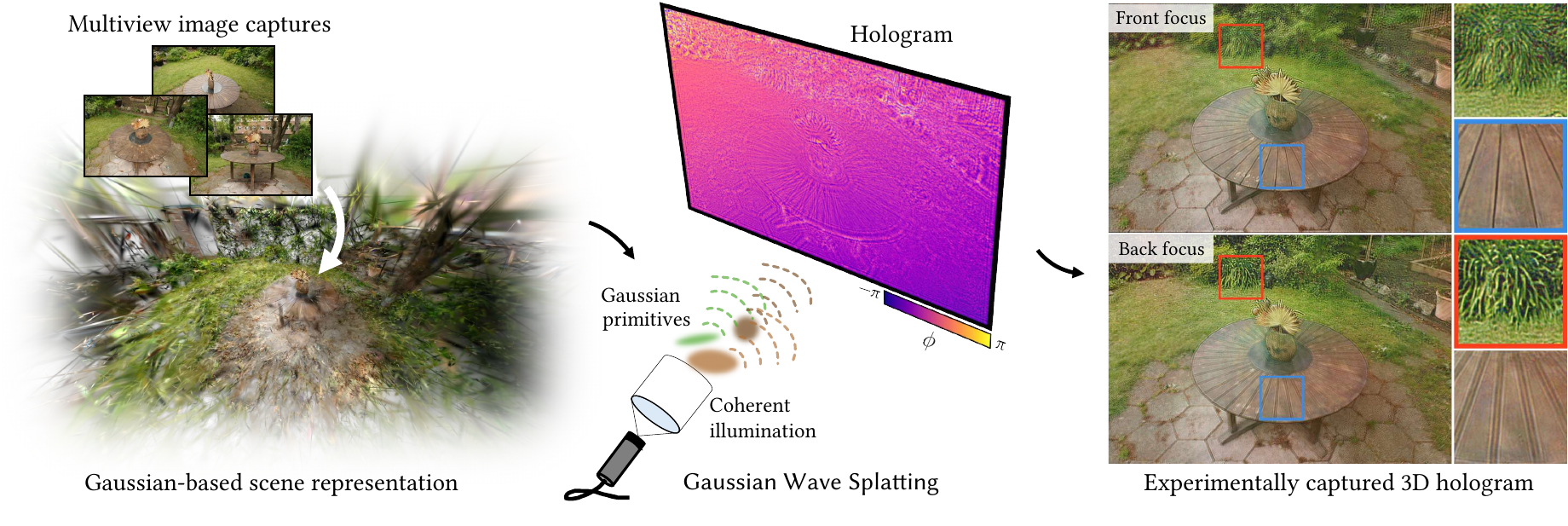}
   \caption{
   Emerging Gaussian splatting techniques take a few casually captured photographs as input and output a Gaussian scene representation (left). Our Gaussian Wave Splatting approach turns these types of representations into holograms that can be directly displayed on emerging holographic displays (center). The color bar below the hologram denotes phase values of the phase-only hologram. Our Gaussian-based computer-generated holography algorithm reconstructs photorealistic 3D scenes with refocusing capabilities as seen in the experimentally captured results on the right. }
   %
  \label{fig:teaser}
\end{teaserfigure} 

\maketitle
\newcommand{\rref}{\mathbf{r}_{\text{0}}}
\newcommand{\xref}{{x}_{\text{0}}}
\newcommand{\yref}{{y}_{\text{0}}}
\newcommand{\zref}{{z}_{\text{0}}}

\newcommand{\rlocal}{\mathbf{r}_{l}}
\newcommand{\xlocal}{{x}_{l}}
\newcommand{\ylocal}{{y}_{l}}
\newcommand{\zlocal}{{z}_{l}}

\newcommand{\cov}{\mathbf{\Sigma}}
\newcommand{\rot}{R}
\newcommand{\rotc}{\mathbf{C}}
\newcommand{\sca}{S}
\newcommand{\kvec}{\mathbf{k}}
\newcommand{\subparatitle}[1]{\textbf{\textit{#1}}\quad}

\definecolor{mg}{rgb}{0.639,0.984,0.722}
\definecolor{my}{rgb}{0.996,0.875,0.643}
\definecolor{mr}{rgb}{0.941,0.561,0.620}
\newcommand{\ccg}{\cellcolor{mg}}  
\newcommand{\ccy}{\cellcolor{my}}  
\newcommand{\ccr}{\cellcolor{mr}}  

\newcommand{\cmark}{\ding{51}}%
\newcommand{\xmark}{\ding{55}}%
\newcommand{\bluecheck}{{\color{blue}\checkmark}}
\newcommand{\blackcheck}{{\color{black}\cmark}}
\newcommand{\redx}{{\Large\color{\red}\xmark} }

\newcommand{\headfontsize}{\scriptsize}
\newcommand{\tablerefsize}{\tiny}
\newcommand{\vertice}{\mathbf{v}}
\newcommand{\normal}{\mathbf{n}}

\newcommand{\roundedtriangle}{%
  \begin{tikzpicture}[baseline=0ex, scale=0.06]
    \draw[rounded corners=0.9pt, thick] (0,0) -- (1.5,2.5) -- (3,0) -- cycle;
  \end{tikzpicture}%
}
\newcommand*\colourcheck[1]{%
  \expandafter\newcommand\csname #1check\endcsname{\textcolor{#1}{\ding{51}}}%
}
\colourcheck{red}
\colourcheck{green}

\newcommand*\colourxx[1]{%
  \expandafter\newcommand\csname #1xx\endcsname{\textcolor{#1}{\ding{55}}}%
}
\colourxx{red}
\colourxx{green}

\newcommand*\colourtriangle[1]{%
  \expandafter\newcommand\csname #1triangle\endcsname{\textcolor{#1}{$\varDelta$}}%
}
\colourtriangle{yellow}

\newcommand{\greenvcell}{\ccg \cmark}
\newcommand{\redxcell}{\ccr \xmark}
\newcommand{\yellowtrcell}{\ccy $\roundedtriangle $  } 
\newcommand{\greenxcell}{\ccg \xmark}
\newcommand{\redvcell}{\ccr \cmark}


\newcommand{\ctcell}{>{\centering\arraybackslash}}
\newcommand{\gw}[1]{\textcolor{purple}{[GW: #1]}}
\newcommand{\sy}[1]{\textcolor{blue}{[SC: #1]}}
\newcommand{\bc}[1]{\textcolor{purple}{[BC: #1]}}
\newcommand{\revision}[1]{\textcolor{black}{#1}}


\newcommand{\citl}{CITL}
\newcommand{\revised}[1]{{#1}}
\newcommand{\editsy}[1]{\textcolor{red}{#1}}
\newcommand{\prop}{\mathcal{P}}
\newcommand{\firstprop}{\prop_1}
\newcommand{\secprop}{\prop_2}
\newcommand{\prophat}{\widehat{g}}
\newcommand{\firstprophat}{\prophat_1}
\newcommand{\secprophat}{\prophat_2}
\newcommand{\target}{a_{target}}
\newcommand{\phaseslm}{\phi}
\newcommand{\efficiency}{\eta}
\newcommand{\fourier}{\mathcal{F}}
\newcommand{\loss}{\mathcal{L}}
\newcommand{\transfer}{\mathcal{H}}
\newcommand{\fx}{f_x}
\newcommand{\fy}{f_y}
\newcommand{\numpixelx}{N_x}
\newcommand{\numpixely}{N_y}
\newcommand{\phasepool}{\mathbb{P}}
\newcommand{\capturedpool}{\mathbb{C}}
\newcommand{\rgbd}{\mathbb{RGBD}}
\newcommand{\lossbp}{\texttt{loss\_bp}}
\newcommand{\fbp}{\texttt{f\_bp}}
\newcommand{\gbp}{\texttt{g\_bp}}
\newcommand{\red}[1]{\textcolor{red}{#1}}
\newcommand{\idx}{i}
\newcommand{\maxidx}{N}
\newcommand{\angularspectrum}{\hat{u}}
\newcommand{\pixelcoord}{\mathbf{r}}

\newcommand{\coordinate}{\mathbf{x}}
\newcommand{\mean}{\bm{\mu}}

\newcommand{\worldspace}{w}
\newcommand{\viewspace}{r}
\newcommand{\canonicalspace}{c}
\newcommand{\objectspace}{o}
\newcommand{\centergaussian}{{\bm{\mu}}}
\newcommand{\projectivemapping}{\mathbf{m}}

\newcommand{\mat}[1]{\mathbb{R}^{{#1}\times{#1}}}
\newcommand{\arr}[1]{\mathbb{R}^{{#1} \times 1}} 

\makeatletter
\newcommand\xleftrightarrow[2][]{%
  \ext@arrow 9999{\longleftrightarrowfill@}{#1}{#2}}
\newcommand\longleftrightarrowfill@{%
  \arrowfill@\leftarrow\relbar\rightarrow}
\makeatother

\section{Introduction}

Holographic near-eye displays are a next-generation technology that offers unprecedented capabilities for smart eyewear. Direct control over the phase or amplitude of a partially coherent light engine of a virtual or augmented reality (VR/AR) display unlocks unique abilities of correcting aberrations of downstream optics, correcting for a user’s prescription, and displaying perceptually realistic focus cues. Conventional emissive or amplitude-modulating displays require perfect (i.e., image-preserving) optics, which places many restrictions on combiner optics designs. The unique capabilities of holographic near-eye displays, on the other hand, enable “imperfect” optical waveguides with ultra-compact VR/AR display form factors where the holographic light engine corrects for the imperfections~\cite{maimone2017holographic,kim:HolographicGlasses,jang2024waveguide,gopakumar2024full}. 

All holographic displays require a computer-generated holography (CGH) algorithm to convert the target content (e.g., an image or a 3D scene) into the hologram that is shown on the spatial light modulator (SLM). Oftentimes, this is a phase retrieval problem for phase-only SLMs, and neural networks have been proposed to accelerate this task (e.g.,~\cite{peng2020neural,shi2021towards,shi2022end}). Existing CGH algorithms support a variety of different input data formats, including RGB images, RGB-D images with per-pixel depth information, meshes, or light fields~\cite{park2017recent, gerchberg1972practical, wakunami2013occlusion, zhang2017computer, matsushima2009extremely, choi2022time, schiffers2023stochastic, chao2024large}. A major challenge for CGH algorithms, however, is that they are either extremely slow and require input data that are not readily available (e.g., densely sampled light fields) or that they use input data which do not natively support view-dependent effects and occlusions (e.g., RGB-D) or photorealistic scenes (e.g., meshes). Therefore, no existing CGH algorithm offers photorealistic image quality at interactive frame rates with readily available content, such as a few casually captured images or a video of a real scene.


Neural rendering techniques have recently enabled novel-view synthesis with photorealistic image quality using just a few images of a scene~\cite{tewari2020state,tewari2022advances}. Among the many proposals for neural scene representations, 3D Gaussians~\cite{kerbl3Dgaussians} have been established as the representation of choice for many applications due to their computational efficiency, high image quality, interpretability, and robust optimization behavior. 



The core contribution of our work is the development of an efficient Gaussian-to-hologram transform, which we call \textit{Gaussian Wave Splatting} (GWS), that inherits the benefits of ray-based Gaussian scene representations for wave-based holographic displays. We derive the exact analytical expression of a Gaussian-to-hologram transform that supports occlusions and view-dependent effects through a wave-based counterpart of the classic volume rendering equation \cite{kajiya1984ray}, which we call \textit{alpha wave blending}. We provide an efficient CUDA implementation for a variant of our algorithm and demonstrate photorealistic holographic image quality in simulation and with experimental captures.


To summarize, our contributions include
\begin{itemize}
    \item Derivation of an exact analytical solution for a Gaussian-to-hologram transform, dubbed \textit{Gaussian Wave Splatting}.
    \item Design of a wave optics counterpart of the classic volume rendering equation, dubbed \textit{alpha wave blending}.
    \item Efficient implementation of a GWS variant using custom CUDA kernels that achieves a $30\times$ speedup.
    \item Validation using numerical simulations and experiments, demonstrating benefits over existing CGH algorithms.
\end{itemize}
Source code is available on \href{https://github.com/computational-imaging/hsplat}{our Github repository}\footnote{\href{https://github.com/computational-imaging/hsplat}{github.com/computational-imaging/hsplat}}.

\section{Related Work}
Our work builds on recent progress in holographic displays~\cite{yaracs2010state,park2017recent,chang2020toward,javidi2021roadmap}, but focuses on developing a new CGH algorithm inspired by recent advances in neural scene representations and rendering. We review these methods in the following.

\paragraph{\bf{CGH Algorithms.}}
CGH refers to an algorithm that converts a target scene into a 2D amplitude, phase, or complex field to be displayed on an SLM, i.e., the hologram, which optically synthesizes the desired target intensity distribution. These algorithms have been developed to accommodate various types of scene representations. One of the classic algorithms is the point-based method~\cite{maimone2017holographic, h2009computer, lucente1993interactive}, where the target content is represented as a collection of points, each propagated to the SLM plane and digitally interfered with a reference wave. The propagation of all points toward the SLM is equivalent to a spatially varying convolution of a point cloud with a depth-dependent point spread function (PSF) and is computationally prohibitive. Moreover, point clouds require an excessive number of primitives to represent a scene with high fidelity and are inherently limited in their ability to model occlusions and view-dependent effects, although look-up table approaches and ray-based occlusion methods have been proposed to partially address these limitations~\cite{kim2008effective, lucente1993interactive, h2009computer, shi2021towards}.

To support view-dependent effects, polygon-based methods \cite{ahrenberg2008computer, kim2008mathematical, matsushima2003fast, matsushima2009extremely, matsushima2014silhouette} and holographic stereograms (HS)~\cite{kang2008accurate, benton1983survey, padmanaban2019holographic, halle1991ultragram} have been proposed. Polygon-based methods convert a mesh into a complex-valued wavefront. However, the wavefront footprint of triangles varies with their orientation, position, and size, requiring substantial memory for look-up table approaches and these methods are often slow~\cite{wang2021acceleration}. Furthermore, numerical stability issues often arise when calculating wavefronts of arbitrarily oriented triangles. Finally, \revision{these methods struggle to render highly detailed appearances unless an excessive number of tiny triangles are used. On the other hand,} holographic stereograms encode densely sampled light fields into holograms and typically result in low image quality because of their heuristic choices of phase distributions of objects~\cite{kang2016fast, padmanaban2019holographic}. 

Recent advances in machine learning have shown promise in tackling these challenges, either by accelerating the CGH algorithm using deep neural networks or by providing more numerically accurate procedures that can adapt to flexible target representations. Several works proposed using convolutional neural networks (CNNs) to approximate the spatially varying convolution, demonstrating real-time performance~\cite{peng2020neural, shi2021towards, shi2022end, yang2022diffraction, horisaki2021three}. While promising, these CNN-based methods rely on RGB, RGB-D, or layered-depth-image (LDI) inputs, which all suffer from depth discontinuity artifacts when rendering from arbitrary viewpoints. Furthermore, such methods require re-training neural networks for different hardware configurations. Gradient-descent-based methods offer more flexibility, accepting custom loss functions and various targets such as RGB and RGB-D images, focal stacks, and light fields~\cite{kuo2023multisource, zhang20173d, chakravarthula2019wirtinger, peng2020neural, choi2021neural, choi2022time, gopakumar2024full, kavakli2023realistic, Chakravarthula2022pupil}. Light field holograms generated via gradient descent~\cite{choi2022time, kim2024holographic, schiffers2023stochastic, chao2024large}, in particular, have shown great potential for reconstructing photorealistic content with view-dependent effects, but increased rendering, memory, input/output (IO), and data streaming costs for light fields as well as their iterative nature make them impractical for real-time applications. Moreover, densely sampled light fields are not readily available, so content has to be specifically curated in this format.

\renewcommand{\arraystretch}{1.1}

\setlength{\tabcolsep}{4pt} 
\setlength{\textfloatsep}{10pt}
\begin{table}[t!]
    \centering
    \captionsetup{aboveskip=0pt, belowskip=3pt}
    {\footnotesize 
    
    \caption{Comparison of CGH algorithms w.r.t. efficiency and visual quality. We compare our GWS with the point-based method (PBM), convolutional neural network (CNN) methods using RGB, RGB-depth (RGB-D), and stochastic gradient descent-based (SGD) methods using focal stacks and light fields (LF), as well as the polygon-based method. Our approach uses readily available input data, such as a few photographs of a real scene, does not require pre-training any networks tied to a specific hardware configuration, and naturally supports occlusions and view-dependent effects (VDEs). 
    }
    \label{tab:cgh_comp_table}
    
    \begin{tabular}{ m{2.2cm}|ccc|cc}
        \multirow{3}{*}{{\centering Methods}} & \multicolumn{3}{c|}{Efficiency} & \multicolumn{2}{c}{Visual quality} \\ 
                              & {\headfontsize  efficient}   & {\headfontsize data } & \headfontsize pretraining & \multirow{2}{*}{VDE} & {\headfontsize photorealistic } \\ 
                                 &               \headfontsize inference                       & \headfontsize  availability    & \headfontsize required &   & \headfontsize  content    \\ 

         \hline
        PBM \tablerefsize{~\cite{h2009computer, maimone2017holographic}}  & \yellowtrcell  & \greenvcell & \greenxcell & \redxcell & \redxcell \\
        \hline
        RGB-D, CNN  \tablerefsize{~\cite{peng2020neural, shi2021towards, shi2022end}} & \greenvcell & \greenvcell & \redvcell & \yellowtrcell & \greenvcell \\
        \hline
        focal stack, SGD   \tablerefsize{~\cite{choi2022time}} & \redxcell & \redxcell & \greenxcell & \yellowtrcell & \greenvcell \\
        \hline
        LF  \tablerefsize{~\cite{kang2008accurate, padmanaban2019holographic, choi2022time}} & \redxcell & \redxcell & \greenxcell  & \greenvcell & \greenvcell \\
        \hline
        Polygons  \tablerefsize{~\cite{matsushima2003fast, kim2008mathematical}} & \redxcell & \greenvcell & \greenxcell & \greenvcell & \redxcell \\
        \hlineB{3}
        GWS (efficient) & \yellowtrcell & \greenvcell & \greenxcell & \yellowtrcell & \greenvcell \\
        \hline
        GWS (exact) & \redxcell & \greenvcell & \greenxcell & \greenvcell & \greenvcell \\
        \hline
    \end{tabular}
    }
\end{table}

Our Gaussian Wave Splatting method inherits benefits of Gaussian scene representations, including photorealistic and 3D-consistent view synthesis and an explicit primitives-based representation, allowing for the generation of holograms that support accurate 3D effects. A comparison with existing CGH methods is shown in Tab.~\ref{tab:cgh_comp_table}.

\paragraph{\bf{Neural Scene Representations and Rendering.}}

Recent advances in neural rendering \cite{tewari2022advances} have made great strides in improving the quality of 3D reconstruction and novel-view rendering. 
Neural rendering algorithms can be roughly classified by the underlying 3D representation, ranging from point clouds~\cite{KPLD21point, aliev2019point, xu2022point}, voxels~\cite{sitzmann2019deepvoxels,yu_and_fridovichkeil2021plenoxels, hu2023multiscale, yu2021plenoctrees}, meshes~\cite{neumesh, kato2018meshrenderer, Lombardi2018deepappearance, hu2021worldsheet, guo2023vmesh}, to implicit representations using MLPs~\cite{mildenhall2020nerf, mueller2022instant, sitzmann2019srns,sitzmann2020implicit}. However, it is non-trivial to convert these 3D representations to high-quality holograms for direct viewing on a 3D holographic display. Almost all of these 3D representations include neural networks, and thus an additional step of extracting RGB-D images, layered depth images (LDI), light fields, or meshes from these neural representations is required for CGH calculation. Among these converted explicit 3D representations, mesh representations retain the most 3D information since each primitive is continuously distributed in 3D space unlike the discrete depth partitioning in RGB-D or LDI representations. However, mesh optimization and extraction processes are either deeply embedded in neural representation training ~\cite{tang2022nerf2mesh, hanocka2020point2mesh} or are based on signed distance function optimization ~\cite{wang2021neus, yariv2021volume} and Marching Cubes \cite{marchingcubes} or Poisson Surface Reconstruction \cite{kazhdan2006poisson}, which all require careful tuning to obtain decent quality meshes.

Recently, 3D Gaussian Splatting \cite{kerbl3Dgaussians} (3DGS) has emerged as the state-of-the-art novel-view synthesis and 3D reconstruction technique. 
3DGS achieves high image quality for novel-view synthesis, is fast to optimize and render, and leverages an explicit Gaussian primitive representation which makes the method suitable for a wide variety of computer vision and graphics applications. The absence of neural networks in the Gaussians representation allows us to directly convert optimized Gaussians into 3D holograms without any preprocessing, eliminating possibilities of image fidelity loss throughout the conversion process. The highly parallelizable nature of splatting Gaussians\revision{~\cite{kerbl3Dgaussians}} is also desirable for building efficient CGH algorithms, as we demonstrate with our novel wave splatting of Gaussians in the Fourier domain. Among the plethora of 3DGS variants~\cite{chen2024survey}, 2DGS \cite{huang20242d} is especially suitable for CGH because an angular spectrum-based analysis of waves is defined on planes, and since 2D Gaussians and meshes are both flat-primitive representations, CGH from 2D Gaussians can therefore greatly benefit from insights made by the pioneering research on polygon-based CGH. Hence, we use 2DGS as our scene representation for CGH calculation.

\section{Background on Gaussian Splatting}
Gaussian-based scene representations have been shown to be expressive and efficiently rendered via differentiable splatting. Our work builds on recent 2DGS methods~\cite{huang20242d}, where a 2D or flat Gaussian is defined by its center $\bm{\mu} \in \mathbb{R}^3$, two principal tangent vectors $\mathbf{t}_u, \mathbf{t}_v \in \mathbb{R}^3$ that control the orientation of the Gaussian, and two scaling factors $s_u, s_v \in \mathbb{R}$ that control the variance of the Gaussian. 


To render the color of a pixel, the colors associated with each Gaussian are alpha-composited from front to back as
%
\begin{equation}
    \mathbf{c} = \sum_{\idx=1}^\maxidx \mathbf{c}_\idx \alpha_\idx \prod_{j=1}^{\idx-1}(1 - \alpha_j), \label{eq:gs_volume_rendering}
\end{equation}
where $\idx$ is the index of the Gaussians, $\mathbf{c}_\idx$ is the color of each Gaussian computed from the per-Gaussian spherical harmonic coefficients and viewing direction, and $\alpha_\idx$ is the alpha value computed from the opacity $o_\idx$ associated with each Gaussian and the standard Gaussian value \revision{$\mathcal{G}_i$} evaluated at the UV coordinate $(u, v) \in \mathbb{R}^2$ corresponding the pixel location in the local 2D Gaussian tangent plane as
\begin{equation}
    \alpha_i = \mathcal{G}_i({u, v})\cdot o_i .
    \label{eq:opacity}
\end{equation}


The attributes (center, rotation, scale, opacity, and spherical harmonic coefficients) of the Gaussians are optimized using gradient descent optimization with photometric losses between rendered and target images as described by Huang et al.~\shortcite{huang20242d}.


\section{Wave Splatting}
In this section, we develop a method to compute holograms that faithfully represent the 3D scene, given a collection of 2D Gaussians representing this scene. To encode the Gaussian primitives as interference patterns based on the theory of wave optics, we need to propagate the wavefront contribution of each Gaussian, namely \revision{$u(\coordinate)$,} in free space to the SLM plane. The optical propagation over a distance $z$ can be described by a free-space wave propagation operator $\prop(\cdot; z)$, such as the angular spectrum method~\cite{goodman2005introduction, matsushima2009band}:
\begin{align}
    \prop\Big(u(\coordinate ); z\Big) = \iint \fourier \left( u(\coordinate )\right) \mathcal{H} \left( \mathbf{k}, z \right) e^{j \left(k_x x + k_y y \right)} d k_x d k_y, \\
    \mathcal{H} \left( \mathbf{k}, z \right) = \begin{cases}
        e^{j k_z z}, & \text{if} \sqrt{k_x^2 + k_y^2} < \frac{2\pi}{
        \lambda},\\
        0, & \text{otherwise},
    \end{cases}, |\mathbf{k}| = \frac{2\pi}{\lambda}
    \label{eq:asm}
\end{align}
where $\lambda$ is the wavelength, $\mathbf{k} = (k_x, k_y, k_z)$ is the wave vector, and $\fourier(\cdot)$ denotes the Fourier transform. Instead of integrating a radiance field along each ray, as in classic volume rendering~\cite{kajiya1984ray}, in our wave-based setting, we splat the wavefront contributions of Gaussians onto a common plane that is orthogonal to the optical axis by simulating their wave propagation through the volume from front to back
%
%
using alpha compositing. Note that this composition can be evaluated at any plane, such as the SLM where $z=0$. Therefore, we obtain the composited \revision{complex-valued} wavefront at the SLM, $u_{\textrm{SLM}}$, as  %
\begin{align}
    u_{\textrm{SLM}} (\coordinate) = \sum_{\idx}^\maxidx \mathbf{c}_\idx o_\idx\mathcal{P}\Big(  u_\idx(\coordinate; z_\idx)T_\idx\left(\coordinate\right) ; -z_\idx\Big), \label{eq:gwsplat}\\
    T_\idx\left(\coordinate\right) = \prod_{j=1}^{\idx-1} (1 - o_{j} |u_j(\coordinate; z_j)|),
    \label{eq:gwsplat_T}
\end{align} 
where \revision{$u_\idx (\coordinate; z_\idx)$} denotes the wavefront contribution of the $i^{th}$ Gaussian at depth $z_i$ (see next subsection, Eq.~\ref{eq:exactgws}) and $o_i$ denotes the opacity of the $i^{th}$ Gaussian. \revision{$T_i(\coordinate)$ is the accumulated transmittance along the wave propagation, from the first Gaussian to the $i^{th}$ Gaussian~\cite{kerbl3Dgaussians, huang20242d}.} We refer to Eq.~\ref{eq:gwsplat} as \textit{alpha wave blending}. Our alpha wave blending technique inherits the advantages of Gaussian splatting, including smooth falloffs for each object opacity (Eq.~\ref{eq:opacity}), to seamlessly merge the Gaussians for high-quality reconstruction. This formulation simplifies to the original ray-based alpha blending if we ignore the wave propagation operator and the phase of all wavefronts (i.e., $u_\idx=\mathcal{G}_i$). 

Previous polygon-based CGH methods, such as silhouette-based approaches~\cite{matsushima2014silhouette}, are inherently slow due to their recursive nature that involves repeatedly masking out the occluded field from the back with a binary aperture. Additionally, the use of binary apertures with translucency to emulate alpha blending~\cite{matsushima2020introduction} results in artifacts, as we discuss and compare further in Sec. ~\ref{sec:results} (see Figs.~\ref{fig:blending} and~\ref{fig:occlusion}). Our alpha wave blending technique, on the other hand, performs exact alpha blending like its ray-based counterpart, and is also compatible with silhouette-based methods for meshes with binary opacity. 

In alpha wave blending, the wavefronts of arbitrarily oriented 2D Gaussians need to be accurately propagated to planes parallel to the SLM plane. One way to achieve this is to place each 2D Gaussian disk on its local tilted plane and propagate it to a parallel plane using tilted propagation, which translates to remapping the angular spectrum~\cite{matsushima2003fast}. However, this approach is inherently slow as it requires performing a Fast Fourier Transform (FFT) for each primitive to compute the angular spectrum on the tilted plane. Additionally, it may introduce numerical errors due to the nonlinear remapping of Fourier coordinates~\cite{zhang2022polygon}. To overcome these limitations, we derive a closed-form solution for the angular spectrum of arbitrary 2D Gaussian disks at the SLM plane (Sec.~\ref{subsec:gaussian-wave-splatting}).

To further speed up alpha wave blending, we develop an approximated alpha blending model inspired by order-invariant opacity \cite{mcguire2013weighted} in traditional computer graphics (Sec.~\ref{subsec:fast-alpha-wave-blending}). This approximated model can be efficiently parallelized in the Fourier domain, and we achieve a 30$\times$ speedup compared to na\"ive alpha wave blending using our custom CUDA CGH kernels.

\subsection{Gaussian Wave Splatting}

\begin{figure}[t]
    \centering
    \includegraphics[width=1.0\columnwidth]{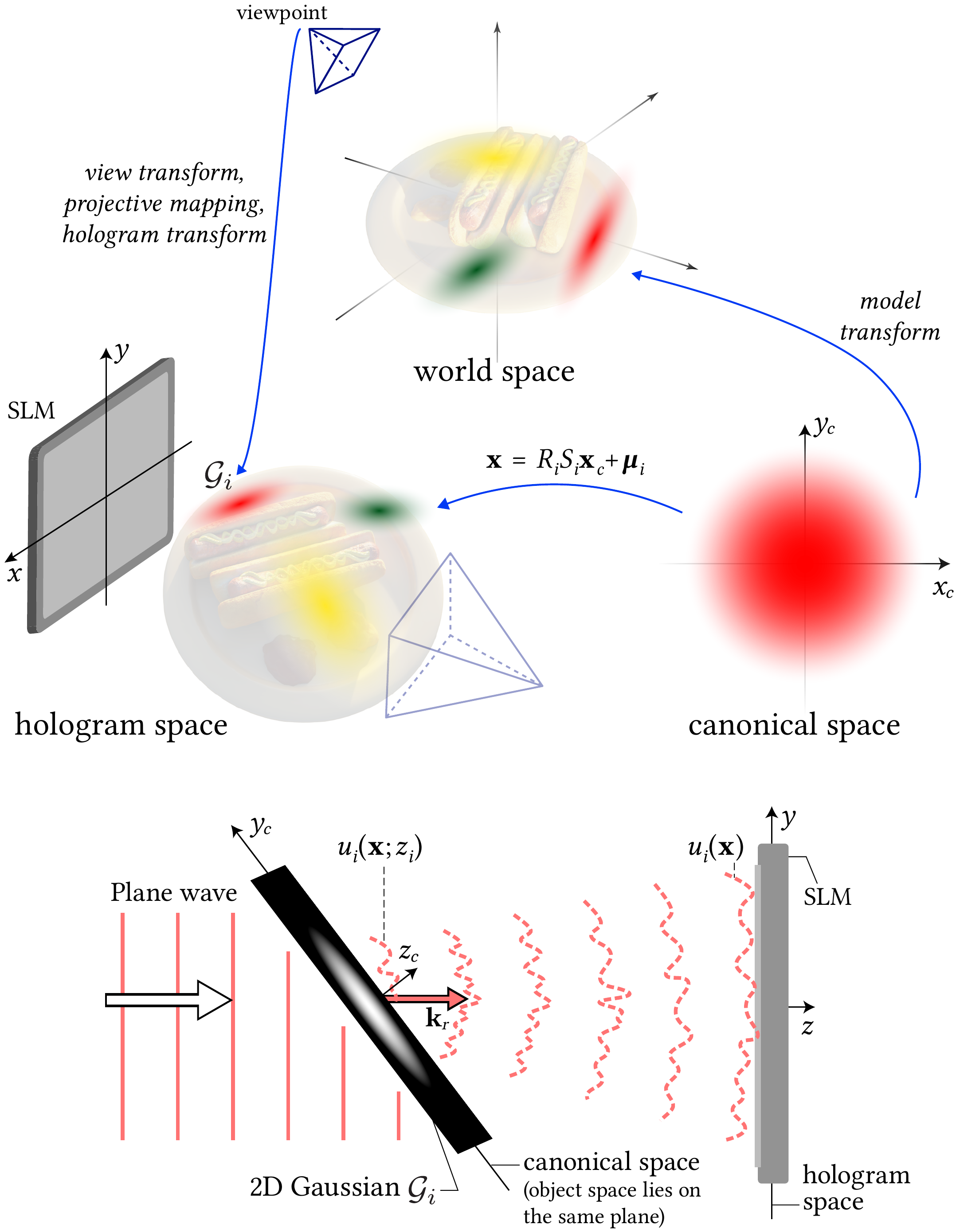}
    \caption{\revision{\textbf{Illustration of coordinate systems and Gaussian Wave Splatting.} \textit{Top:} We take a set of optimized primitives from off-the-shelf neural rendering frameworks in \textit{world space} and convert them into into \textit{hologram space} through a sequence of coordinate transformations. Each Gaussian in hologram space is represented by its own affine transformation from \textit{canonical space}. \textit{Bottom:} In hologram space, GWS computes the analytical solution of the wavefront footprint at the SLM generated by a Gaussian wavefront---with arbitrary rotation and position---illuminated by a plane wave.}}
    \label{fig:diagram}
\end{figure}

\label{subsec:gaussian-wave-splatting} 
In this section, we derive the closed-form solution for the wavefront contribution of a 2D Gaussian to the SLM. To focus on the relationship between the 2D Gaussian's position, rotation, and the SLM plane, we assume that the SLM lies centered on the $xy-$plane. The position, rotation, and scale of the 2D Gaussian, discussed in the following section, are defined in this \textit{hologram space} \revision{(see Fig.~\ref{fig:diagram})}. To render holograms of the optimized 2DGS model defined in \textit{world space} from any given viewpoint, we can simply rotate and project the 2D Gaussians with the corresponding \textit{view transform} and \textit{\revision{projective mapping}} as we would in the traditional graphics pipeline to transform the off-the-shelf Gaussians to the \textit{ray space}\revision{~\cite{zwicker2001ewa}}. The Gaussians in ray space are sequentially transformed into hologram space, based on the hardware setup geometry, including the relative position between eyepiece, SLM, and \revision{the target volume, as illustrated in Fig.~\ref{fig:diagram}. We refer interested readers to the Supplementary Materials for the complete holographics pipeline used to render holograms from optimized Gaussian splatting models.}

To calculate the wavefront of the $\idx^{th}$ Gaussian $u_i(\coordinate)$ in hologram space, we begin with a zero-mean, unit-variance standard Gaussian $u_\canonicalspace$ defined in the \textit{canonical space} $\coordinate_\canonicalspace = (x_\canonicalspace, y_\canonicalspace, \revision{z_\canonicalspace = 0})$. We note that Gaussians have the neat property for CGHs that their Fourier transform, $\angularspectrum_\canonicalspace$, is also a zero-mean Gaussian
\begin{align}
u_\canonicalspace(\coordinate_\canonicalspace) =e^{-\frac{1}{2}(x_\canonicalspace^2 + y_\canonicalspace^2)} & \xleftrightarrow{\fourier} \angularspectrum_\canonicalspace(\mathbf{k}_\canonicalspace) = 2 \pi e^{-2\pi^2 \mathbf{k}_\canonicalspace^\mathsf{T} \mathbf{k}_\canonicalspace}.\label{eq:ref_gaussian}  
\end{align}

From this canonical Gaussian, our Gaussian in hologram space is defined by its center $\bm{\mu}_i \in \arr{3}$ and covariance matrix $\cov_i = \rot_i \sca_i \sca_i^\mathsf{T} \rot_i^\mathsf{T} \in \mat{3}$, which can be factorized into a scaling matrix $\sca_i \in \mat{3}$ and a rotation matrix $\rot_i = \rot_x\rot_y\rot_z \in \mat{3}$. Since we work with 2D Gaussian disks, we assume that the $z$-scales of all Gaussians are zero. With these parameters, any 2D Gaussian can be transformed from the canonical space to the hologram space as $\coordinate = \rot_i \sca_i \coordinate_\canonicalspace + \bm{\mu}_i$. We are interested in how such a Gaussian contributes to the SLM plane in hologram space \revision{(see Fig.~\ref{fig:diagram})}, where they will be composited using alpha wave blending. To this end, we derive the closed-form solution for the angular spectrum of 2D Gaussian primitives $\angularspectrum_\idx (\mathbf{k})$ in hologram space. Note that, while we proceed with the angular spectrum in the following equations, this is equivalent to transformations between wavefronts, which can also be obtained using the Fourier transform.
First, we transform the wavefront in the canonical space to the \textit{object space}, which lies on the same plane as the canonical space and is defined by a 2D affine transform using the scaling matrix $ \coordinate_\objectspace = \sca_i \coordinate_\canonicalspace$. From the affine theorem for two-dimensional Fourier transforms~\cite{bracewell1993affine}, we obtain the angular spectrum in object space $\angularspectrum_\objectspace(\mathbf{k}_\objectspace)$ as
\begin{align}    
u_\objectspace(\coordinate_\objectspace) = u_\canonicalspace(\sca_\idx^{-1} \coordinate_\objectspace) & \xleftrightarrow{\fourier} \angularspectrum_\objectspace(\mathbf{k}_\objectspace) = \det(\mathbf{\sca} )\,\angularspectrum_\canonicalspace(\mathbf{(\sca}_i^{-1})^{-\mathsf{T}} \mathbf{k}_\objectspace) .\label{eq:affine_2d}  
\end{align} 

Next, we propagate the wavefront in the object space to the \textit{hologram space}, which is translated and tilted relative to the object space: $\coordinate = \rot_\idx \coordinate_\objectspace + \centergaussian_\idx$. We observe that translation by $\centergaussian_\idx$ in the spatial domain is equivalent to applying a phase ramp $e^{j\mathbf{k} \cdot \centergaussian_\idx}$ in the Fourier domain, as observed in Eq.~\ref{eq:asm}. The propagation of wavefronts between tilted planes can be described by the rotational transform of the wavefield~\cite{matsushima2003fast}. The main insight behind this approach is that the wave vector also follows the same rotation, meaning a plane wave of wavenumber $\mathbf{k}$ in hologram space corresponds to a plane wave of wavenumber $\rot_\idx^{-1}\mathbf{k}$ in object space. Thus, the angular spectrum in hologram space can be obtained by remapping the angular spectrum in the local space, $\mathbf{k}_\objectspace = \rot_\idx^{-1} \mathbf{k}$:
\begin{align}
    \angularspectrum_i(\mathbf{k}) &=  \textrm{det}( \mathbf{J}) \textrm{det}(\mathbf{S}_\idx) \angularspectrum_\canonicalspace \left(\mathbf{S}_\idx \rot_\idx^{-1}\mathbf{k} \right) e^{j  \mathbf{k}\cdot \mean_\idx} \label{eq:gaussian_to_hologram_new} \\
    &= 2 \pi \textrm{det}( \mathbf{J}) \textrm{det}(\mathbf{S}_\idx)  e^{-2\pi^2 \mathbf{k}^\mathsf{T} \cov_i \mathbf{k}}e^{j  \mathbf{k}\cdot \mean_\idx} \label{eq:exactgws} . 
\end{align} 
Here, $\textrm{det} (\mathbf{J}) = \frac{k_{\objectspace_z}}{k_z}$ is the Jacobian determinant of the transformation of Fourier coordinates. We dub Eq.~\ref{eq:exactgws} the \textit{\revision{G}aussian \revision{W}ave \revision{S}platting} equation, as it describes the exact angular spectrum of the splatted wavefront at the SLM plane. The wavefront profile at the SLM $u_\idx(\coordinate)$ can be obtained by applying the inverse Fourier transform to $\angularspectrum_i(\mathbf{k})$. \revision{Similarly, the wavefront profile at depth \( z_\idx \), \( u_\idx(\coordinate; z_\idx) \), can be obtained by setting the depth \( z \) component in \( \mean_\idx \) to zero in Eq.~\ref{eq:exactgws}, if we want to apply the attenuation factor \( T_\idx \) before splatting the wavefront onto the SLM plane.} Note that Eq.~\ref{eq:gaussian_to_hologram_new} is not limited to 2D Gaussians but can be generalized to any primitives that have a closed-form solution for the angular spectrum in canonical space~\cite{kim2008mathematical, ahrenberg2008computer}.



This closed-form solution offers various benefits, including an efficient CUDA implementation in the Fourier domain as well as support for occlusions and view-dependent effects. Due to the bandwidth of the SLM, we assume all Gaussians are illuminated by a reference plane wave propagating in the direction of $\mathbf{k}_r$. This involves a shift in the angular spectrum by $\mathbf{k}_r$ in the Fourier domain.




\subsection{Fast CUDA Implementation of Approximated Gaussian Wave Splatting using Order-Invariant Transparency}
\label{subsec:fast-alpha-wave-blending}

We can apply the Fourier transform to both sides of Eq. \ref{eq:gwsplat} and rewrite alpha wave blending in the Fourier domain:
\begin{align}
    \angularspectrum_{\textrm{SLM}} (\mathbf{k}) 
    & = \sum_{\idx}^\maxidx \mathbf{c}_\idx o_\idx\mathcal{F}\Big(  u_\idx(\coordinate)T_\idx\left(\coordinate\right)\Big)\mathcal{H}(\mathbf{k}, -z_\idx) \notag \\
    & = \sum_{\idx}^\maxidx \mathbf{c}_\idx o_\idx \Big(  \angularspectrum_\idx(\mathbf{k})\ast \mathcal{F}\Big(T_\idx\left(\coordinate\right)\Big)\Big)\mathcal{H}(\mathbf{k}, -z_\idx),     
    \label{eq:gwsplat_fourier}
\end{align} 
where $\ast$ denotes the convolution operator. The formulation of Eq. \ref{eq:gwsplat_fourier} is equivalent to the Gaussian splatting volume rendering equation in the spatial domain defined in Eq. \ref{eq:gs_volume_rendering}, the only difference being that Eq. \ref{eq:gwsplat_fourier} is defined for each frequency location $\mathbf{k}$ instead of pixel location $\coordinate$. However, unlike its spatial-domain counterpart, it is difficult to  utilize per-pixel parallelization to speed up the calculation of Eq. \ref{eq:gwsplat_fourier} due to the presence of a convolution, which cannot be efficiently calculated for each frequency location $\mathbf{k}$. This is an inherent computational challenge in CGH where a single pixel on a plane affects all pixels within its subhologram on other planes.

To fully utilize CUDA parallelization for real-time CGH calculation, we design an approximated version of alpha wave blending to allow for efficient per-frequency parallelization. Specifically, we discard the transparency term, i.e., $T_i(\coordinate) \approx 1$, in Eq. \ref{eq:gwsplat_fourier} and therefore remove the costly convolution operation: 
\begin{align}
    \angularspectrum_{\textrm{SLM}} (\mathbf{k}) 
    & \approx \sum_{\idx}^\maxidx \mathbf{c}_\idx o_\idx\angularspectrum_\idx(\mathbf{k})\mathcal{H}(\mathbf{k}, -z_\idx).       
    \label{eq:gwsplat_fourier_approx}
\end{align} 

It is now straightforward to parallelize the computation of this approximated CGH model since each term in Eq. \ref{eq:gwsplat_fourier_approx} can be independently calculated per-frequency location $\mathbf{k}$. Therefore, we implement custom CUDA kernels to perform efficient CGH calculation using this approximated alpha wave blending model.

The spatial-domain counterpart of Eq. \ref{eq:gwsplat_fourier_approx} can be described as:
\begin{align}
    \mathbf{c} \approx \sum_{i=1}^N c_i \alpha_i = \sum_{i=1}^N c_i \mathcal{G}_i o_i,
\label{eq:OIT}
\end{align}
where $\mathbf{c}$ is the final rendered pixel color, $c_i, \alpha_i, o_i \in \mathbb{R}$ are the color, alpha, and opacity of the $i^\text{th}$ Gaussian, respectively, and $\mathcal{G}_i$ is the splatted 2D Gaussian value evaluated at the location of the pixel currently being rendered. Note that the transparency term is omitted compared to the original volume rendering equation (Eq.~\ref{eq:gs_volume_rendering}). Interestingly, this image formation model is known as order-invariant transparency (OIT) in traditional computer graphics \cite{mcguire2013weighted}, and is mainly designed for speeding up alpha compositing operations. As the name suggests, OIT ignores the ordering of primitives, discards the front-to-back alpha compositing procedure in Eq. \ref{eq:gs_volume_rendering}, and represents the final composited color simply as a weighted sum of the Gaussian colors and opacities. 

Since this image formation model is different from the alpha compositing model in 2DGS, we cannot directly apply the efficient alpha wave blending method to the optimized 2DGS models. Therefore, we replace the volume rendering equation (Eq. \ref{eq:gs_volume_rendering}) used by 2DGS with Eq. \ref{eq:OIT}, and refer to this 2DGS variant as \textit{2DGS-OIT}. We optimize 2DGS-OIT models for the efficient version of our alpha wave blending algorithm, implement custom CUDA kernels to perform forward and backward passes of Eq. \ref{eq:OIT}, and integrate them with existing 2DGS libraries. Please refer to the supplemental materials for mathematical derivations and implementation details.

\begin{figure}
    \centering
    \includegraphics[width=1.0\linewidth]{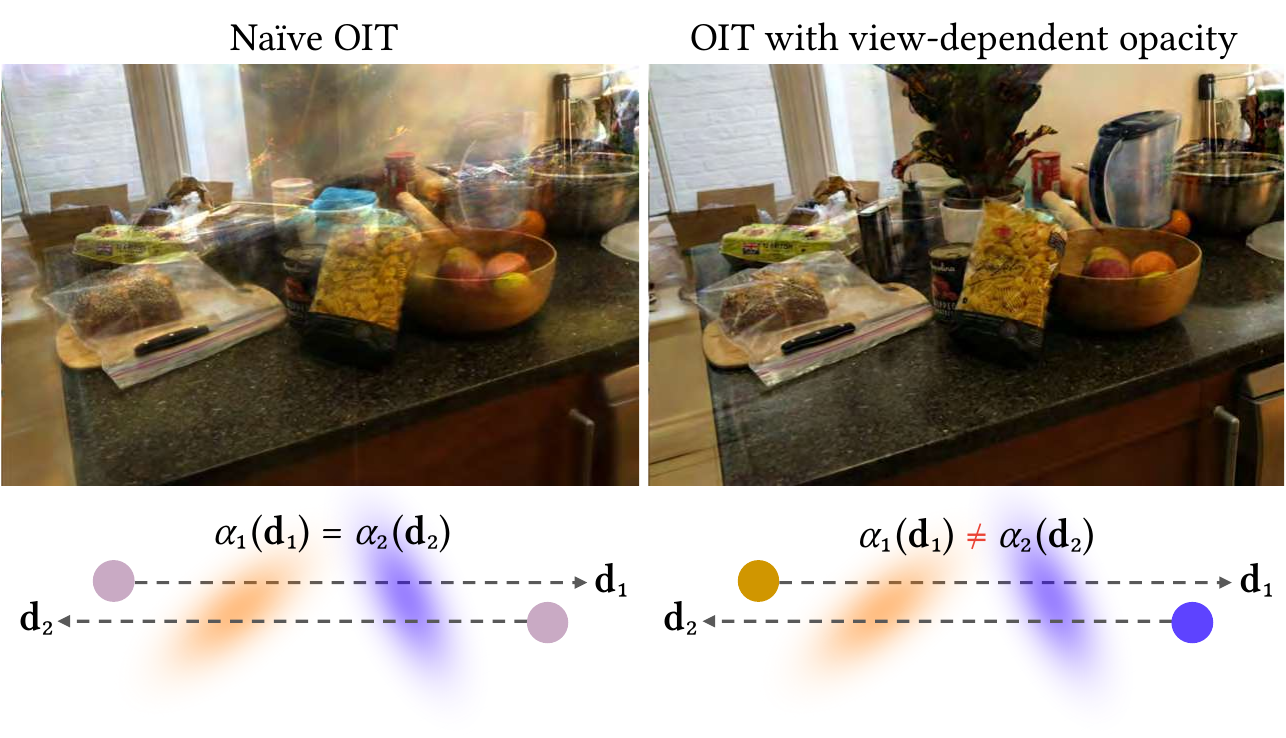}
    \caption{
    \textbf{Order-invariant transparency with view-dependent opacity.} 
    We ablate novel-view synthesis of our 2DGS-OIT model using view-agnostic opacity (clamping opacity to $(0, 1)$ using a sigmoid function as in naive Gaussian splatting) and view-dependent opacity represented with spherical harmonics. The na\revision{\"i}ve OIT model suffers from severe color leakage or even complete disappearance of objects. Conversely, the OIT model with view-dependent opacity greatly mitigates such artifacts, achieving image quality comparable to 2DGS.}
    \label{fig:OIT_reconstruction}
\end{figure}

A problem with Eq.~\ref{eq:OIT} is that, although easily parallelizable in the Fourier domain, the absence of ordering information among primitives prevents the Gaussian splatting model from accurately representing occlusion. This is illustrated in Fig. \ref{fig:OIT_reconstruction}, where the rendered colors from two opposite viewpoints along the same line of sight are exactly the same, yet intuitively the rendered color from each viewpoint should be more similar to the color of the closer Gaussian. In order to re-introduce occlusion modeling and ordering information back to Eq. \ref{eq:OIT}, we leverage \textit{view-dependent opacity} by defining the opacity of each Gaussian as a set of spherical harmonics (SH) coefficients $\mathbf{o}_\idx \in \mathbb{R}^K$: $\alpha_\idx = \mathcal{G}_\idx \cdot \text{SH}(\mathbf{o}_\idx, \mathbf{d})$, where $K$ is the number of SH coefficients, $\text{SH}$ is the linear spherical harmonics operator and $\mathbf{d}\in \mathbb{R}^3$ is the viewing direction. This leads to significant improvements in 3D reconstruction quality, as shown in Fig.~\ref{fig:OIT_reconstruction}.


\if 
Hence, to fully utilize CUDA parallelization for real-time CGH calculation, we enforce additional constraints on 2DGS optimization such that the final optimized 2DGS models are well-suited for CUDA-parallelizable CGH calculation. Specifically, we introduce an additional opacity regularization term (Eq. \ref{eq:opacity_reg}) in 2DGS to enforce the Gaussians to become either fully transparent or opaque, and clamp the opacity values to 0 or 1 after optimization. This regularization term and its variants are also widely used in prior surface reconstruction from Gaussians works ~\cite{Dai2024GaussianSurfels, jiang2024gaussianshader, xu2024texturegs, guedon2024sugar}, where low-opacity Gaussians are pruned and opaque Gaussians are optimized to be well-aligned with object surfaces for efficient mesh extraction.

\begin{align}
    \mathcal{L}_\text{o} = \frac{1}{N} \sum_{i=1}^N (\ln(o_i) + \ln(1 - o_i))
    \label{eq:opacity_reg}
\end{align}
Here, $N$ is the number of Gaussians and $o_i \in (0, 1) $ is the opacity of the $i^\text{th}$ Gaussian.

When calculating holograms from an optimized 2DGS model and a specific viewpoint, we first quantize the Gaussian opacities to zero and one and prune the transparent Gaussians. Next, we perform back-face culling to remove Gaussians on the opposite side of the viewpoint using normals of the 2D Gaussian disks and the viewing direction. Finally, we simply add up the wavefront contributions of the remaining 2D Gaussians at the hologram plane, which can be easily parallelized over Gaussians and pixels in the Fourier domain based on Eq.~\ref{eq:gwsplatting}. Please refer to the supplemental materials for more details on the CUDA kernel implementation and 2DGS optimization.
\fi


\section{Experiments}
\label{sec:results}

\subsection{Implementation Details}
\noindentparagraph{\bf{Datasets and 3D Scene Representations. }}
We create 3D holograms of selected scenes from the synthetic Blender dataset \cite{mildenhall2020nerf} and the MipNeRF-360 dataset \cite{barron2022mipnerf360} using different 3D scene representations, including 2D Gaussians, meshes, and point clouds.

We use the open-source Gaussian splatting software library \texttt{gsplat} \cite{ye2024gsplat} to optimize naive 2DGS models used for our \textit{exact} Gaussian Wave Splatting (Sec~\ref{subsec:gaussian-wave-splatting}). Additionally, we implement custom kernels and integrate them into \texttt{gsplat} to optimize our 2DGS-OIT models for the \textit{fast} variant of GWS (Sec~\ref{subsec:fast-alpha-wave-blending}). To flexibly control the final number of optimized Gaussians, we leverage the 3DGS-MCMC densification strategy \cite{kheradmand20243dgsmcmc} and cap the maximum number of Gaussians.

To be able to apply polygon-based CGH on meshes, the meshes are required to have per-face colors since it is unclear how to apply texture mapping in polygon-based CGH.  However, a disproportionate number of triangles is required to reconstruct detailed appearance when using per-face colors, therefore almost all high-quality mesh models do not use per-face colors and leverage texture mapping. Therefore, we start from optimized textured mesh models released by the NeRF2Mesh \cite{tang2022nerf2mesh} authors and calculate per-face colors by averaging the per-vertex colors to acquire the textured meshes that are applicable for polygon-based CGH. 

For point cloud representations, we used provided SfM points for the MipNeRF-360 scenes and points uniformly sampled on the NeRF2Mesh meshes for the Blender scenes.

\noindentparagraph{\bf{Algorithm Implementations.}} We implement GWS and all baseline methods in PyTorch, and additionally implement fast GWS in CUDA. Since most prior primitive-based CGH methods lack reliable open-source reference implementations, one of the major goals of this project is to create a unified primitive-based CGH algorithms development framework that allows for hologram generation from the various 3D representations, such as point clouds, meshes, and Gaussian splats. To this end, we create the \href{https://github.com/computational-imaging/hsplat}{\texttt{hsplat} PyTorch library} for primitive-based CGH. Please refer to the Supplementary Materials for pseudocode and implementation details.

\begin{figure*}[ht!]
    \centering
    \includegraphics[width=1\textwidth]{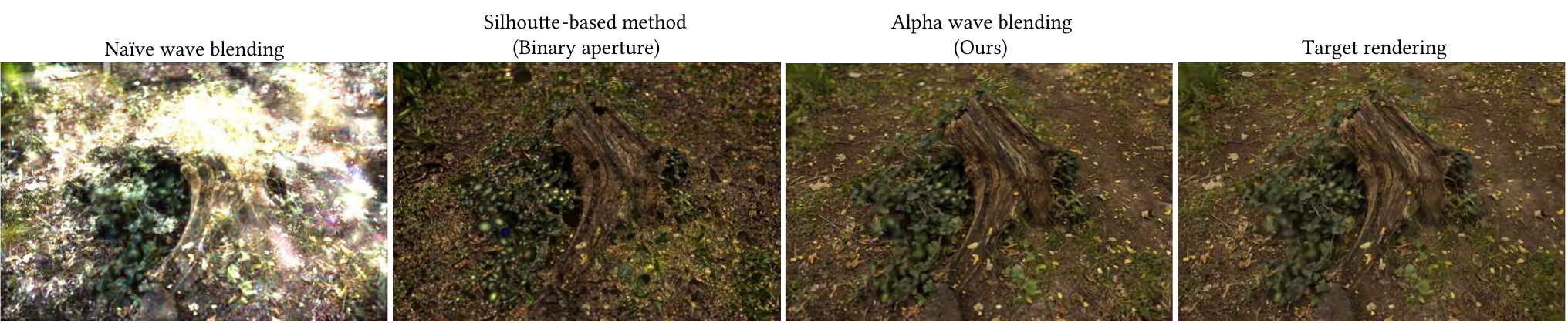}
    \caption{\textbf{\revision{Simulated comparisons} on wavefront blending \revision{strategies}.} We ablate wavefront blending strategies to reconstruct a scene represented by a collection of Gaussians \revision{in simulation}. Na\"ively adding up all the Gaussians without considering occlusion leads to oversaturated colors (\textit{left}). State-of-the-art occlusion handling techniques ~\cite{matsushima2014silhouette} employ a binary aperture to attenuate wavefront contributions from the rear based on the opacity of the front object, and do not appropriately handle smooth Gaussian falloffs and alpha blending (\textit{center left}). Our alpha wave blending method (\textit{center right}) performs exact alpha blending and faithfully reconstructs the 3D scene. On the right, we show the target image rendered with ray-based alpha blending.}  \label{fig:blending}
\end{figure*}

\begin{table}[ht!]
   \footnotesize
   \centering
    \captionsetup{aboveskip=0pt, belowskip=3pt}
   \caption{\textbf{Quantitative performance of different CGH algorithms. } We evaluate the image quality of simulated holograms generated using different CGH methods in terms of PSNR $(\uparrow)$ / SSIM $(\uparrow)$ / LPIPS $(\downarrow)$. Our methods achieve 5--11dB PSNR improvement over baseline methods.}
   \renewcommand{\arraystretch}{1.2} 
   \begin{tabular}{@{}P{0.48\linewidth}P{0.22 \linewidth}P{0.22 \linewidth}@{}}
   \toprule
   \makecell[l]{Method - 3D representation} &
   Blender &
   Mip-NeRF 360 \\
   \midrule
   \makecell[l]{\scalebox{0.95}{\citet{chen2009computer} - Point Cloud}} & 18.34 / 0.77 / 0.20 & 9.45 / 0.074 / 0.70 \\
   \makecell[l]{\scalebox{0.95}{\citet{matsushima2014silhouette} - Mesh}} & 20.79 / 0.81 / 0.18 & 15.77 / 0.27 / 0.65 \\
   \makecell[l]{Ours (fast) - Gaussians} & 26.55 / 
   \textbf{0.92} / \textbf{0.10} & 20.70 / \textbf{0.59} / 0.42 \\
   \makecell[l]{Ours (exact) - Gaussians} & \textbf{27.30} / 0.91 / 0.11 & \textbf{22.37} / 0.58 / \textbf{0.40}\\
   \bottomrule
   \end{tabular}

\label{tab:quan_metrics}
\end{table}

\noindentparagraph{\bf{Experimental Setup.}} \revision{To experimentally validate our algorithms, we build a holographic display benchtop prototype using a FISBA READYBeam™ Bio2 fiber-coupled module and a Holoeye PLUTO phase-only liquid crystal on silicon (LCoS) SLM. Color images are captured with separate exposures for each wavelength and combined in post-processing. In our setup, the depth range in hologram space is approximately 1cm, corresponding to a total depth range in view space from optical infinity to approximately 0.5m. To address hardware imperfections, we account for laser beam intensity variations and perform spatially varying lookup table calibration~\cite{choi2021neural, shi2022end}.} We refer readers to the Supplementary Materials for details on the experimental setup and its calibration procedures.


\noindentparagraph{\bf{\revision{Displaying GWS holograms with Phase-only SLMs.}}}
\revision{
Our GWS algorithms compute a complex-valued hologram $u_{\textrm{SLM}}$, similar to other primitive-based methods. To encode the complex-valued wavefront splatted onto the SLM plane into a phase-only pattern, we use the double phase--amplitude coding method (DPAC) described in the literature~\cite{hsueh1978computer, maimone2017holographic}. DPAC directly encodes a complex-valued wavefront $ae^{j\phi}$ into two phase values, $P_\pm = a \pm cos^{-1} \phi,$ and interlaces $P_+$ and $P_-$ like a checkerboard pattern~\cite{maimone2017holographic}. 
}



\subsection{Simulation and Experimental Results}
\noindentparagraph{\bf{\revision{Occlusion Handling.}}}

In Gaussian splatting, occlusion is handled via alpha compositing, and we evaluate occlusion handling for the wave counterpart in Fig.~\ref{fig:blending}. The left column shows simulated reconstructions using holograms generated without occlusion handling, where all rear Gaussians contribute to the hologram. The second column presents results with a silhouette-based method emulating alpha blending~\cite{matsushima2020introduction}, which attenuates rear wavefront contributions with a translucent mask of the front object. Our alpha wave blending (third column) uniquely achieves accurate scene reconstruction from a collection of Gaussians, \revision{which is shown in the rightmost column}.

\noindentparagraph{\bf{Baseline Comparisons with Simulations and Experiment\revision{s.}}} We compare our exact GWS algorithm and our fast variant of GWS with various primitive-based CGH methods, including point-based methods \cite{chen2009computer} and polygon-based methods \cite{matsushima2014silhouette, matsushima2009extremely}. Please refer to the supplemental materials for comparisons with a more extensive set of baselines.

\begin{figure*}[ht!]
    \centering
    \includegraphics[width=1.0\textwidth]{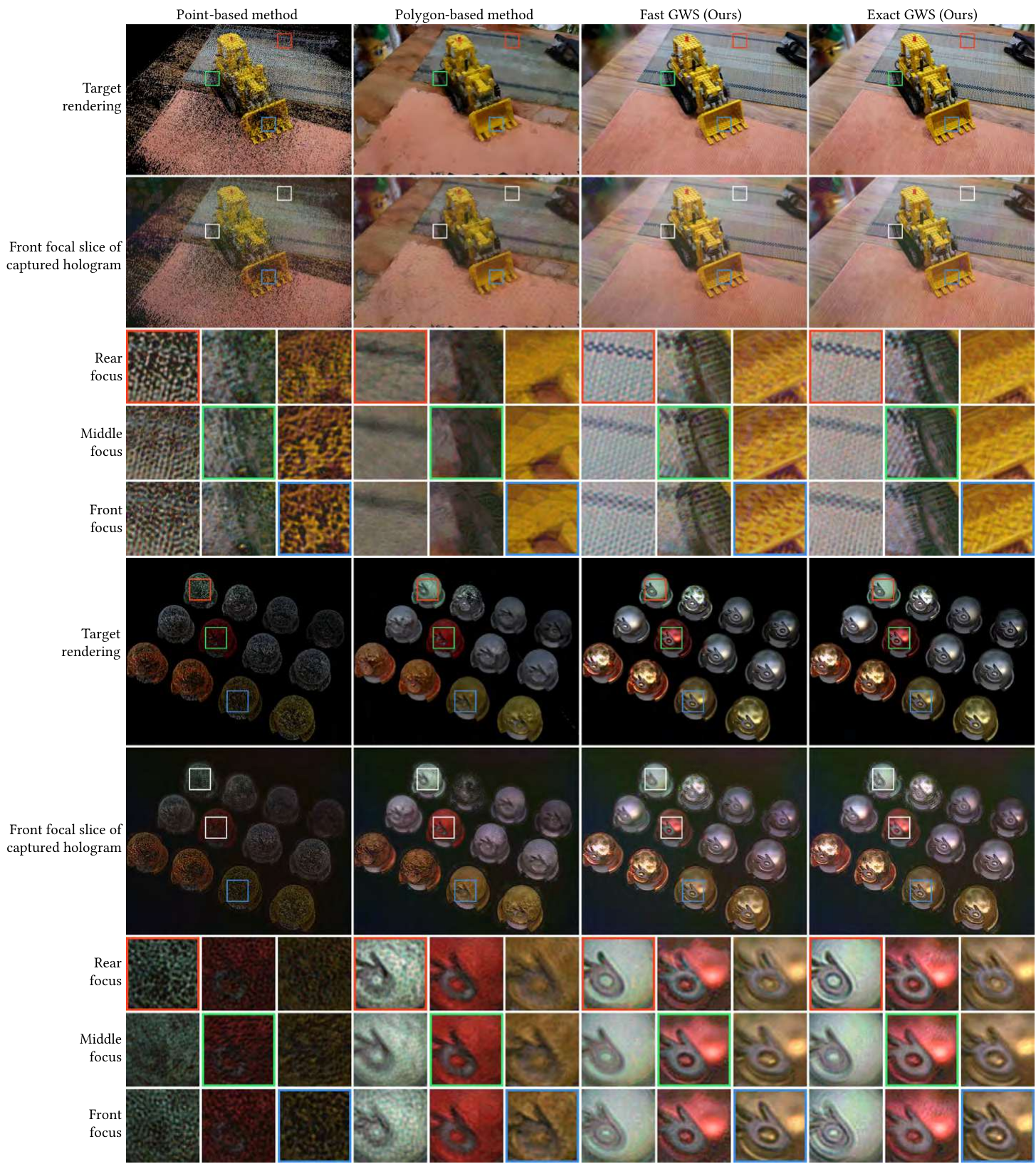}
    \caption{\textbf{Experimentally captured focal stacks of holograms generated using different primitive-based CGH algorithms. } For each CGH baseline, we show the target rendering of its input 3D representation using a pinhole camera model (\textit{top row}). Sparse point-cloud representations used in point-based methods~\cite{chen2009computer} inevitably lead to holes in the rendered target image. Polygon-based methods~\cite{matsushima2009extremely} produce overly smooth results unless an excessive number of polygons are used. Our Gaussian Wave Splatting methods achieve the best results with sharp in-focus regions and more accurate blur in defocus regions. Please refer to the supplementary materials for extensive baseline comparisons.}
    \label{fig:baseline_simulation}
\end{figure*}

We show experimentally captured 3D focal stacks of holograms generated from different 3D representations in Fig. \ref{fig:baseline_simulation}. Point clouds are inherently sparse, thus visible holes are present in both the target rendering and the synthesized 3D hologram and the point sizes need to be heuristically tuned to achieve a merely passing image quality. Meshes with per-face colors lead to overly smooth colors and cannot reconstruct high-frequency details unless paired with a texture map or using a disproportionate number of primitives. However, texture mapping is infeasible in the context of polygon-based CGH and a large number of primitives leads to infeasible processing times. Our GWS method achieves superior 3D hologram quality with sharp reconstruction in focused regions as shown in the insets with colored borders. Furthermore, GWS delivers superior image quality despite using an equal or reduced number of geometric primitives compared to the mesh models.

\begin{figure*}[ht!]
    \centering
    \includegraphics[width=1.0\textwidth]{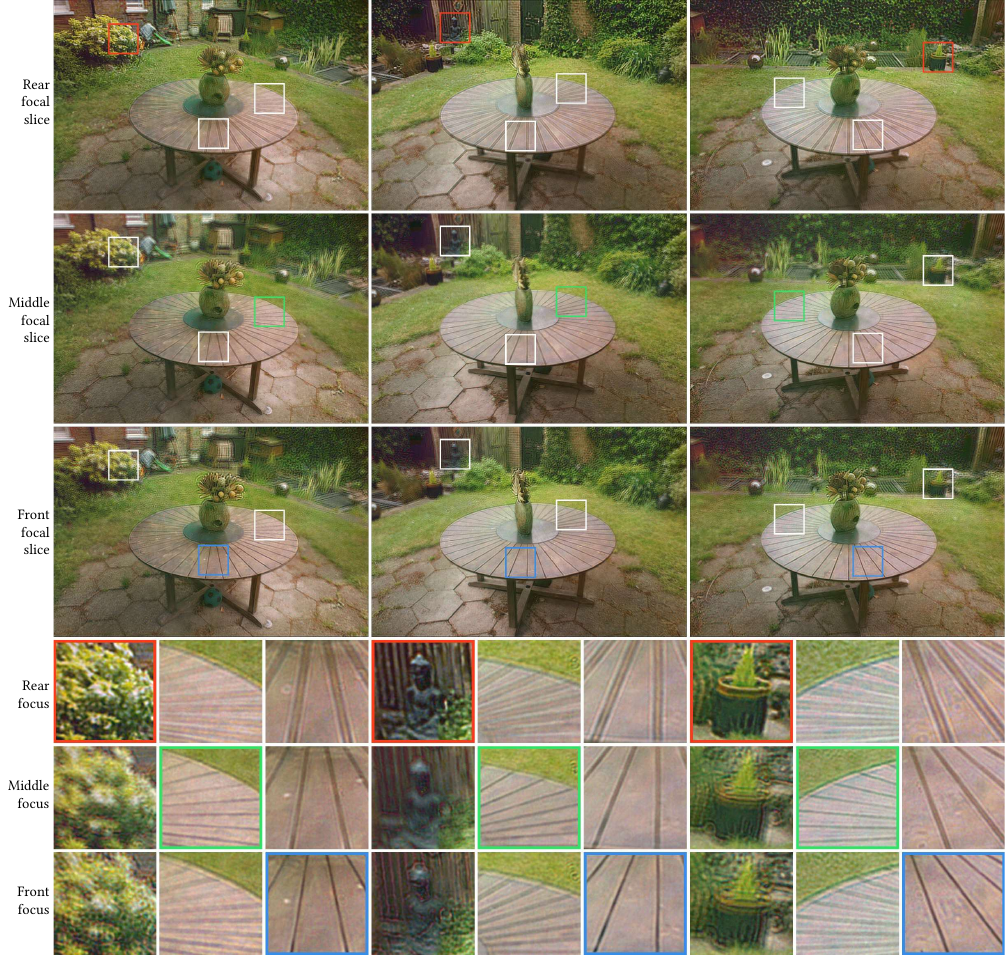}
    \caption{\textbf{Experimentally captured focal stacks of exact GWS holograms from different novel viewpoints.} We generate holograms from \revision{multiple} novel viewpoints \revision{using our exact GWS method applied to a single scene representation} and capture the corresponding 3D focal stacks at each viewpoint. \revision{The results show that }GWS achieves accurate 3D refocusing effects \revision{across} all viewpoints, demonstrating the robustness of our method. \revision{See the supplementary video for an additional visualization.}}
    \label{fig:nvs_fs_captured}
\end{figure*}

We additionally show experimentally captured focal stacks of exact GWS holograms generated from various novel viewpoints of an optimized 2DGS model in Fig. \ref{fig:nvs_fs_captured}. GWS achieves accurate refocusing effects at \textit{every} viewpoint, demonstrating the robustness of our method. Please refer to the supplemental video for the full circular trajectory novel-view rendering and captured 3D focal stacks at each novel viewpoint. \revision{To generate the hologram for each individual frame, we transform the Gaussians into the hologram space using the view transform corresponding to each video frame, following the \textit{holographics} pipeline described in the supplemental materials.}

Finally, we report quantitative results of simulated baseline comparisons in terms of various image quality metrics in Table~\ref{tab:quan_metrics}. We evaluate the in-focus image quality by merging simulated focal stacks rendered from the generated holograms into an all-in-focus image using RGB-D masks computed from depth maps rendered from optimized 2DGS models, and compare this all-in-focus image with ground-truth images from the training dataset. Both the fast and exact version of our Gaussian Wave Splatting algorithm consistently outperform other primitive-based CGH algorithms in terms of all image quality metrics. \revision{Note that since Gaussian splatting uses a weighted photometric loss that prioritizes pixel-wise reconstruction (PSNR) over structural similarity and perceptual quality, fast GWS achieves slightly better performance in SSIM and LPIPS compared to exact GWS.}



\noindentparagraph{\bf{Exact and Fast Gaussian Wave Splatting.}} In the rightmost two columns of Fig. \ref{fig:baseline_simulation}, we show experimentally captured focal stacks of holograms generated using our GWS algorithms. The experimentally captured results of the fast variant suffers from slight contrast reduction and background color leakage artifacts in some regions due to the lack of accurate order information among Gaussians, although these artifacts are already greatly mitigated with the view-dependent opacities described in Section 4 as shown in the target rendering. The exact version achieves better results due to the accurate modeling of occlusions through alpha wave blending at the cost of longer CGH calculation time.

\begin{table}[t!]
   \footnotesize
   \centering
    \captionsetup{aboveskip=0pt, belowskip=3pt}
   \caption{\textbf{Runtime comparison of primitive-based CGH.} We report the runtime of other primitive-based algorithms and both variants of GWS. Since prior methods lack open-source implementations, we directly borrow the reported runtime from the corresponding papers, and evaluate GWS using the same target resolution and number of primitives specified in the papers. Given the same target hologram resolution and number of primitives, fast GWS achieves a runtime comparable to prior methods while greatly exceeding their image quality with a 30$\times$ speedup compared to exact GWS. }
   \renewcommand{\arraystretch}{1.2} 
    \resizebox{\columnwidth}{!}{
   \begin{tabular}{@{}l l l l@{}}
   \toprule
   \makecell[l]{Method} & \makecell[l]{Resolution} &
   \makecell[l]{Number of primitives} &
   \makecell[l]{Runtime} \\
   \midrule
   \citet{chen2009computer} & 1024 $\times$ 1280 & 15k points & 2.47s \\
   \citet{matsushima2014silhouette} & 2000 $\times$  2000 & 15k triangles & 13.5s \\
   \midrule
   Ours (fast) & 1024 $\times$  1280 & 15k Gaussians & 6.84s \\
    & 2000 $\times$  2000 & 15k Gaussians & 18.06s \\
    \midrule
   Ours (exact) & 1024 $\times$  1280 & 15k Gaussians & 180s \\
    & 2000 $\times$  2000 & 15k Gaussians & 470s \\
   \bottomrule
   \end{tabular}
   }

\label{tab:runtime_baselines}
\end{table}

We show runtime comparisons of exact and fast GWS and prior primitive-based CGH methods in Table \ref{tab:runtime_baselines}. Fast GWS achieves a nearly $30\times$ speedup over exact GWS at the cost of a slight drop in image quality, and achieves a runtime comparable to prior methods while greatly exceeding their image quality. Please refer to the supplemental materials for a more detailed runtime analysis.

\noindentparagraph{\bf{Partially Coherent GWS For View-Dependent Effects}.}

\begin{figure}[t]
    \centering
    \includegraphics[width=1.0\columnwidth]{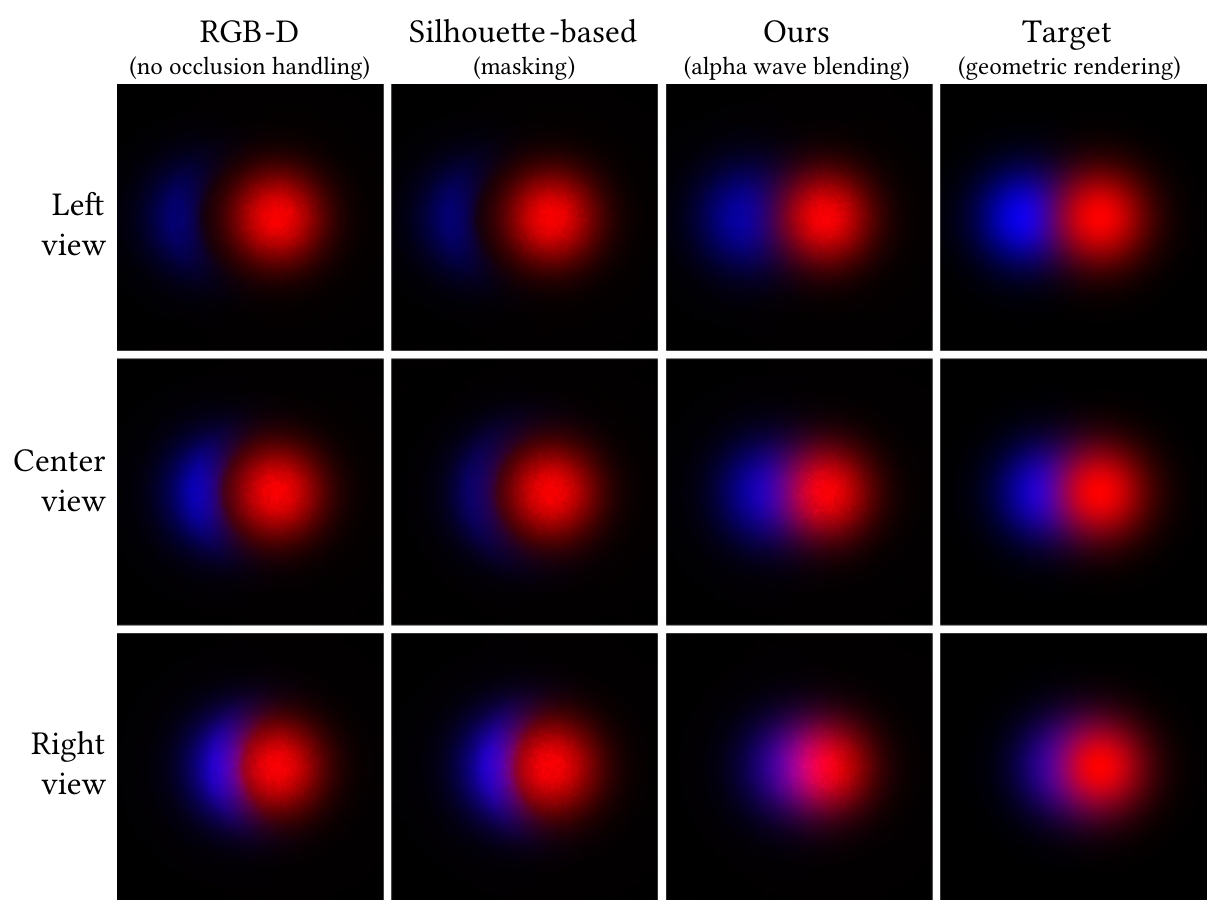}
    \caption{\textbf{Simulated comparisons on parallax.} RGB-D data lacks information about objects behind occluders, resulting in blank regions when viewed from other viewpoints. The silhouette-based method~\cite{matsushima2014silhouette}, designed for binary apertures, masks the wavefronts from the rear, leading to inaccurate wavefront masking and blending of continuous Gaussian profiles. Our alpha wave blending method effectively handles the occlusion of Gaussians, naturally blends them according to their smooth falloffs from all viewpoints, and best matches the target shown on the right.}
    \label{fig:occlusion}
\end{figure}

\revision{
Directional effects in primitive-based CGHs have primarily been discussed in the context of representing the reflectance distribution of surfaces~\cite{cuypers2012reflectance, park2017recent}. Pioneering work has attempted to control the angular emission profile using subtriangles~\cite{kim2008mathematical} and appropriate phase profiles for them. Yeom et al. demonstrated that the desired reflectance distribution could be approximated as a convolution between the angular spectrum of the mesh and the angular distribution~\cite{askari2017occlusion, yeom2016calculation}. In our general, partially coherent GWS formulation, Eq.~\ref{eq:gaussian_to_hologram_new} can be adapted as follows:
\begin{align}
    \angularspectrum_{\textrm{SH}_l^m} ( \mathbf{k} ) = \textrm{det} ( \mathbf{J}) \textrm{det}(\mathbf{S}_\idx) \Big( \angularspectrum_\canonicalspace \left( \mathbf{S}_\idx \rot_\idx^{-1}\mathbf{k} \right) \ast_{\mathbf{k}} Y_l^m(\mathbf{k}) e^{j\phi (\mathbf{k})} \Big) e^{j  \mathbf{k} \cdot \mean_\idx}, \label{eq:convolution}
\end{align}
where $\ast_{\mathbf{k}}$ is the convolution operator in the spatial frequency domain and $Y_l^m(\mathbf{k}) e^{j\phi (\mathbf{k})}$ denotes the angular kernel, having the spherical harmonics function $Y_l^m$ of degree $l$ and order $m$ as its amplitude.}

Importantly, the phase profile of this kernel $\phi (\mathbf{k})$ plays a crucial role as it might affect the intensity profile in the spatial domain, and the degree of freedom to represent arbitrary spatio-angular light field is inherently limited by the rank-1 constraint of coherent wavefronts~\cite{zhang2011analysis}. To address this, one can rely on partial coherence, expecting the average of the amplitude of the kernel to be unity in the spatial domain to keep the amplitude profile of our Gaussians. To achieve this, we sample multiple phases for \revision{$\phi (\mathbf{k})$} from a uniform distribution in the range $[-\pi, \pi]$. Please refer to the Supplementary Materials for more details.

In Fig.~\ref{fig:occlusion}, we \revision{compare various CGH methods based on} the parallax effects observed from different viewpoints. To this end, we generate partially coherent wavefronts using \revision{24 frames} for a toy scene\revision{,} in which a red Gaussian is located at the zero-disparity plane and a blue Gaussian is positioned 2 cm behind it. As the viewpoint shifts from right to left, the blue Gaussian moves as expected, and our alpha wave blending, shown in the third column, faithfully represents the desirable occlusion effects corresponding to the viewpoints. The previous state-of-the-art occlusion handling technique, i.e., the silhouette-based method~\cite{matsushima2014silhouette}, and the basic point-based method using RGB-D~\cite{shi2021towards}, fail to model the occlusions for Gaussians due to incorrect masking or incomplete scene representation. Our method is the only one that takes continuous falloffs into account, uniquely reconstructing the blended purple color in the overlapped region (right view)\revision{, whereas other binary-mask-based methods, or data representations which do not support alpha maps, cannot represent such blends}. Both baselines require a heuristic threshold to discretize the depth map or mask the Gaussian~\cite{chen2009computer}, which we set to 0.1. In contrast, our alpha wave blending eliminates the need for such hyperparameter tuning while accurately alpha blending Gaussians from all perspectives. 

\section{Discussion}
In summary, we develop CGH algorithms that encode Gaussian-based scene representations into holograms. To this end, we derive an alpha wave blending formulation as well as analytical expressions that map Gaussian primitives to a complex-valued wavefronts. Unlike conventional Gaussian splatting, where the rendered image can be rasterized per pixel, the wave nature of holographic rendering increases computational cost, because each SLM pixel is affected by an extended sub-hologram region on another, parallel plane. However, we observe that our analytical expression can be efficiently implemented per frequency in the Fourier domain and implement custom CUDA kernels that accelerate this process, achieving a $30\times$ speedup. We validate our framework through simulations and experiments, demonstrating photorealistic image quality using off-the-shelf and widely used novel-view synthesis datasets. 

\noindentparagraph{\bf{\revision{Limitations and Future Work.}}}


\revision{Our work is a first step towards seamlessly connecting emerging neural graphics pipelines to CGH algorithms, leaving many exciting future directions to be explored.}

\revision{
First, our work focuses on encoding a collection of Gaussians into a complex wavefront while placing less emphasis on the phase distribution of the Gaussians, which governs defocus behavior. As a result, our method, except Fig.~\ref{fig:occlusion}, only ensures correct in-focus imagery, exhibiting unnatural defocus blur and a limited eyebox. In contrast, \textit{random-phase holograms} can maximize the bandwidth utilization of the SLM and have been demonstrated to achieve accurate occlusion and parallax, synthesize natural defocus blur, reduce the depth of field, and enlarge the eyebox~\cite{amako1995speckle, lohmann1967binary, yoo2021optimization}. Extending GWS to generate random-phase holograms is therefore an interesting future direction. We note that DPAC typically provides good image quality for \textit{smooth-phase holograms} with certain filtering in the Fourier domain. Thus, alternative phase encoding methods, such as a learned phase encoder~\cite{peng2020neural} that accounts for optical setup imperfections, or a gradient-descent-based method to encode complex GWS wavefronts into a phase-only pattern, need to be developed to potentially improve image quality, bandwidth utilization, and compactness~\cite{Gopakumar2021unfiltered, Chen2021complex}. }

\revision{
Next, although our CUDA implementation demonstrates a significant speedup over a na\"ive implementation, it does not yet achieve real-time hologram generation. Incorporating priors from closed-form solutions of other CGH methods could further accelerate runtimes~\cite{im2015accelerated, wang2023high}, for example, by reducing the region of interest for propagation~\cite{matsushima2014silhouette} or by developing efficient CUDA implementations that directly handle non-localized convolutions in the Fourier domain. Furthermore, integrating neural networks~\cite{shi2021towards}, leveraging custom hardware~\cite{Yamamoto2022, Makowski2022}, or designing end-to-end holographic camera systems~\cite{Wang2024liquidholo, Yu2023holocam} could also facilitate real-time CGH.}

\revision{
Finally, our current holographic display hardware supports only a limited \'etendue due to the small diffraction angle of the SLM~\cite{kuo2020high}. This limitation motivates our current approximation of view-dependent effects using only the zeroth order of spherical harmonics. Demonstrating higher-order spherical harmonics with partially coherent, \'etendue-expanded hardware systems represents another exciting direction for future work.
}

\noindentparagraph{\bf{\revision{Conclusion.}}}
By integrating emerging neural rendering techniques with holographic display technology, our algorithms pave the way toward efficient holographic rendering with photorealistic content.

\begin{acks}


We thank Seung-Woo Nam and Gun-Yeal Lee for fruitful discussions. Brian Chao and Jacqueline Yang are supported by Stanford Graduate Fellowships and the NSF Graduate Research Fellowship Program (GRFP). We also thank EssilorLuxottica and Meta for their support.

\end{acks}

\bibliographystyle{ACM-Reference-Format}
\bibliography{bibs}



\end{document}


\title{Gaussian Wave Splatting for Computer-Generated Holography --- Supplementary Material}
\begin{abstract}
    This document serves as supplementary material to the paper ``\textit{Gaussian Wave Splatting for Computer-Generated Holography}''. The sections of this supplementary material are organized as follows. 
    
    In Section~\ref{sec:holographics}, we elaborate on the transformations of between various spaces, starting from the space where the collection of optimized Gaussians reside, to the space where the SLM is defined. We then explain in detail how Gaussian Wave Splatting takes into account all the aforementioned transformation parameters and splats a single Gaussian onto the SLM plane. 
    
    In Section~\ref{sec:software}, we describe in detail how to splat a collection of Gaussians using different occlusion-handling techniques to generated a hologram. We also elaborate on the optimization of 2DGS and 2DGS-OIT models used for GWS.

    In Section ~\ref{sec:hardware}, we describe the experimental holographic display setup and its corresponding calibration process we used to validate our algorithm.

    In Section ~\ref{sec:simulation}, we provide extensive simulated experiments to validate our method, including extended quantitative and qualitative baseline comparisons, runtime analysis, additional discussion on view-dependent effects, and occlusion-handling techniques.

    In Section ~\ref{sec:experiments}, we show an extensive set of experimentally captured results. 

    In Section ~\ref{sec:misc}, we provide additional details of prior methods, including tilted plane propagation and deriving spectrums of triangles in polygon-based CGH.
    
\end{abstract}

\author{Suyeon Choi}
\email{suyeon@stanford.edu}
\authornote{denotes equal contribution.}
\orcid{0000-0001-9030-0960}
\affiliation{
  \institution{Stanford University}
  \country{USA}
}
\author{Brian Chao}
\email{brianchc@stanford.edu}
\orcid{0000-0002-4581-6850}
\authornotemark[1]
\affiliation{
  \institution{Stanford University}
  \country{USA}
}
\author{Jacqueline Yang}
\email{jyang01@stanford.edu}
\orcid{0009-0002-3101-3026}
\affiliation{
  \institution{Stanford University}
  \country{USA}
}
\author{Manu Gopakumar}
\email{manugopa@stanford.edu}
\orcid{0000-0001-9017-4968}
\affiliation{
  \institution{Stanford University}
  \country{USA}
}
\author{Gordon Wetzstein}
\email{gordon.wetzstein@stanford.edu}
\orcid{0000-0002-9243-6885}
\affiliation{
  \institution{Stanford University}
  \country{USA}
}


\graphicspath{ {../} }

\renewcommand{\shortauthors}{Choi, Chao et al.}

\begin{teaserfigure}
  \centering
	\includegraphics[width=0.85\columnwidth]{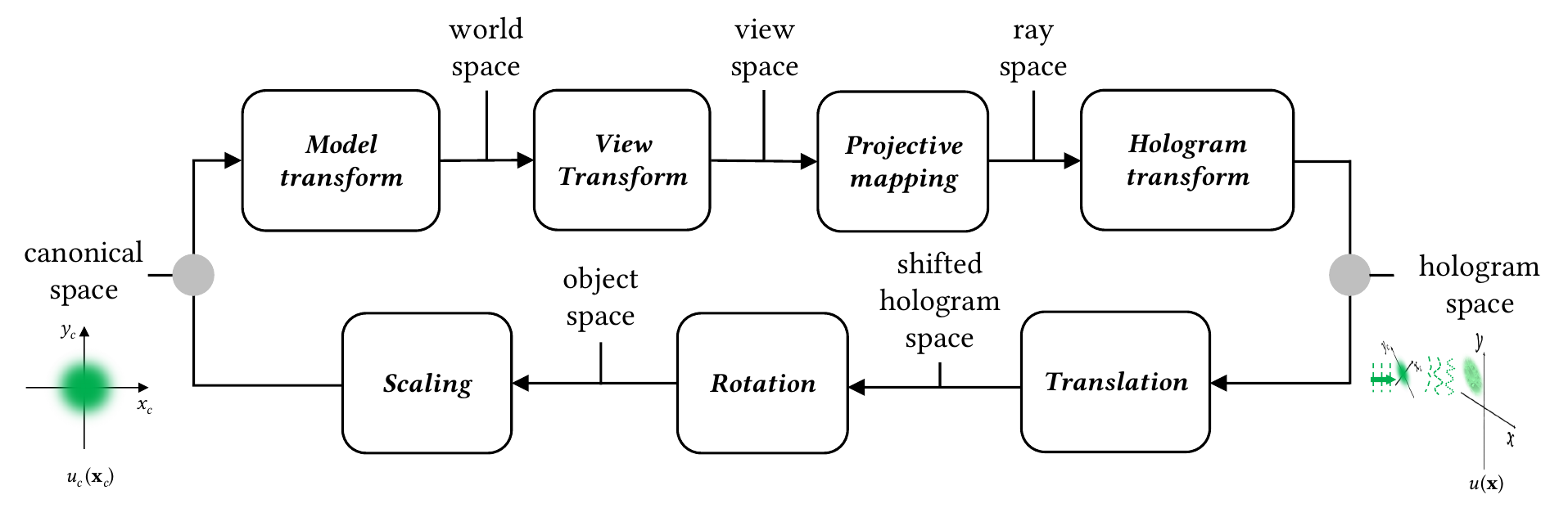}
   \caption{\textbf{The \textit{Holographics} pipeline. } We devise a graphics pipeline tailored for primitive-based CGH, dubbed the \textit{holographics} pipeline. This pipeline takes in a set of optimized primitives from off-the-shelf neural rendering frameworks as input and converts them into primitives suitable for CGH calculation through a sequence of coordinate transformations. Our Gaussian Wave Splatting algorithms then use the transformed primitives as input to compute the complex-valued wavefront of the hologram. }
  \label{fig:holographics}
\end{teaserfigure} 

\maketitle

\section{Holographics Pipeline}
\label{sec:holographics}
\label{sec:holographics}

Here, we outline our pipeline of primitive transformations, dubbed the \textit{holographics} pipeline (Fig. \ref{fig:holographics}). Just like the traditional graphics pipeline, we transform geometric primitives between various spaces in holographics, and we describe in detail these conversions in this section.
\subsection{Geometric pipeline}
We elaborate on the coordinate systems and transformations between various spaces in holographics such that our framework is compatible with the traditional computer graphics pipeline~\cite{foley1996computer, blinn1996jim}. While we primarily define these transformations for Gaussian primitives, it can easily be extended to other primitive-based CGHs~\cite{matsushima2003fast, chen2009computer}.

\paragraph{\textbf{Model transform}}
To begin with, we assume the optimized Gaussians from an off-the-shelf neural rendering framework~\cite{tancik2023nerfstudio, ye2024gsplat} are defined by a full 3D covariance matrix $\Sigma_\worldspace \in \mat{3}$ in \textit{world space} centered at a point $\centergaussian_\worldspace \in \arr{3}$:
\begin{align}
    \mathcal{G} ( \coordinate_w ) = e^{-\frac{1}{2} (\coordinate_\worldspace - \centergaussian_\worldspace)^\mathsf{T}\Sigma_\worldspace^{-1}(\coordinate_\worldspace - \centergaussian_\worldspace)}.
\end{align}
The covariance matrix can be factorized into a scaling matrix $S_\worldspace \in \mat{3}$ and a rotation matrix $R_\worldspace \in \mat{3}$ as $\Sigma_\worldspace = R_\worldspace S_\worldspace S_\worldspace^\mathbf{T} R_\worldspace^\mathbf{T} \in \mat{3}$. These Gaussians in world space are transformed from the \textit{canonical space} where we have a standard Gaussian centered at the origin as Eqs.~\ref{eq:canonicalgaussian} and ~\ref{eq:modeltransform} (see Fig.~\ref{fig:holographics}).
\begin{align}
    \mathcal{G} (\coordinate_\canonicalspace) &= e^{-\frac{1}{2} \coordinate_\canonicalspace^\mathsf{T} \coordinate_\canonicalspace}, \label{eq:canonicalgaussian}\\
     \coordinate_w &= 
        R_\worldspace S_\worldspace\coordinate_c + \centergaussian_w , \label{eq:modeltransform} \\ & \coordinate_c, \coordinate_w \in \mathbb{R}^{3 \times 1}, \nonumber
\end{align}
where $\coordinate_c$ and $\coordinate_w$ denote the coordinates in canonical space and world space, respectively. This can be thought as \textit{model transform}. 



\paragraph{\textbf{View transform and projective mapping}}

To render these Gaussians in world space from a specific camera viewpoint, we apply a view transformation $W \in \mat{3}$ that transforms Gaussians from world space into \textit{view space}, where the camera is assumed to be looking into the negative z direction. This is followed by a projective transform defined by camera intrinsics (focal length, principal point, and projection type such as orthographic, perspective, fisheye, etc.) that converts the view space into the \textit{ray space}, where the z-axis is parallel to the outgoing rays from all camera pixels. While this projective transform from view space to ray space is not affine, Zwicker et al. \cite{zwicker2001ewa} introduced a \textit{local} affine approximation of this mapping, allowing the covariance matrix $\Sigma_\viewspace \in \mat{2}$ and the center of the Gaussian $\centergaussian_\viewspace \in \arr{2}$ in ray space to be obtained by~\cite{kerbl3Dgaussians, zwicker2001ewa}:
\begin{align} 
\Sigma_\viewspace &= J W \Sigma_\worldspace W^\mathsf{T} J^\mathsf{T}, \\
    \centergaussian_\viewspace &= \projectivemapping( W \centergaussian_\worldspace), \label{eq:projectivemapping}
\end{align}
where $J \in \mathbb{R}^{2\times3}$ is the Jacobian of the affine approximation of the projective transformation $\projectivemapping$~\cite{zwicker2001ewa}.

Finally, we make a minor modification to the aforementioned projective transform by ``lifting'' the projected 2D covariance and center back to 3D. This is necessary because the depth values are required for CGH calculation. Specifically, we augment $\centergaussian_\viewspace$ with the the depth value of the Gaussian in view space and redefine it as:
\begin{align}
    \centergaussian_\viewspace = \begin{bmatrix}
        \projectivemapping( W \centergaussian_\worldspace) \\
        (W \centergaussian_\worldspace)_2
    \end{bmatrix}
\end{align}

To lift 2D covariance back to 3D, we perform Eigenvalue decomposition on the projected 2D covariance matrix to get the 2D rotation and scales $R_\viewspace' \in \mat{2}$ and $S_\viewspace' \in \mat{2}$:
\begin{align}
    J W \Sigma_\worldspace W^\mathsf{T} J^\mathsf{T} = R_\viewspace' S_\viewspace'^2 {R_\viewspace'}^\mathsf{T}
\end{align}

Then, we augment the 2D rotation matrix $R_\viewspace'$ with zero rotation around the x and y axes and set the z-scale of the Gaussian to be zero:
\begin{align}
    R_\viewspace & = \begin{bmatrix}
        R_\viewspace' & \mathbf{0} \\
        \mathbf{0} & 1
    \end{bmatrix} \\
    S_\viewspace & = \begin{bmatrix}
        S_\viewspace' \\
        0
    \end{bmatrix}
\end{align}
and finally acquire the lifted 3D covariance $\Sigma_\viewspace \in \mat{3}$:
\begin{align}
    \Sigma_\viewspace = R_\viewspace S_\viewspace S_\viewspace^\mathsf{T} R_\viewspace^\mathsf{T}
\end{align}

Note that ray space coordinates $\coordinate_\viewspace$ also can be obtained from the world coordinate using the same transformation as in Eq.~\ref{eq:projectivemapping}:
\begin{align}
    \coordinate_\viewspace = \projectivemapping (W\coordinate_\worldspace)
\end{align}

\paragraph{\textbf{Hologram transform}}
To convert Gaussians in \textit{ray space} to holograms, we must transform them into the \textit{hologram space}, which is determined by the geometry of the optical parameters of the holographic display. For example, we can remap the $z$-values of Gaussians in ray space to the desired depth range determined by the focal length of the eyepiece. We set the origin of the hologram coordinate system at the center of the SLM and align the $xy-$plane with the SLM plane. In other words, the SLM is positioned at $z=0$ plane. We refer to the transform from ray space to hologram space as \textit{hologram transform}. The covariance matrix $\Sigma$ and the center $\centergaussian$, coordinate vector $\coordinate$ in hologram space are given by a corresponding 3D affine transformation matrix between these spaces $T_{\viewspace \rightarrow h}$:
\begin{align}
    \Sigma &= T_{\viewspace \rightarrow h} \Sigma_\viewspace T_{\viewspace \rightarrow h}^\mathsf{T}, \\
    \centergaussian &= T_{\viewspace \rightarrow h} \centergaussian_\viewspace, \,\, \coordinate = T_{\viewspace \rightarrow h}  \coordinate_\viewspace
\end{align}

\paragraph{\textbf{Unified transformation}}
By combining these transformations as described in Fig.~\ref{fig:holographics}, the entire process from canonical space to hologram space results in the covariance matrix $\Sigma$ and the translation vector $\centergaussian$ in hologram space as follows:
\begin{align}
  \Sigma &= (T_{\viewspace \rightarrow h} J W R_\worldspace S_\worldspace) S_\worldspace^\mathsf{T} R_\worldspace^\mathsf{T} W^\mathsf{T} J^\mathsf{T} T_{\viewspace \rightarrow h}^\mathsf{T} \label{eq:combine}\\
  &= TT^\mathsf{T} \\
  &= RSS^\mathbf{T}R^\mathsf{T}, \\
    \centergaussian &= T_{\viewspace \rightarrow h} \projectivemapping (W\centergaussian_\worldspace), \label{eq:hologramposition}
\end{align}
where $S$ and $R$ are the scaling matrix and the rotation matrix that defines the covariance matrix $\Sigma$.

In summary, the geometric pipeline in holographics starts with the parameters of an optimized Gaussian in world space, which are the covariance matrix $\Sigma_\worldspace$ and the mean of the Gaussian $\centergaussian_\worldspace$. These parameters are used to establish a geometric relationship between the hologram space and the canonical space. The parameters $J$ and $W$ are determined by the camera intrinsics and extrinsics, while the optical configurations of the holographic display specifies $T_{\viewspace \rightarrow h}$. Using these parameters ($R$, $S$, $\centergaussian$) in Eqs.~\ref{eq:combine}---\ref{eq:hologramposition}, we derive the angular spectrum of each Gaussian in the hologram space in the next section.

\subsection{Hologram pipeline}

To perform wave splatting in hologram space from a collection of optimized Gaussians in world space, we derive a closed-form solution for the Gaussian wavefront in hologram space, enabling efficient parallel implementation. The key insight of our approach is that a simple closed-form solution exists in canonical space, which we leverage by mapping the plane wave (angular spectrum) in hologram space from its counterpart in canonical space. To this end, we first establish the geometric relationship within this pipeline (between these spaces) and derive the closed-form solution of the angular spectrum in the Fourier domain, $\hat{u} (\mathbf{k})$, in hologram space. In other words, we map the spatial frequency component (i.e., the value of $\hat{u} (\mathbf{k})$) through a geometrical transformation (translation, rotation, and scaling) from hologram space back to canonical space, focusing on each parameter $R$, $S$, and $\centergaussian$, obtained in the previous section. In the following, we present the local 3D affine transform obtained in the previous section, and the closed-form solution of the angular spectrum in canonical space, which is a simple Gaussian:


\begin{align}
    \coordinate = RS \coordinate_c + \centergaussian \,\, \leftrightarrow \,\, \coordinate_c = S^{-1}R^{-1}( \coordinate - \centergaussian) 
\end{align}
\begin{align}
  \mathcal{G} (\coordinate_c) &= e^{-\frac{1}{2} \coordinate_c^\mathsf{T}\coordinate_c }  \xleftrightarrow{\fourier} \hat{\mathcal{G}}_c(\mathbf{k}_c) = 2 \pi e^{-2\pi^2 \mathbf{k}_c^\mathsf{T} \mathbf{k}_c},
\end{align}
where $\coordinate_c$ and $\mathbf{k}_c$ are the spatial coordinate and the wave vector in canonical space and $\hat{\mathcal{G}}$ denotes the Fourier transform of $\mathcal{G}$.

\paragraph{\textbf{Translation: $\coordinate_t = \coordinate - \centergaussian$}}

First, we translate the coordinate system by $\centergaussian$, aligning the orientation of the coordinate system with that of the Gaussian. Translation in the spatial domain is equivalent to applying a phase ramp in the Fourier domain: 
\begin{align} 
\coordinate_t = \coordinate - \centergaussian \leftrightarrow \angularspectrum (\mathbf{k}) = \angularspectrum_t (\mathbf{k}_t) e^{j\mathbf{k}\cdot\coordinate}, 
\label{eq:translation}
\end{align} 
where the subscript $t$ denotes the coordinate after the translation.

\paragraph{\textbf{Rotation: $\coordinate_o = R^{-1} \coordinate_t$}}
Next, we map the angular spectrum at the $x_t$--$y_t$ plane to that in the object space, where our Gaussian lies on a $x_o$--$y_o$ plane. Geometrically, this local space can be obtained by rotating shifted hologram space, $\coordinate_t$, in the Euler angle order of $x$, $y$, and $z$, such that $R^{-1} = R_z^{-1} R_y^{-1} R_x^{-1}$, and the coordinates are governed by the given rotation matrix $\coordinate_o = R^{-1} \coordinate_t$. Since the wave vectors also follow the same rotational transformation, the plane wave corresponding to the wave vector $\mathbf{k}_t$ corresponds to the plane wave corresponding to $\mathbf{k}_o = R^{-1} \mathbf{k}_t$ in the object space. Thus, we can nonlinearly map the angular spectrum from the object space to the shifted hologram space as: 
\begin{align} 
\coordinate_o = R^{-1} \coordinate_t \leftrightarrow \angularspectrum_t (\mathbf{k}_t) &= \textrm{det}( \mathbf{J}) \angularspectrum_o\left( R^{-1} \mathbf{k} \right), \label{eq:rotation}
\end{align} 
where $\textrm{det} (\mathbf{J}) = \frac{k_{o_z}}{k_{t_z}}$ is the Jacobian determinant of the transformation of Fourier coordinates. Note that this is essentially the celebrated result of tilted wave propagation~\cite{matsushima2003fast}.

\paragraph{\textbf{Scaling: $\coordinate_c = S^{-1}\coordinate_o$}}
The final step to arrive at the canonical space is to scale the Gaussian accordingly. This is a simple 2D affine transformation without bias though, and we describe the full formulation, including the bias term $\mathbf{b}$, which is zero in our case, for flexibility and completeness.
\begin{align}
\coordinate_c &= S^{-1}\coordinate_o + \mathbf{b}.
\end{align}
Using the affine theorem for two-dimensional Fourier transforms~\cite{bracewell1993affine}, we obtain
\begin{align}    
u_o(\coordinate_o) = \mathcal{G}(S^{-1}\coordinate_o + \mathbf{b}) & \xleftrightarrow{\fourier} \angularspectrum_o(\mathbf{k}_o) = \frac{1}{\det S^{-1}} \hat{\mathcal{G}}_c(S^{\mathsf{T}} \mathbf{k}_0) e^{j \mathbf{b}^{\mathsf{T}}S^{\mathsf{T}} \mathbf{k}_o}, \label{eq:scaling} 
\end{align}

where $\mathbf{k}$ is the wave vector and $\angularspectrum_o$ and $\hat{\mathcal{G}}$ denotes the Fourier transform of $u_o$ and $\mathcal{G}$, respectively

Combining Eqs.~\ref{eq:translation}, ~\ref{eq:rotation}, and ~\ref{eq:scaling}, we obtain the final closed-form solution in hologram space presented in the manuscript, from the angular spectrum derived in canonical space:
\begin{align}
\angularspectrum (\mathbf{k}) =     \textrm{det}( \mathbf{J}) \textrm{det}(S) \hat{\mathcal{G}} \left(SR^{-1}\mathbf{k} \right) e^{j  \mathbf{k}\cdot \centergaussian}  \label{eq:final}.
\end{align}

Note that while we set $\mathbf{b} = 0$ in this formulation, this becomes ill-posed with the translation $\mathbf{d}$ we performed ($\mathbf{d} = \centergaussian$ in Eq.~\ref{eq:translation}). We can set the amount of displacement $\mathbf{d}$ arbitrarily, not limited to $\centergaussian$. For instance, we can use $\mathbf{b} = -S^{-1} R^{-1} \centergaussian_{xy}$ and $\mathbf{d} = C\centergaussian_z$, where $\centergaussian_{xy}$ represents the transverse components of the vector $\centergaussian$, and $\centergaussian_z$ represents the $z$ component of $\centergaussian$, retaining its vector form. We show that how this formulation is compatible with the polygon-based methods in  Section~\ref{sec:misc}.



\clearpage
\newpage
\section{Additional Details on Software}
\label{sec:software}

In this section, we provide pseudocodes of the various algorithms described in the paper. In fact, all algorithms presented in the paper can be classified into three classes based on how occlusion is handled for primitives: 
\begin{itemize}
    \item \textbf{Na\"ive Wave Splatting (WS) Algorithm}: simply adding up the wavefront contribution of each primitive without any occlusion handling. Baselines that fall into this category include: Na\"ive (sparse point cloud), Na\"ive (dense point cloud, RGBD), and our fast Gaussian Wave Splatting (Gaussians). 
    \item  \textbf{Exact Wave Splatting (WS) Algorithm}: the exact wave splatting algorithm performs CGH using \textit{alpha wave blending} described in the main paper. Since occlusion is explicitly described by alpha wave blending, prior methods that perform occlusion handling can also be easily realized using WS, as we will show in \ref{subsec:silhouette}. Baselines that fall into this category include: Point cloud WS (sparse point cloud), Polygon WS (mesh), and our exact Gaussian Wave Splatting (Gaussians). 
    \item \textbf{Silhouette Method}: The algorithms that lies in this category are all prior methods that implement visibility tests with binary apertures, including Chen \& Wilkinson \cite{chen2009computer} (sparse point cloud) and the silhouette method (mesh) proposed by Matsushima et al. \cite{matsushima2014silhouette}. However, we show that with some rearrangement of equations, this can be exactly realized using Wave Splatting. We note that this classical method encompasses most occlusion-handling methods, which reject or attenuate wavefronts from the rear using binary or translucent masks~\cite{chen2009computer, shi2021towards}.  
\end{itemize}

These three classes of algorithms sequentially processes each primitive and adds up their wavefront contribution at the SLM plane. Within these three, the Na\"ive Wave Splatting algorithm is specifically suited for per-pixel parallelization since no order information and no alpha masking (which corresponds to a convolution in Fourier domain as described in the main paper) is required. Therefore, we additionally describe how to implement the per-pixel parallelizable version of the Na\"ive Wave Splatting algorithm, which also serves as  the basis for our efficient CUDA implementation of the fast variant of our Gaussian Wave Splatting algorithm.

\subsection{Na\"ive Wave Splatting Algorithm}
\label{subsec:naive_ws}
The Na\"ive Wave Splatting (WS) algorithm simply calculates the wavefront contribution of each primitive at the SLM plane and sums them up. We outline the sequential version of the Na\"ive WS algorithm using Gaussian primitives in Algorithm \ref{alg:naive_ws}, and describe the per-pixel parallelized version in Section~\ref{subsec:fast_naive_ws}.

\begin{algorithm}[t]
\caption{Na\"ive Gaussian Wave Splatting, sequential version}
\label{alg:naive_ws}
\SetKwComment{Comment}{$\blacktriangleright$\ }{}
\SetCommentSty{textnormal}
\DontPrintSemicolon
\Comment{No need to sort primitives.}
\KwIn{$K$: Number of primitives}
\KwIn{$z_k, o_k, c_k, S_k, \textrm{J}_k, R_k, \mu_k$: The depth, opacity, color, scale, rotation, center, and Jacobian determinant of the Fourier coordinate transformation of the $k^\text{th}$ Gaussian}
\KwIn{$N_x, N_y$: Image dimensions}
\KwOut{$u_\text{SLM}$: Wavefront at the SLM plane}
$\angularspectrum_\text{SLM} \gets \mathbf{0}_{N_x \times N_y}$ \Comment{Angular spectrum of wavefront at SLM plane}
\ForEach{$k$ \textbf{in} $1 \ldots K$}{
    $\angularspectrum_k \gets  \textrm{det}( \mathbf{J}_k) \textrm{det}(S_k) \hat{\mathcal{G}} \left(S_kR_k^{-1}\mathbf{k} \right) e^{j  \mathbf{k}\cdot \centergaussian_k}$ \Comment{Query the angular spectrum of the $k^\text{th}$ Gaussian at the SLM plane using Eq.~\ref{eq:final} }
    $\angularspectrum_\text{SLM} \gets \angularspectrum_\text{SLM} + \angularspectrum_k \cdot c_k \cdot o_k$ \Comment{multiply by the color and opacity of the primitive, and add to the SLM angular spectrum}
}
$u_\text{SLM} = \mathcal{F}^{-1}(\angularspectrum_\text{SLM})$ \\
\Return $u_\text{SLM}$
\end{algorithm}

\begin{figure*}[!ht]
    \centering
    \includegraphics[width=\textwidth]{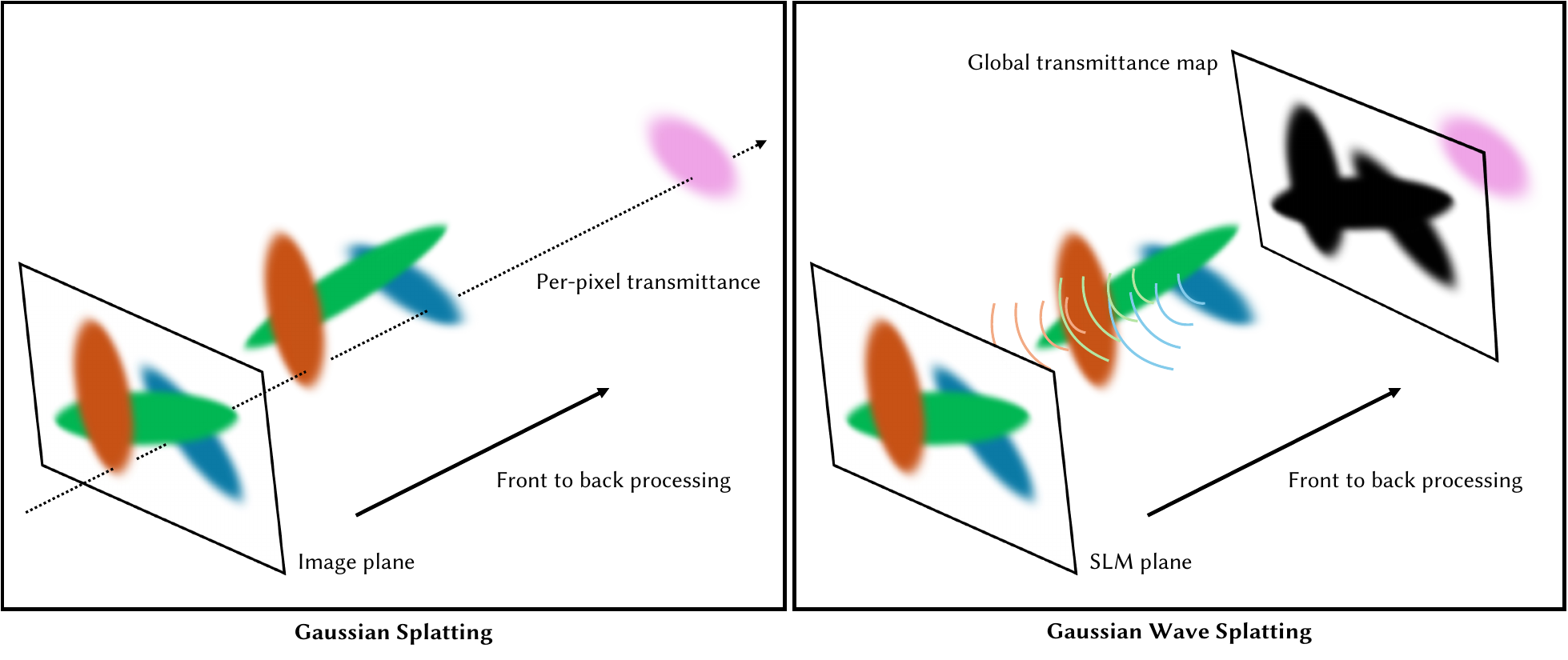}
    \caption{\textbf{Gaussian Splatting and Gaussian Wave Splatting.} Gaussian Splatting performs per-pixel alpha blending for all pixels in the rendered image, which is highly suitable for parallelization. Conversely, Gaussian Wave Splatting requires alpha blending to be performed on a global scale with the full image resolution since wave propagation can only be simulated with the full spatial extent of the SLM, and global transmittance map needs to be actively maintained throughout the CGH process. However, both methods share the exact same concept of alpha blending. }
    \label{fig:gs_vs_gws}
\end{figure*}

\subsection{Exact Wave Splatting Algorithm}
\label{subsec:full_ws}
At the core of the Exact Wave Splatting (WS) algorithm is \textit{alpha wave blending}, which first sorts the primitive from front (near SLM) to back (far from SLM) and sequentially processes each primitive. The wavefront of each primitive is sequentially ``splatted'' onto the SLM plane from near to far and a global transmittance map is simultaneously updated with the transmittance of each primitive. This transmittance map is a crucial component in alpha wave blending as it contains all the occlusion information. 


The above splatting procedure can also be thought of as propagating a wave starting from the SLM plane, and sequentially accumulating the wavefront contributions of primitives that are encountered along the process one by one. This can be formulated as a recursive equation, relating the accumulated wavefront up to the $\idx$-th Gaussian, $u_{1 \cdots \idx} (\mathbf{r})$, by adding the attenuated wavefront of the $\idx$-th Gaussian to the previously accumulated wavefront, $u_{1 \cdots (\idx -1)} (\mathbf{r})$, after propagation:

\begin{align}
u_{1 \cdots \idx} (\coordinate) = \mathcal{P}\left(u_{1 \cdots (\idx-1)}\left(\coordinate\right) ; z_{\idx} - z_{\idx-1}\right) + \mathbf{c}_\idx o_\idx T_\idx\left(\coordinate\right)  u_\idx\left(\coordinate\right) , \\ u_{1 \cdots 1}\left(\mathbf{r}\right) = u_1\left(\coordinate\right), T_\idx\left(\coordinate\right) = \prod_{j=1}^{\idx-1} (1 - o_{j} |u_j(\coordinate)|),
\end{align} 
where $\mathbf{c}_\idx$ is the color of the $i^\text{th}$ primitive By expanding the recursive equation and collapsing the sequential propagations into a single propagation distance using the depth-invariant property of the wave propagation operator, we can reformulate the recursive equation into a summation form: 

\begin{align}
u_{1 \cdots \idx} (\coordinate) &= \mathcal{P}\left(u_{1 \cdots (\idx-1)}\left(\coordinate\right) ; z_{\idx} - z_{\idx-1}\right) + \mathbf{c}_\idx o_\idx T_\idx\left(\coordinate\right)  u_\idx\left(\coordinate\right) \nonumber \\
&= \mathcal{P}\left( \mathcal{P}\left(u_{1 \cdots (\idx-2)}\left(\coordinate\right); z_{\idx-1} - z_{\idx-2} \right)  ; z_{\idx} - z_{\idx-1}\right) \nonumber \\
&\quad + \mathcal{P} \left( \mathbf{c}_{\idx-1} o_{\idx-1} T_{\idx-1} \left( \coordinate\right)  u_{\idx-1} \left( \coordinate \right); z_{\idx} - z_{\idx-1} \right) \nonumber \\ 
&\quad + \mathbf{c}_\idx o_\idx T_\idx\left(\coordinate\right)  u_\idx\left(\coordinate\right) \nonumber \\
&= \cdots = \sum_{j=1}^{\idx} \mathcal{P} \left( \mathbf{c}_j o_j T_j \left(\coordinate\right)  u_j \left(\coordinate\right); z_\idx - z_j\right) .\label{eq:awb_up_to_idx}
\end{align}

We would like to highlight the significance of this collapsible property of wave propagation, which enables this simple summation formulation. After summing all $\maxidx$ Gaussians, the accumulated wavefront will be positioned at a depth of $z_\maxidx$. We can then simply propagate the entire wavefront back to the SLM plane, where $z=0$, adding only a marginal computational cost with a single FFT-based wave propagation~\cite{shimobaba2009simple}, arriving at our alpha blending equation described in Eq.~5 of the manuscript:
\begin{align}    
    u_{\textrm{SLM}} (\coordinate) &= \mathcal{P} \left(\sum_{\idx=1}^{\maxidx} \mathcal{P} \left( \mathbf{c}_\idx o_\idx T_\idx \left(\coordinate\right)  u_\idx \left(\coordinate\right); z_\maxidx - z_\idx\right); -z_\maxidx \right) \\
    &=\sum_{\idx}^\maxidx \mathbf{c}_\idx o_\idx\mathcal{P}\Big(  u_\idx(\coordinate)T_\idx\left(\coordinate\right) ; -z_\idx\Big). \label{eq:gwsplat_supp}
\end{align}

The formulation of exact WS is almost exactly identical as its ray space counterpart, which is the volume rendering equation widely used in traditional graphics and recent neural rendering works \cite{mildenhall2020nerf, kerbl3Dgaussians}. The major difference is that since we are frequently converting between the frequency domain and spatial domain for wave propagation, per-pixel alpha blending is not possible in exact WS but instead requires a global transmittance map to be maintained throughout the process, as illustrated in Figure \ref{fig:gs_vs_gws}. We specifically outline the pseudocode of Gaussian Wave Splatting in Algorithm \ref{alg:full_ws}, but it can be trivially adapted to support other primitive types such as points or triangles simply by replacing the Gaussian angular spectrum in the first line of the for loop with angular spectrums of other primitives. 

\noindentparagraph{\bf{Other implementation details.}}
Under plane wave illumination in the direction $\mathbf{k}_r$, the angular spectrum derived in Eq.~\ref{eq:final} can be shifted by $\mathbf{k}_r$~\cite{park2017recent}. With this intuition, supporting directional effects using spherical harmonics can be understood as a convolution of the angular spectrum with an angular kernel defined by spherical harmonics~\cite{yeom2016calculation, askari2017occlusion}. This is discussed in detail in Sec.~\ref{subsec:occlusion}. Additionally, we match the phase of each wavefront from the primitives to produce multiple coherent wavefronts that constructively interfere, following practices in previous CGH methods~\cite{park2015removal, maimone2017holographic, padmanaban2019holographic}.

\begin{algorithm*}[h]
\caption{Exact Gaussian Wave Splatting}
\label{alg:full_ws}
\SetKwComment{Comment}{$\blacktriangleright$\ }{}
\SetCommentSty{textnormal}
\DontPrintSemicolon
\Comment{First sort primitives from front (close to SLM) to back (far from SLM) based on depth values}
\Comment{$\mathcal{P}(u, z, \lambda) = \mathcal{F}^{-1}(\mathcal{F}(u)\mathcal{H}(\lambda, z))$ is the ASM wave propagation operator and $\mathcal{H}$ is the angular transfer function. }
\KwIn{$K$: Number of primitives}
\KwIn{$z_k, o_k, c_k, S_k, \textrm{J}_k, R_k, \mu_k$: The depth, opacity, color, scale, rotation, center, and Jacobian determinant of the Fourier coordinate transformation of the $k^\text{th}$ Gaussian}
\KwIn{$N_x, N_y$: Image dimensions}
\KwIn{$\lambda$: Illumination wavelength}
\KwIn{$t_\text{eps}$: Threshold for low alpha value, used to ignore low transmittance primitives. We set  $t_\text{eps} = \frac{1}{255}$ following 3DGS ~\cite{kerbl3Dgaussians}.}
\KwIn{$t_\text{binarize}$: Threshold to binarize alpha. Used to realize the disk-based occlusion handling described in Chen \& Wilkinson \cite{chen2009computer}}
\KwOut{$u_\text{SLM}$: Wavefront at the SLM plane}
$u_\text{SLM} \gets \mathbf{0}_{N_x \times N_y}$ \Comment{Wavefront at SLM plane}
$T \gets \mathbf{1}_{N_x \times N_y}$ \Comment{Initialize transmittance map} 
\ForEach{$k$ \textbf{in} $1 \ldots K$}{
    $\angularspectrum_k \gets  \textrm{det}( \mathbf{J}_k) \textrm{det}(S_k) \hat{\mathcal{G}} \left(S_kR_k^{-1}\mathbf{k} \right) $ \Comment{Query the angular spectrum of the $k^\text{th}$ Gaussian at the shifted hologram space using Eq.~\ref{eq:final} }
    
    $u_\text{SLM} \gets u_\text{SLM} + \mathcal{P}(T\cdot u_k, z_k, \lambda)\cdot c_k \cdot o_k $ \Comment{Splat the wavefront onto the SLM plane by propagating the masked wavefront the SLM plane, multiplying the propagated wavefront by the the primitive color and opacity, and add to the wavefront on the SLM plane}
    
    $\alpha_k \gets o_k \cdot |u_k|$ \Comment{Compute the alpha map from the primitive opacity and the primitive wavefront amplitude}

    $\alpha_k \gets \alpha_k \cdot (\alpha_k > t_\text{eps})$ \Comment{Ignore low alpha values like in Gaussian Splatting \cite{kerbl3Dgaussians}}

    \If{\text{use disk-based occlusion handling}}{
        $\alpha_k \gets \alpha_k > t_\text{binarize}$ \Comment{Exact GWS can also be used to implement visibility tests using binary apertures as described in Chen \& Wilkinson \cite{chen2009computer}}
    }

    $T \gets T \cdot (1 - \alpha_k)$ \Comment{Update the transmittance map}
}
\Return $u_\text{SLM}$
\end{algorithm*}

\subsection{Silhouette-based Method}
\label{subsec:silhouette}
Perhaps the most accurate method to address occlusion in prior CGH literature is the silhouette-based method~\cite{matsushima2014silhouette}, and we outline its pseudocode in Algorithm \ref{alg:silhouette}. The silhouette-based method starts from the rearmost polygon $\mathbf{P}_N$ and accumulates the wavefront contributions that are not attenuated by the front objects, continuing the propagation until it reaches the frontmost polygon $\mathbf{P}_1$. The wavefront contribution up to the $\idx$th polygon $u_{\idx \cdots N}$, starting from $u_{N \cdots N} = u_N\left(\mathbf{r}\right)$, can be inductively formulated as:


\begin{align}
u_{\idx \cdots \maxidx} (\coordinate) = M_\idx(\coordinate) \mathcal{P}\left(u_{\idx+1 \cdots \maxidx}\left(\coordinate  \right); z_\idx - z_{\idx + 1} \right) + o_i u_\idx\left(\coordinate \right),
 \label{eq:silhoutte}   
\end{align}
where $M_\idx(\coordinate) = 1 - o_\idx A_\idx(\coordinate)$ is the silhouette mask of a translucent polygon $\mathbf{P}_\idx$, defined to emulate alpha blending in computer graphics~\cite{matsushima2020introduction}. In this context, $A_\idx$ represents the binary polygon aperture, and $o_\idx$ denotes the alpha values. This method has shown promising results for high-definition polygon CGH~\cite{matsushima2009extremely}. However, its sequential nature, especially the mask profile on the accumulated wavefront, complicates intuitive use of the collapse property, and the reliance on discrete apertures makes it difficult to apply to Gaussians. Notably, computational cost can be significantly reduced by limiting the region of interest for propagation using Babinet's principle~\cite{matsushima2020introduction}. Please refer to  Sec.~\ref{sec:misc} for the full analytical solution for polygon-based methods~\cite{zhang2022polygon, park2017recent, kim2008mathematical, ahrenberg2008computer}.

\noindentparagraph{\bf{From Alpha Wave Blending to Silhouette-based Methods.}}
Our alpha wave blending formulation that directly inherits the continuous aperture property of Gaussians and composites front to back, generalizes these silhouette-based methods and is compatible with polygons that use binary apertures, as derived in this section and demonstrated in Figs.~\ref{fig:extended_baselines_1} --- ~\ref{fig:extended_baselines_12}.

To derive the connection between the general alpha wave blending formulation and silhouette-based methods, we start from Eq.~\ref{eq:awb_up_to_idx}, after propagating the final wavefront to where the first polygon is located (to match the final depth) and consider color and opacity as translucency applied to the binary aperture profile $o_\idx A_\idx (\coordinate)$~\cite{matsushima2020introduction}. In this setting, the attenuation map $T_i$ is equivalent to $\prod_{j=1}^{\idx-1} (1 - o_{j} A_j(\coordinate))$. 

\begin{align}
 u_{1 \cdots \maxidx} (\coordinate) &= \mathcal{P} \left(\sum_{\idx=1}^{\maxidx} \mathcal{P} \left( o_\idx A_\idx(\coordinate) T_\idx (\coordinate)  u_\idx (\coordinate); z_\maxidx - z_\idx \right); z_1 - z_N \right) \\
 &= \sum_{\idx=1}^{\maxidx} \mathcal{P} \left( o_\idx A_\idx(\coordinate) T_\idx (\coordinate)  u_\idx (\coordinate); z_1 - z_\idx \right) \\
 &= \sum_{\idx=1}^{\maxidx} \mathcal{P} \left( o_\idx A_\idx(\coordinate) u_\idx (\coordinate)
 \prod_{j=1}^{\idx-1} (1 - o_{j} A_j(\coordinate)) ; z_1 - z_\idx \right) \\
 &= \sum_{\idx=1}^{\maxidx} \mathcal{P} \left( o_\idx A_\idx(\coordinate) u_\idx (\coordinate) \prod_{j=1}^{\idx-1} M_j (\coordinate); z_1 - z_\idx \right).
\end{align}

We can expand the summation propagating from front to back, following the order of silhouette-based method, and the mask term can be pulled out at the depth of the polygon sequentially, allowing us to arrive at the same recursive formulation to Eq.~\ref{eq:silhoutte}.


\begin{algorithm*}[t!]
\caption{Silhouette-based Method}
\label{alg:silhouette}
\SetKwComment{Comment}{$\blacktriangleright$\ }{}
\SetCommentSty{textnormal}
\DontPrintSemicolon

\Comment{First sort primitives from back (far from SLM) to front (close to SLM) based on depth values.}
\Comment{$\mathcal{P}(u, z, \lambda) = \mathcal{F}^{-1}(\mathcal{F}(u)\mathcal{H}(\lambda, z))$ is the ASM wave propagation operator and $\mathcal{H}$ is the angular transfer function. }
\KwIn{$K$: Number of primitives}
\KwIn{$z_k, o_k, c_k$: The depth, opacity, and color of the $k^\text{th}$ primitive}
\KwIn{$N_x, N_y$: Image dimensions}
\KwIn{$\lambda$: Illumination wavelength}
\KwOut{$u_\text{SLM}$: Wavefront at the SLM plane}

$u \gets 0$ \Comment{Initialize accumulated wavefront}

\ForEach{$k$ \textbf{in} $1 \ldots K$}{
    \If{$k = 1$}{
        $u' \gets u$ \Comment{Processing first primitive so previous accumulated wavefront is zero}
    }
    \Else{
        $u' \gets \mathcal{P}(u, z_k, \lambda)$ \Comment{Propagate previous accumulated wavefront to current depth}
    }
    
    $u_k \gets \texttt{Generate object wavefront(k)}$ 
    
    $u_k \gets \mathcal{P}(u_k, z_k, \lambda)$ \Comment{Propagate object wavefront to its depth}
    
    $u \gets (\mathbf{1} - |u_k|\cdot o_k) \cdot u' + u_k \cdot c_k \cdot o_k$ \Comment{Combine previous accumulated wavefront and current object wavefront to get new accumulated wavefront}

    $u \gets \mathcal{P}(u, -z_k, \lambda)$ \Comment{Propagate accumulated wavefront back to SLM plane}
}

$u_\text{SLM} \gets u$\;
\Return $u_\text{SLM}$
\end{algorithm*}

\subsection{Fast Na\"ive Wave Splatting Algorithm}
\label{subsec:fast_naive_ws}
Since no order information and alpha blending is need in the Na\"ive WS algorithm, it is trivial to parallelize and can be efficiently implemented in CUDA. 

Whether to parallelize over pixels or primitives, however, is purely a design choice that depends on the number of primitives and the target hologram resolution. Since we are given a 3D model with a fixed number of primitives and the number of the pixels in the hologram is usually larger than the number of primitives, we decided to parallelize over pixels to achieve maximum gain.

We roughly outline our how the kernel launches and memory allocation is defined in our CUDA implementation of the Fast Na\"ive WS algorithm in Algorithm \ref{alg:cuda_wavefront_perpixel}.

\begin{algorithm*}[ht!]
\caption{Fast Na\"ive Gaussian Wave Splatting, parallelized over pixels}
\label{alg:cuda_wavefront_perpixel}
\SetKwComment{Comment}{$\blacktriangleright$\ }{}
\SetCommentSty{textnormal}
\DontPrintSemicolon

\SetKwFunction{KernelFunc}{ComputePixelWavefrontKernel}

\KwIn{$K$: Number of primitives}
\KwIn{$z_k, o_k, c_k, S_k, \textrm{J}_k, R_k, \mu_k$: The depth, opacity, color, scale, rotation, center, and Jacobian determinant of the Fourier coordinate transformation of the $k^\text{th}$ Gaussian stored in device memory.}
\KwIn{$N_x, N_y$: Image dimensions}
\KwOut{$u_\text{SLM}$: Wavefront at the SLM plane}
$\angularspectrum_\text{SLM} \gets \mathbf{0}_{N_x \times N_y}$ \Comment{Angular spectrum of wavefront at SLM plane}
\BlankLine
\Comment{Host code}
$\text{gridDim} \gets (N_x/B_x, N_y/B_y)$ \Comment{Each thread processes one pixel}
$\text{blockDim} \gets (B_x, B_y)$ 

$\angularspectrum_\text{SLM} \gets 0$ \Comment{Initialize output array in device memory}
\KernelFunc{$\ll\!\!\ll \text{gridDim}, \text{blockDim} \gg\!\!\gg$}($z_k, o_k, c_k, ..., \mu_k$, $K$, $\angularspectrum_\text{SLM}$)

\BlankLine
\SetKwProg{Fn}{Function}{:}{}
\Fn{\KernelFunc{..., $K$, $U_\text{SLM}$}}{
    \Comment{Calculate global pixel coordinates}
    $x \gets \text{blockIdx.x} \times \text{blockDim.x} + \text{threadIdx.x}$\;
    $y \gets \text{blockIdx.y} \times \text{blockDim.y} + \text{threadIdx.y}$\;
    
    \If{$x < N_x$ \textbf{and} $y < N_y$}{
        \For{$k \gets 1$ \KwTo $K$}{
            \Comment{Process each primitive sequentially for this pixel}   $\angularspectrum_k[x, y] \gets  \textrm{det}( \mathbf{J}_k) \textrm{det}(S_k) \hat{\mathcal{G}} \left(S_kR_k^{-1}\mathbf{k} \right) e^{j  \mathbf{k}\cdot \centergaussian_k}[x, y]$ \Comment{Query the angular spectrum of the $k^\text{th}$ primitive at $[x, y]$ pixel location at the SLM plane using Eq.~\ref{eq:final} }
            
            $\angularspectrum_\text{SLM}[x, y] \gets \angularspectrum_\text{SLM}[x, y] + \angularspectrum_k[x, y]\mathcal{H}(z_k, \lambda)[x, y] \cdot c_k \cdot o_k$\; \Comment{Propagate the angular spectrum of the primitive by its depth using the angular transfer function $\mathcal{H}$ in ASM}
        }
    }
}

\BlankLine
\Comment{Back on host}
$u_\text{SLM} = \mathcal{F}^{-1}(\angularspectrum_\text{SLM})$ \Comment{Inverse Fourier transform of result}
\Return $u_\text{SLM}$
\end{algorithm*}

\subsection{Details on 2DGS Models}
We use the open source \texttt{gsplat}~\cite{ye2024gsplat} library to optimize 2DGS models for Gaussian Wave Splatting CGH calculation. All models are optimized using default parameters defined in the open source code base.

To flexibly control the number of the optimized Gaussians in the final model to match the number of optimized Gaussians with the number of triangles in the mesh models for fair comparison, we used the 3DGS-MCMC densification strategy \cite{kheradmand20243dgsmcmc} which is also supported in \texttt{gsplat}. We additionally optimized 2DGS models with a maximum of 1M Gaussians for video results of focal stacks from various novel views.  

\subsection{Details on 2DGS-OIT Models}
We implement custom CUDA forward and backward kernels and integrate them in \texttt{gsplat} to optimize 2DGS-OIT models that are used for the fast variant of Gaussian Wave Splatting CGH. Since the OIT image formation is different than that of 2DGS, the gradients of the model parameters need to be re-derived. Here, we show derivations of several intermediate gradient functions that can be easily integrated with existing Gaussian Splatting software frameworks, and refer the readers to the extensive writeup provided by the \texttt{gsplat} authors for full gradient derivations \cite{ye2024gsplat}.

In 2DGS-OIT, the color of each pixel being rendered  is represented as:
\begin{align}
    \mathbf{c} \approx \sum_{i=1}^N \mathbf{c}_i \alpha_i = \sum_{i=1}^N \mathbf{c}_i \mathcal{G}_i o_i,
\label{eq:OIT}
\end{align}
where $\mathbf{c}$ is the final rendered pixel color, $\mathbf{c}_i, \alpha_i, o_i \in \mathbb{R}$ are the color, alpha, opacity, and opacity of the $i^\text{th}$ Gaussian, respectively, and $\mathcal{G}_i$ is the splatted 2D Gaussian value evaluated at the location of the pixel currently being rendered. 

Since $c_i$ and $\alpha_i$ are independent of each other, the gradient formulas of $\alpha_i$ and $c_i$ with respect to the loss $\mathcal{L}$ is trivial:

\begin{align}
    \frac{\partial \mathcal{L}}{\partial \alpha_i} = \mathbf{c}_i \cdot \ \frac{\partial \mathcal{L}}{\partial \mathbf{c}} \\
    \frac{\partial \mathcal{L}}{\partial \mathbf{c}_i} = a_i \cdot \ \frac{\partial \mathcal{L}}{\partial \mathbf{c}}
\end{align}

Other gradients, such as gradients with respect to opacities and spherical harmonics coefficients, are the same as the gradients in original 2DGS, and we refer the readers to the \texttt{gsplat} writeup \cite{ye2024gsplat} for more details.

To optimize a 2DGS-OIT model, we first pretrain a 2DGS model with the 3DGS-MCMC strategy and initialize the 2DGS-OIT model with means and rotations of the optimized Gaussians from 2DGS. We do not densify Gaussians throughout the 2DGS-OIT optimization process. We observe that 2DGS-OIT models converge much faster and is more stable to optimize with this pretrained 2DGS initialization. We use degree 3 spherical harmonics (16 SH coefficients) to model the view-dependent opacity of each Gaussian. All other optimization configurations are the same as that of 2DGS.

\subsection{Details on the Efficient CUDA Implementation of Fast Gaussian Wave Splatting}
\label{subsec:details}
We implement the fast version of our Gaussian Wave Splatting using custom CUDA kernels following the logic described in Algorithm \ref{alg:cuda_wavefront_perpixel}. We spawn 1024 threads for each block (32 in both height and width dimensions of the hologram), and sequentially process \textit{all} Gaussians in each kernel call. This CUDA kernel is integrated in our \texttt{hsplat} library so that it works seamlessly with \texttt{PyTorch}.

\newpage

\clearpage
\newpage
\section{Additional Details on Hardware}
\label{sec:hardware}

\begin{figure}[h]
    \centering
    \includegraphics[width=0.8\linewidth]{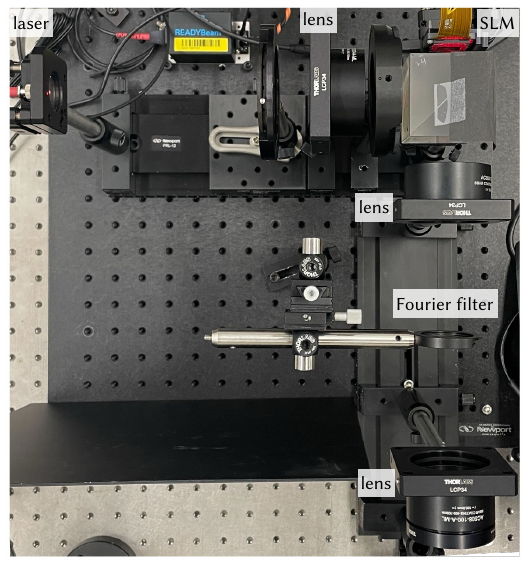}
    \caption{\textbf{Holographic display prototype used for the experimental results.} }
    \label{fig:setup}  
\end{figure}
\noindentparagraph{\bf{Experimental setup.}}  
To experimentally validate our algorithms, we build a holographic display benchtop prototype using a FISBA READYBeam™ Bio2 fiber-coupled module with three optically aligned laser diodes with wavelengths of 638, 520, and 488 nm, respectively, and a Holoeye PLUTO phase-only liquid crystal on silicon (LCOS) SLM with a resolution of $1920 \times 1080$ pixels, a pixel pitch of 8~$\mu$m, and a precision of 8 bits. We follow an optical configuration similar to those used in Choi et al.~\cite{choi2021neural} and Shi et al.
~\cite{shi2021towards}, which is shown in Fig.~\ref{fig:setup}. Color images are captured with separate exposures for each wavelength and combined in post-processing. All images are captured with a FLIR Grasshopper3 2.3 MP color USB3 vision sensor. In our setup, the depth range in hologram space is approximately 1cm, which is equivalent to total depth range of 2D in view space, from optical infinity to approx. 0.5 m.

\noindentparagraph{\bf{Phase encoding.}}  
Our CGH algorithm encodes Gaussian-based scene representations into a complex-valued wavefront, similar to other primitive-based methods. While one can build a hardware setup with multiple SLMs to modulate complex-valued wavefronts~\cite{shi2017near, reichelt2012full}, most holographic display setups use phase-only SLMs, requiring phase encoding techniques. To encode the complex-valued wavefront splatted onto the SLM plane into a phase-only pattern, we use the double-phase amplitude coding method (DPAC) as described in recent literature~\cite{hsueh1978computer, maimone2017holographic, shi2021towards}. The double-phase method directly encodes amplitude and phase into two phase values. Specifically, this method converts a complex-valued wavefront $ae^{j\phi}$ into two phase values, $P_1$ and $P_2$:
\begin{align}
    P_1 = a - cos^{-1} \phi \\
    P_2 = a + cos^{-1} \phi,
\end{align}
and interlaces $P_1$ and $P_2$ like a checkerboard pattern~\cite{maimone2017holographic}. 

\begin{figure}[h]
    \centering
    \includegraphics[width=\linewidth]{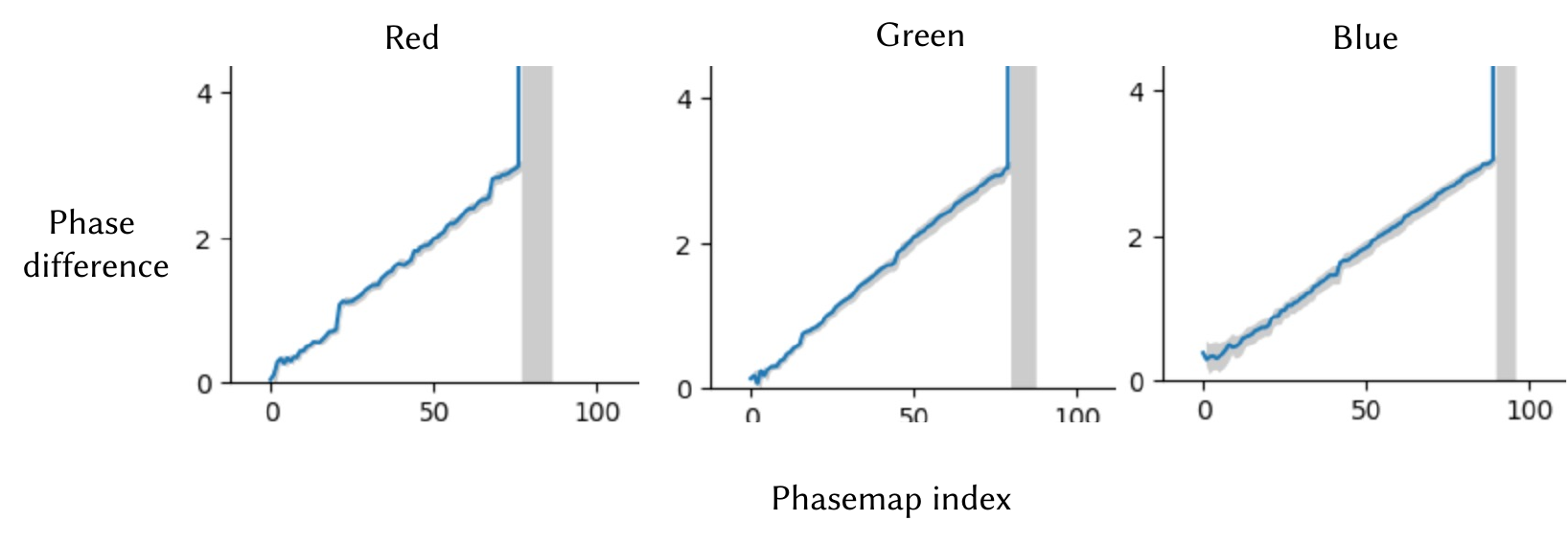}
    \caption{\textbf{Spatially varying, calibrated LUTs.} We visualize the spatially varying, calibrated lookup table (LUT) for red, green, and blue channel using the described procedure. This LUT maps the phase difference to the 8-bit image index that is dictated by the SLM. The blue line represents the mean values across all pixels, while the gray shades indicate the standard deviation over all pixels. For all pixels, we rectify the LUT by assigning a very large value to ensure that the nearest neighbor search does not return values exceeding $\pi$. Note that discontinuities indicate an inability to represent certain phase differences, resulting in image quality degradation. For a better illustration of this, we show an unideal LUT for the red channel; however, in our experiments, we use a better-calibrated LUT for red as well as illustrated for the other channels.}
    \label{fig:lut}  
\end{figure}

\noindentparagraph{\bf{Hardware calibration}}

To address hardware imperfections, we account for laser beam intensity variations and perform spatially varying lookup table (LUT) calibration~\cite{choi2021neural, shi2021towards, shi2022end}. The calibration procedure is as follows: first, we carefully align the setup, ensuring that the camera sensor is focused on the Fourier plane, and match the filter size optimal for the double-phase method~\cite{shi2021towards}. Next, we set the initial reference phase level to 127, construct checkerboard patterns, and incrementally increase the difference between $P_2$ and $P_1$, and vice versa, measuring the corresponding intensity per $2\times 2$ superpixel to generate the phase LUT (see Fig.~\ref{fig:lut}). To avoid phase-wrapping artifacts, we use $3\pi$ range modulation for each wavelength~\cite{maimone2017holographic}.

We observed that adding additional LUT entries for reference levels ranging from 42 (corresponding to $-\pi$ in $3\pi$ modulation) to 212 (corresponding to +$\pi$ in $3\pi$ modulation) and addressing anti-aliasing also marginally improves image quality. While we do not claim the calibration techniques as our contribution, we open-source the code used for calibration along with \texttt{hsplat}\footnote{\href{https://github.com/computational-imaging/hsplat}{https://github.com/computational-imaging/hsplat}}.

\section{Additional Simulation Results}
\label{sec:simulation}

\begin{table*}[!ht]
   \footnotesize
   \centering
   \renewcommand{\arraystretch}{2} 
   \begin{tabular}{@{}P{0.15\textwidth}P{0.1\textwidth}P{0.7\textwidth}@{}}
   \toprule
   \makecell[l]{Method} & \makecell[l]{3D\\representation} & \makecell[l]{Description}  \\
   \midrule
   \makecell[l]{Na\"ive} & \makecell[l]{Sparse \\ point cloud} & \makecell[l]{Simply treats each point as a pixel and adding up wavefront at the SLM plane without any occlusion handling.} \\
   \makecell[l]{Chen \& Wilkinson \cite{chen2009computer} \\ 
   \textbf{(in main paper)}} & \makecell[l]{Sparse \\ point cloud} & \makecell[l]{Treats each point as a fully opaque disk that can potentially block points behind it based on visibility tests. \\ The disk size is a tunable parameter. } \\
   \makecell[l]{Wave Splatting} & \makecell[l]{Sparse \\ point cloud} & \makecell[l]{Treats each point as an isotropic Gaussian and uses Wave Splatting and alpha wave blending to splat points onto the SLM. \\ The Gaussian scale is a tunable parameter. } \\
   \makecell[l]{Na\"ive} & \makecell[l]{Dense point \\ cloud / RGBD} & \makecell[l]{Each point exactly corresponds to one pixel and the depth of each point is extracted from ground truth depth maps. } \\
   \makecell[l]{Matsushima et al. \cite{matsushima2014silhouette} \\ 
   \textbf{(in main paper)}} & \makecell[l]{Mesh} & \makecell[l]{Each triangle is fully opaque and can block triangles behind it. } \\
   \makecell[l]{Wave Splatting} & \makecell[l]{Mesh} & \makecell[l]{A different realization of the same algorithm proposed by Matsushima et al. \cite{matsushima2014silhouette} using Wave Splatting. } \\
   \midrule
   \makecell[l]{Ours (fast) \\
   \textbf{(in main paper)}} & \makecell[l]{Gaussians} & \makecell[l]{The fast variant of Gaussian Wave Splatting that converts 2DGS-OIT models to holograms using approximated alpha wave \\ blending (order-invariant transparency).} \\
   \makecell[l]{Ours (exact) \\
   \textbf{(in main paper)}} & \makecell[l]{Gaussians} & \makecell[l]{The exact variant of Gaussian Wave Splatting that converts 2DGS models to holograms using exact alpha wave blending. } \\
   \bottomrule
   \end{tabular}
\caption{\textbf{Baseline methods. } In this table, we describe the input 3D representation and algorithm overview for each baseline method. We compare our Gaussian Wave Splatting algorithm with six other primitives-based CGH baselines.}
\label{tab:baseline_explanation}
\end{table*}

In this section, we report and extensive set of baseline comparisons and ablation studies to verify the effectiveness of Gaussian Wave Splatting. In addition to the three default baselines presented in the main paper, here we compare our method against another three, resulting in total of six baseline comparisons. We provide an overview of each baseline in Table. \ref{tab:baseline_explanation}.

\begin{table*}[!ht]
   \scriptsize
   \centering
   \renewcommand{\arraystretch}{1.2} 
   \begin{tabular}{@{}P{0.23\textwidth}P{0.09 \textwidth}P{0.09 \textwidth}P{0.09 \textwidth}P{0.09 \textwidth}P{0.09 \textwidth}P{0.09 \textwidth}P{0.09 \textwidth}@{}}
   \toprule
   \makecell[l]{Method - 3D representation} &
   Lego & Chair & Hotdog & Materials & Mic & Ship & Avg.\\
   \midrule
   \makecell[l]{Na\"ive - Point Cloud} & 10.55 / 0.66 / 0.34 & 14.67 / 0.77 / 0.20 & 14.59 / 0.67 / 0.31 & 15.89 / 0.72 / 0.26 & 22.63 / 0.90 / 0.11 & 15.75 / 0.59 / 0.35 & 15.68 / 0.72 / 0.26\\
   \makecell[l]{Chen \& Wilkinson \cite{chen2009computer} - Point Cloud} & 12.50 / 0.68 / 0.31 & 15.86 / 0.79 / 0.18 & 16.55 / 0.69 / 0.28 & 16.77 / 0.75 / 0.23 & 24.26 / 0.91 / 0.09 & 17.45 / 0.63 / 0.32 & 17.23 / 0.74 / 0.23 \\
   \makecell[l]{Wave Splatting - Point Cloud} & 12.97 / 0.69 / 0.29 & 16.39 / 0.80 / 0.17 & 16.84 / 0.70 / 0.27 & 16.90 / 0.75 / 0.22 & 24.82 / 0.92 / 0.08 & 17.76 / 0.64 / 0.31 & 17.61 / 0.75 / 0.22 \\
   \makecell[l]{Nai\"ive - RGBD} & 25.84 / \textbf{0.90} / 0.12 & 24.42 / \textbf{0.92} / \textbf{0.08} & 28.75 / 0.94 / 0.13 & 29.19 / \textbf{0.93} / \textbf{0.10} & \textbf{30.25} / \textbf{0.97} / \textbf{0.04} & 24.31 / 0.85 / 0.22 & 27.13 / \textbf{0.92} / 0.11 \\
   \makecell[l]{Matsushima et al. \cite{matsushima2014silhouette} - Mesh} & 19.27 / 0.75 / 0.24 & 18.29 / 0.82 / 0.15 & 22.92 / 0.79 / 0.22 & 19.67 / 0.81 / 0.19 & 24.16 / 0.94 / 0.07 & 20.32 / 0.71 / 0.27 & 20.77 / 0.80 / 0.19 \\
   \makecell[l]{Wave Splatting - Mesh} & 18.89 / 0.74 / 0.25 & 17.93 / 0.81 / 0.15 & 22.65 / 0.77 / 0.23 & 19.41 / 0.80 / 0.20 & 23.19 / 0.93 / 0.08 & 20.24 / 0.70 / 0.27 & 20.38 / 0.79 / 0.20 \\
   \midrule
   \makecell[l]{Ours (fast) - Gaussians} & 25.24 / \textbf{0.90} / 0.12 & 25.16 / 0.91 / \textbf{0.08} & 29.82 / \textbf{0.95} / \textbf{0.08} & 27.93 / \textbf{0.93} / \textbf{0.10} & 28.79 / 0.95 / 0.05 & \textbf{26.43} / \textbf{0.86} / \textbf{0.19} & 27.23 / \textbf{0.92} / \textbf{0.10} \\
   \makecell[l]{Ours (exact) - Gaussians} & \textbf{27.67} / \textbf{0.90} / \textbf{0.11} & \textbf{26.43} / \textbf{0.92} / \textbf{0.08} & \textbf{30.15} / 0.94 / 0.12 & \textbf{28.49} / 0.92 / \textbf{0.10} & 29.40 / 0.96 / \textbf{0.04} & 25.64 / 0.84 / 0.22 & \textbf{27.96} / 0.91 / 0.11 \\
   \bottomrule
   \end{tabular}
\caption{\textbf{Quantitative performance of different CGH algorithms on the Blender \cite{mildenhall2020nerf} dataset. } We evaluate the image quality of 3D holograms generated using our Gaussian Wave Splatting algorithm and various baseline primitives-based CGH methods in terms of PSNR $(\uparrow)$, SSIM $(\uparrow)$, and LPIPS $(\downarrow)$. All point cloud methods suffer from poor image quality due to the overly dim images with black holes. CGH from mesh outperforms point cloud methods, but high frequency details still cannot be reconstructed with per-face colors. CGH from RGBD images achieves much better quality, but still trails Gaussian Wave Splatting due to the unnatural ringing artifacts at depth discontinuities. Both exact and fast variants of Gaussian Wave Splatting achieve superior results over all baseline methods. }
\label{tab:per_scene_quan_metrics_blender}
\end{table*}

\begin{table*}[!ht]
   \scriptsize
   \centering
   \renewcommand{\arraystretch}{1.2} 
   \begin{tabular}{@{}P{0.23\textwidth}P{0.09 \textwidth}P{0.09 \textwidth}P{0.09 \textwidth}P{0.09 \textwidth}P{0.09 \textwidth}P{0.09 \textwidth}P{0.09 \textwidth}@{}}
   \toprule
   \makecell[l]{Method - 3D representation} &
   Kitchen & Counter & Stump & Room & Garden & Bicycle & Avg.\\
   \midrule
   \makecell[l]{Na\"ive - Point Cloud} & 6.67 / 0.06 / 0.72 & 7.87 / 0.12 / 0.75 & 11.48 / 0.03 / 0.72 & 9.85 / 0.10 / 0.76 & 7.39 / 0.02 / 0.73 & 11.07 / 0.04 / 0.74 & 9.05 / 0.06 / 0.74 \\
   \makecell[l]{Chen \& Wilkinson \cite{chen2009computer} - Point Cloud} & 10.20 / 0.15 / 0.72 & 7.86 / 0.12 / 0.74 & 11.60 / 0.03 / 0.72 & 9.41 / 0.03 / 0.74 & 6.60 / 0.00 / 0.75 & 10.46 / 0.01 / 0.69 & 9.35 / 0.06 / 0.73 \\
   \makecell[l]{Wave Splatting - Point Cloud} & 10.79 / 0.26 / 0.59 & 7.86 / 0.12 / 0.74 & 11.61 / 0.03 / 0.72 & 9.41 / 0.03 / 0.74 & 6.60 / 0.00 / 0.75 & 10.46 / 0.01 / 0.69 & 9.45 / 0.07 / 0.70 \\
   \makecell[l]{Nai\"ive - RGBD} & 17.79 / 0.67 / 0.36 & 22.13 / \textbf{0.76} / 0.40 & 17.33 / 0.38 / 0.54 & 24.26 / 0.74 / 0.44 & 16.14 / 0.20 / 0.49 & 17.44 / 0.44 / 0.48 & 19.18 / 0.53 / 0.45 \\
   \makecell[l]{Matsushima et al. \cite{matsushima2014silhouette} - Mesh} & 14.74 / 0.23 / 0.65 & 12.47 / 0.35 / 0.68 & 18.69 / 0.26 / 0.65 & 17.15 / 0.51 / 0.64 & 15.18 / 0.18 / 0.62 & 16.80 / 0.20 / 0.65 & 15.84 / 0.29 / 0.65 \\
   \makecell[l]{Wave Splatting - Mesh} & 14.66 / 0.21 / 0.65 & 12.36 / 0.32 / 0.68 & 18.64 / 0.25 / 0.65 & 17.11 / 0.49 / 0.64 & 15.09 / 0.16 / 0.62 & 16.79 / 0.19 / 0.64 & 15.77 / 0.27 / 0.65 \\
   \midrule
   \makecell[l]{Ours (fast) - Gaussians} & 22.63 / \textbf{0.82} / \textbf{0.19} & 20.25 / 0.72 / 0.41 & 21.68 / 0.48 / 0.47 & 23.19 / \textbf{0.82} / 0.42 & 15.29 / \textbf{0.25} / 0.52 & 21.18 / \textbf{0.49} / 0.49 & 20.70 / \textbf{0.60} / 0.42 \\
   \makecell[l]{Ours (exact) - Gaussians} & \textbf{23.99} / 0.77 / 0.23 & \textbf{23.33} / 0.74 / \textbf{0.38} & \textbf{22.16} / \textbf{0.49} / \textbf{0.44} & \textbf{25.12} / 0.78 / \textbf{0.41} & \textbf{18.11} / 0.23 / \textbf{0.47} & \textbf{21.50} / 0.48 / \textbf{0.47} & \textbf{22.37} / 0.58 / \textbf{0.40} \\
   \bottomrule
   \end{tabular}
\caption{\textbf{Quantitative performance of different CGH algorithms on the Mip-NeRF 360 \cite{barron2022mipnerf360} dataset. }  We evaluate the image quality of 3D holograms generated using our Gaussian Wave Splatting algorithm and various baseline primitives-based CGH methods in terms of PSNR $(\uparrow)$, SSIM $(\uparrow)$, and LPIPS $(\downarrow)$. All point cloud methods suffer from poor image quality due to the overly dim images with black holes. CGH from mesh outperforms point cloud methods, but high frequency details still cannot be reconstructed with per-face colors. CGH from RGBD images achieves much better quality, but still trails Gaussian Wave Splatting due to the unnatural ringing artifacts at depth discontinuities. Both exact and fast variants of Gaussian Wave Splatting achieve superior results over all baseline methods. }
\label{tab:per_scene_quan_metrics_mipnerf360}
\end{table*}

\subsection{Extended Quantitative Baseline Comparisons}

We show per-scene quantitative comparisons with the the extended set of baseline CGH methods in terms of PSNR, SSIM, and LPIPS in Table \ref{tab:per_scene_quan_metrics_blender} and \ref{tab:per_scene_quan_metrics_mipnerf360}. We evaluate in-focus image quality by merge simulated focal stacks rendered from the generated holograms into an all-in-focus image using RGBD masks computed from depth maps rendered from optimized 2DGS models, and compare this all-in-focus image with ground truth images from the training dataset. Both variants of our Gaussian Wave Splatting algorithm consistently outperform other primitive-based CGH algorithms across all image quality metrics and all scenes. 

All sparse point cloud methods suffer from low image quality due to the extreme sparsity of primitives, leading to overly dark images or images with holes. CGH from mesh cannot reconstruct sharp details unless a disproportionate number of triangles is used. Although RGBD CGH is expected to be the upper bound of all CGH methods since the hologram is directly synthesized from a ground truth all-in-focus RGB image and its corresponding depth map, it trails Gaussian Wave Splatting (GWS) since there significant ringing artifacts at depth discontinuities, while GWS achieves a smoother blur due to the summation of different Gaussian kernels and effectively hides the artifacts. Both of our GWS variants greatly outperforms all other baselines, achieving a superior 5-11 dB average PSNR improvement. 

\subsection{Extended Qualitative Baseline Comparisons}
We show focal stacks and zoom-in patches of focal slices generated from holograms synthesized using different CGH baselines and 3D representations in Figures 
\ref{fig:extended_baselines_1} to \ref{fig:extended_baselines_12}.

Without occlusion handling, applying the the Na\"ive algorithm on sparse point clouds results in over-saturated colors since the wavefront of all points are simply added up. With occlusion handling using disk-based visibility tests as described in Chen \& Wilkinson \cite{chen2009computer} or using alpha wave blending, the colors are more accurate. However, the inherent sparseness of point clouds significantly limits the image quality of point cloud-based CGH, leading to overly dim images with holes. This issue is especially prominent for in-the-wild captured 3D reconstruction datasets such as Mip-NeRF 360 \cite{barron2022mipnerf360}, where the SfM point clouds are simply too sparse to provide dense coverage of object surfaces, such as the \textit{bicycle} or \textit{garden} scene. With dense point cloud representations like RGBD images, image quality greatly improves, but still suffers from ringing artifacts at depth discontinuities.

CGH from meshes using the silhouette method proposed by Matsushima et al. \cite{matsushima2014silhouette}
and alpha wave blending achieves decent image quality, yet high frequency details cannot be reconstructed due to the per-face color representation that is used for polygon-based CGH. 

Both variants of our Gaussian Wave Splatting (GWS) algorithm achieves the best image quality compared to all six other baselines. From the zoomed-in crops from focal slices, we see that GWS accurately reconstructs sharp details in focused regions and synthesizes more natural blur in defocus regions. Compared with polygon-based models, GWS also achieves superior image quality while using an equal or reduced number of geometric primitives compared to the mesh models.

\subsection{3D Focal Stacks of Novel Views}
We show focal stacks generated from holograms synthesized at different camera viewpoints using our exact Gaussian Wave Splatting algorithm in Figure \ref{fig:nvs_fs}. We follow the Holographics pipeline described in Section \ref{sec:holographics} to generate a hologram from an optimized 2DGS model given a specific viewpoint and camera intrinsics. We see that both high quality novel view synthesis and natural 3D focus effects are accurately reproduced at \textit{all} views, demonstrating the robustness and effectiveness of Gaussian Wave Splatting. Please refer to the supplemental video for the circular smooth trajectory of all novel viewpoints and corresponding experimentally captured focal stacks.

\subsection{Sparse Point Cloud CGH Ablation Studies}
We perform various ablation studies on CGH from sparse point clouds. First, we ablate the disk size described in Chen \& Wilkinson \cite{chen2009computer}. This is a crucial parameter that determines the spatial extent of the mask that is used to perform visibility tests of points. In our reimplementation of the method, we use the parameter $t_\text{eps}$ described in Algorithm \ref{alg:full_ws} to control the disk size. From Fig \ref{fig:pc_disk_ablation}, we see that the CGH result is highly sensitive to the disk size as a larger disk (smaller threshold) more aggressively blocks points when performing visibility tests, leading to dimmer images. Smaller disk sizes, on the other hand, could lead to over-saturated images due to lenient visibility tests and light leakage.

Secondly, we ablate the Gaussian scale parameter when applying Wave Splatting to sparse point clouds and treating each point as an isotropic Gaussian. We describe the scale of each isotropic Gaussian as a multiple of the pixel pitch of the SLM. From Figure \ref{fig:pc_awb_ablation}, we see that images are dim when the Gaussian scale is small and becomes brighter as the Gaussian scale increases  due to larger spatial coverage with bigger Gaussians.

From these two ablation studies, we see that sparse point cloud representations are not ideal for photorealistic CGH due to its bad 3D spatial coverage nature and the need of heuristically tuning point sizes to achieve barely passing image quality. The only way to potentially achieve photorealistic quality is by using RGBD images or dense point clouds.

\begin{figure}[!t]
    \centering
    \includegraphics[width=\linewidth]{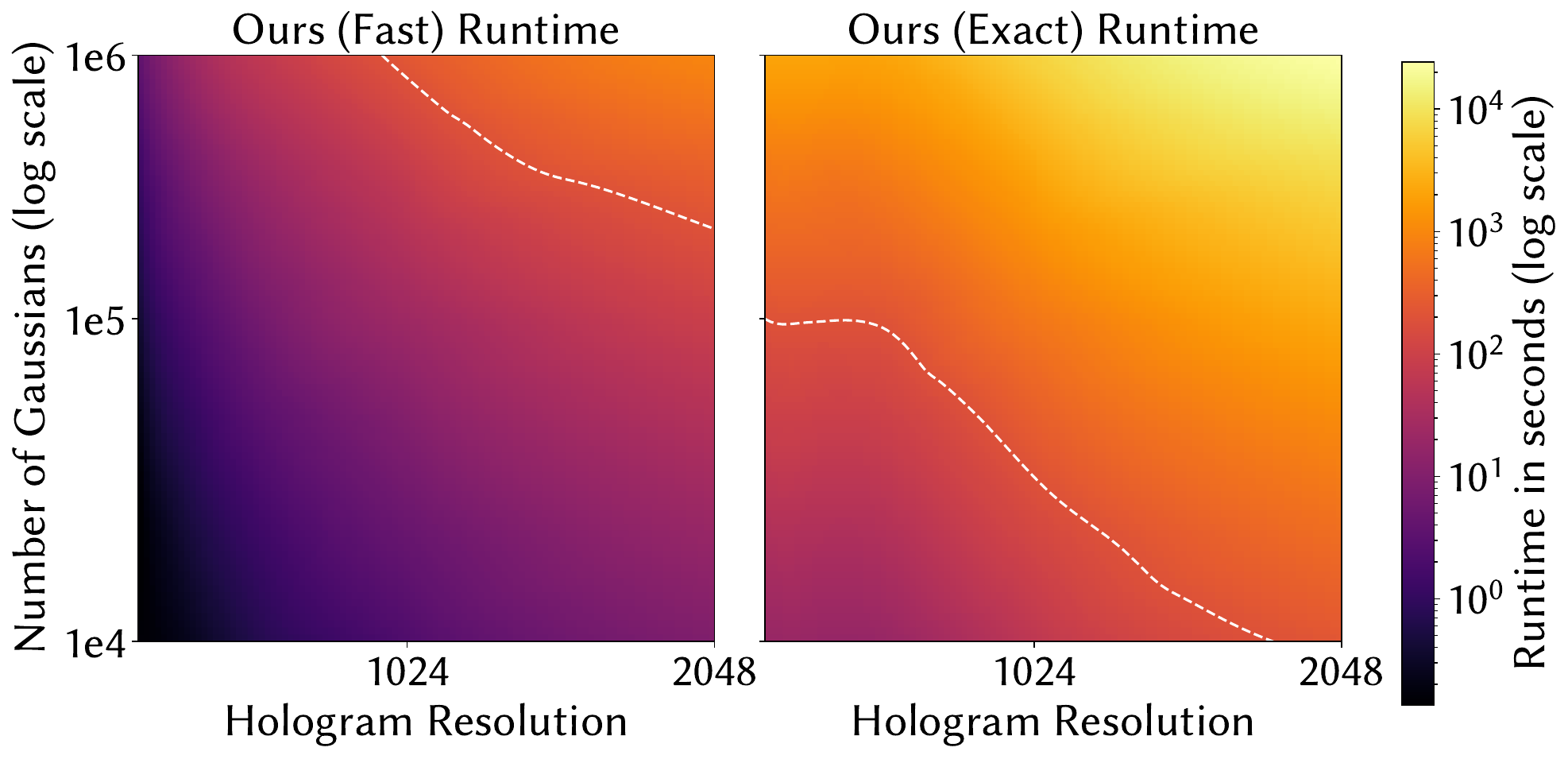}
    \caption{\textbf{Runtime comparisons of fast and exact GWS.} We show Gaussian Wave Splatting CGH calculation time using varying target hologram resolutions and number of Gaussians as a 2D heatmap. An isocurve representing the runtime of 100 seconds is plotted as a white dashed line. Fast GWS achieves nearly 30$\times$ speedup over exact GWS, and can generate holograms of much higher resolution and fidelity, which is positively correlated to the number of Gaussians, compared to the exact method given the same runtime. }
    \label{fig:runtime}
\end{figure}

\subsection{Runtime Comparisons}
We compare the runtime of fast and exact Gaussian Wave Splatting in Fig. \ref{fig:runtime}. We vary the number of Gaussian primitives used and the target hologram resolution and plot the runtimes as a 2D heatmap. An isocurve corresponding to a runtime of 100 seconds is plotted as a white dashed line. Fast GWS is able to synthesize higher resolution holograms using more primitives compared to exact GWS given the same runtime, which can be observed from Fig. \ref{fig:runtime}  as the fast GWS isocurve is to the top-left of the exact GWS isocurve. Given the same number of primitives and target hologram resolution, fast GWS achieves a nearly $30\times$ speedup over exact GWS.

\subsection{Additional Discussion on View-Dependent Effects}
\label{subsec:occlusion}

\begin{figure}[!h!]
    \centering
    \includegraphics[width=\linewidth]{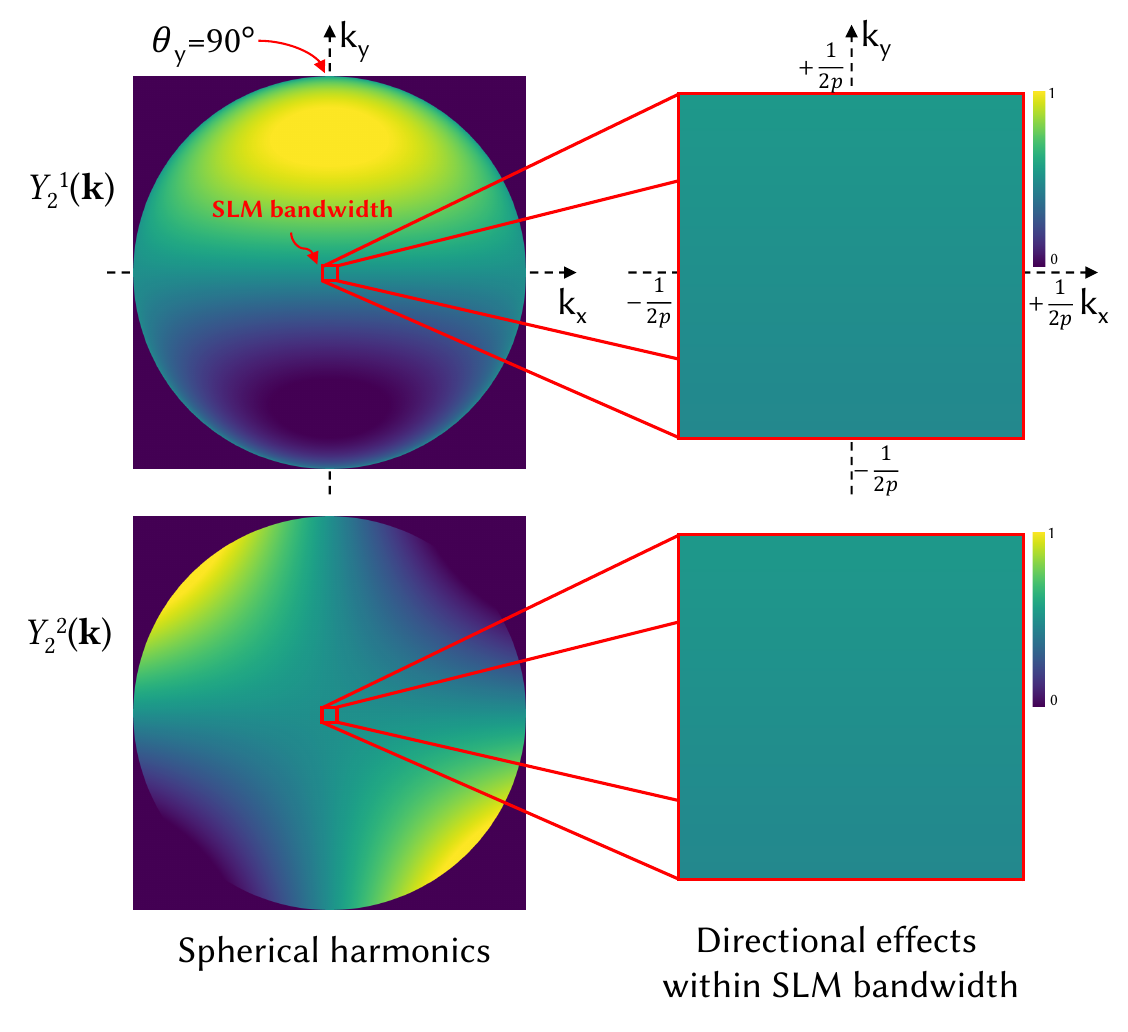}
    \caption{\textbf{Limited SLM bandwidth.} On the left, we show the angular profiles of spherical harmonics, i.e., $Y_1^2(\mathbf{k})$ and $Y_2^2(\mathbf{k})$. These are queried from all angles over the hemisphere, ranging from -90$\deg$ to $+90\deg$ for both the $x-$ and $y-$axes. The SLM (with 8$\mu$m pixel pitch in our case) supports only a limited range of angular support, which we found to be very uniform across the entire region. Colormaps on the right denote the normalized amplitude value.}
    \label{fig:justify}
\end{figure}
In Fig.~\ref{fig:justify}, we visualize the full angular emission profile of a spherical harmonics basis (second degree, the first two orders) over the hemisphere. This map can be interpreted as the angular emission profile, or the amplitude of the desired angular spectrum for holograms. Unfortunately, most current holographic displays only support a limited space-bandwidth product, restricting the angular range due to the pixel pitch size, which is typically a few degrees. This is visualized as the red box in the center. From this, we observe that the angular emission within the range of angles supported by the SLM is relatively uniform, motivating us to approximate the directional effects using zeroth-order spherical harmonics.

Nevertheless, here we discuss an approach to incorporate full directional effects in our Gaussian Wave Splatting. The classical approach to control the angular profile, such as random phase methods, or pioneering work in polygon-based method iterature~\cite{kim2008mathematical, ahrenberg2008computer, park2015removal, askari2017occlusion, zhang2018fast} can be understood in a more principled manner with this discussion. Gaussian Splatting uses spherical harmonics (SH) to represent view dependence~\cite{basri2003lambertian, yu_and_fridovichkeil2021plenoxels, kerbl3Dgaussians}. They form a complete set of orthonormal bases, and typically, the first few coefficients are optimized for each color channel. Here, we discuss how these coefficients can be incorporated into our holograms per basis. 

\begin{figure}[!h!]
    \centering
    \includegraphics[width=\linewidth]{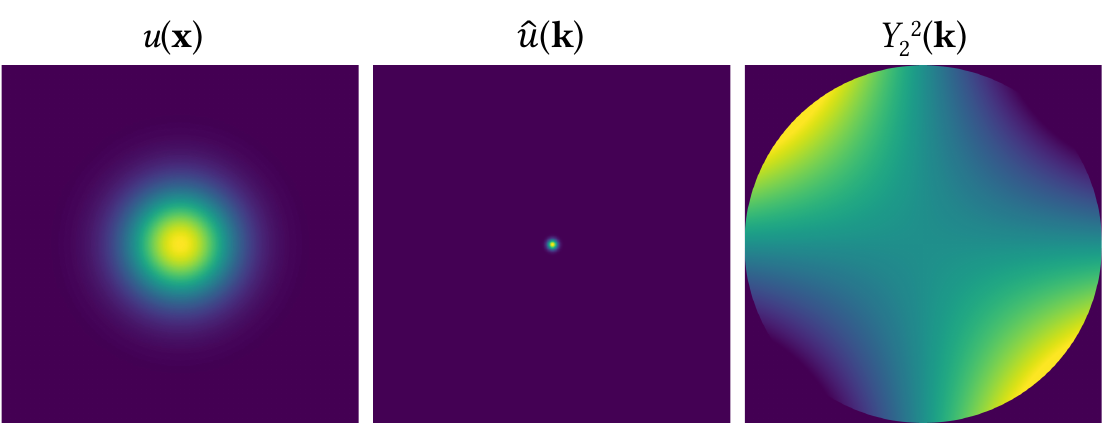}
    \caption{\textbf{Visualization of Gaussians in spatial and Fourier domain.} We visualize the Gaussians in spatial and Fourier domain.}
    \label{fig:justify2}
\end{figure}

Directional effects in primitive-based CGHs have primarily been discussed in the context of representing the reflectance distribution of surfaces~\cite{park2017recent}. Pioneering works have attempted to control the angular emission profile using sub-triangles~\cite{kim2008mathematical} and appropriate phase profiles for them. Yeom et al. demonstrated that the desired reflectance distribution could be approximated as a convolution between the angular spectrum of the mesh and the angular distribution~\cite{askari2017occlusion, yeom2016calculation}. In our general GWS formulation, this concept can be adapted as follows:
\begin{align}
    \angularspectrum_{\textrm{SH}_l^m} ( \mathbf{k} ) =  \textrm{det} ( \mathbf{J}) \textrm{det}(\mathbf{S}_\idx) \left( \angularspectrum_\canonicalspace \left(\mathbf{S}_\idx \rot_\idx^{-1}\mathbf{k} \right) \ast_{\mathbf{k}} Y_l^m(\mathbf{k}) e^{j\phi (\mathbf{k})} \right) e^{j  \mathbf{k} \cdot \mean_\idx} \label{eq:convolution},
\end{align}
where $\ast_{\mathbf{k}}$ is the convolution operator in the spatial frequency domain and $Y_l^m(\mathbf{k}) e^{j\phi (\mathbf{k})}$ denotes the angular kernel, having the spherical harmonics function of degree $l$ and order $m$ as its amplitude. For simplicity, we consider an example Gaussian without rotation, translation, scaling, and a basis of the first-degree and first-order, but incorporating these parameters and the extension to other basis are straightforward. The spatial and angular amplitude profile of an example Gaussian with a reasonable size, on the order of tens of microns, is shown in Fig.~\ref{fig:justify2}. Ideally, we want $\angularspectrum_{\textrm{SH}_l^m} ( \mathbf{k})$ to have the angular emission profile $Y_l^m(\mathbf{k})$, while retaining the Gaussian amplitude profile of the original $u(\coordinate)$ from our Gaussian wave splatting. However, note that the convolution in the Fourier domain is equivalent to element-wise multiplication in the spatial domain. The amplitude of $Y_l^m(\mathbf{k}) e^{j\phi (\mathbf{k})}$ in the spatial domain (i.e., the amplitude of the Fourier transform of $Y_l^m(\mathbf{k}) e^{j\phi (\mathbf{k})}$) highly depends on the phase profile $\phi (\mathbf{k})$. This task essentially boils down to achieving the spatio-angular light field profile $l(\coordinate, \mathbf{k})$ from a coherent wavefront.

It has been noted in the literature that a coherent wavefront can only represent a light field corresponding to rank-1 mutual intensity~\cite{zhang2011analysis, choi2022time, hamann2018time}. Due to this constraint, one might not be able to perfectly achieve both the desired spatial and angular emission profiles simultaneously with a coherent wavefront modulated by a bandlimited SLM. However, it has been demonstrated that one can rely on partial coherence to achieve high-quality light field holograms with time-multiplexing~\cite{choi2022time, kim2024holographic}. Inspired by this work, we discuss a heuristic sampling strategy in a time-multiplexed setting to achieve the desired angular emission profile (e.g., the first-order SH basis) while preserving the desired spatial amplitude.

\begin{figure*}[!h!]
    \centering
    \includegraphics[width=\linewidth]{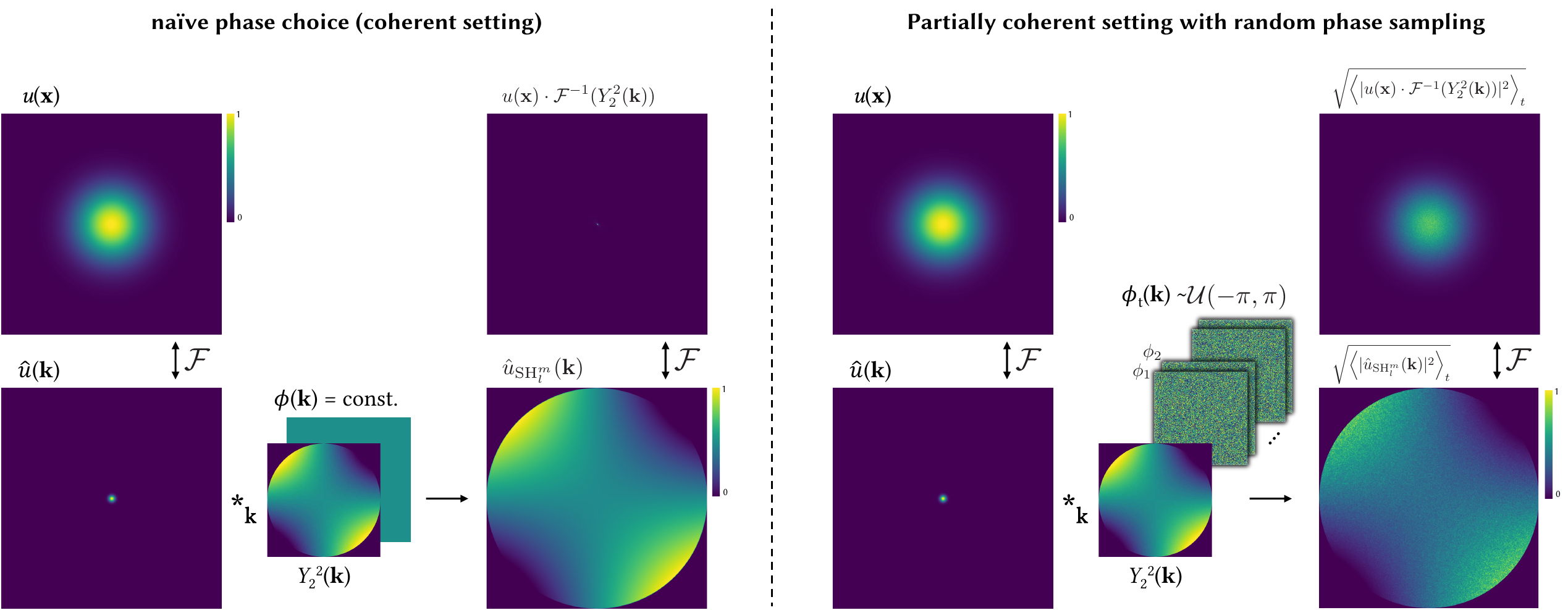}
    \caption{\textbf{The effect of phase sampling and partial coherence for directional effects.} Here, we visualize a virtual setting where we have control over a very large angular support. On the left, we show that the naïve choice of phase for the angular kernel leads to very narrow support in the spatial domain, causing the wavefront to lose spatial information. On the right, we show the partially coherent setting, where we sample random phases to convolve the angular kernel with a Gaussian. The average intensity over time in the Fourier domain demonstrates that it achieves the desired angular spectrum while maintaining the Gaussian shape. Colormap denotes the normalized amplitude value.}
    \label{fig:justify3}
\end{figure*}

For simplicity, we consider the unit Gaussian, and this problem can be formulated as finding a set of phase profiles $\{\phi (\mathbf{k}) \}_{1, \cdots, T}$ that achieve the desired splatted Gaussian amplitude profile $u$ and angular emission profile $Y_l^m$. While this is an interesting direction for future exploration, we observe that random phase profiles sampled from the uniform distribution range of $[ -\pi, \pi ]$ can achieve the desired spatial profile, as the time-averaged spatial amplitude of Fourier spectrum of these samples would be unity. Note that this is the classical choice of a fully random phase that mimics a diffuser,~\cite{matsushima2020introduction} which corresponds to the zeroth-order spherical harmonics in our setting. Fig.~\ref{fig:justify3} shows that the spatial and angular amplitudes of $\angularspectrum_{\textrm{SH}_l^m}$ using this heuristic sampling, demonstrating that the time-averaged amplitude achieves a Gaussian in the spatial domain and a spherical harmonics basis in the angular domain. We use this strategy for Fig.~3 in the manuscript and its extended version in Fig.~\ref{fig:vde_extended}.

In this toy example, we compare occlusion handling methods for Gaussians. For all methods, we partially sample the pupil shape of a circle, translating it horizontally from -0.8 mm to 0.8 mm. This range covers the full exit pupil in this setting, with the pupil diameter being slightly smaller than the exit pupil. In Fig.~\ref{fig:vde_extended}, the first two columns show the results of the phase-matched RGBD method (\textit{first column}) and our method with a single illumination wave (\textit{second column}), which concentrates energy around the center of the exit pupil. Consequently, when pupil moves to near the edge of the eye box, the energy is filtered out. For RGBD and silhouette methods, since they require a hyperparameter (threshold) to define the depth map boundaries or the disk used to mask out waves from the rear, we test an extensive set of hyperparameters. The optimal hyperparameter would depend on the scene geometry. Notably, in all cases, the reconstruction using our method using multiple reference illuminations (sampled from a random phase distribution in the Fourier domain with zero-order spherical harmonics emission) is the closest to the target images geometrically rendered from different viewpoints.

\clearpage
\newpage
\begin{figure*}
    \centering
    \includegraphics[height=0.93\textheight]{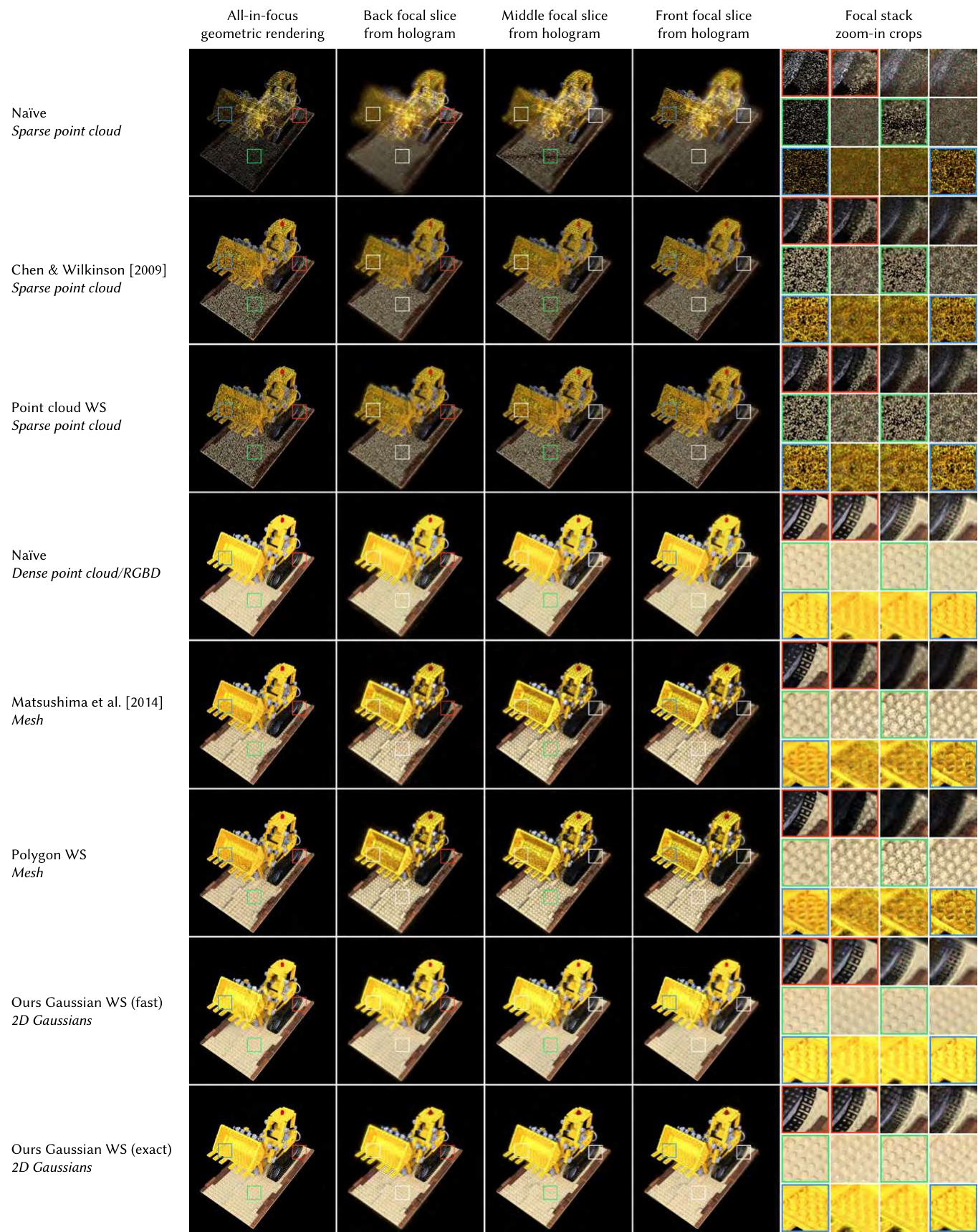}
    \caption{\textbf{Extended simulated qualitative baseline comparisons.} Point cloud CGH methods produce dim images due to sparse 3D representations. Dense point clouds like RGBD improve results but still have ringing artifacts at depth discontinuities. Polygon-based CGH lacks high-frequency details due to the per-face color constraint. In contrast, our Gaussian Wave Splatting algorithm delivers superior image quality through accurate occlusion handling and novel view synthesis enabled by alpha wave blending.}
    \label{fig:extended_baselines_1}  
\end{figure*}

\begin{figure*}
    \centering
    \includegraphics[height=0.93\textheight]{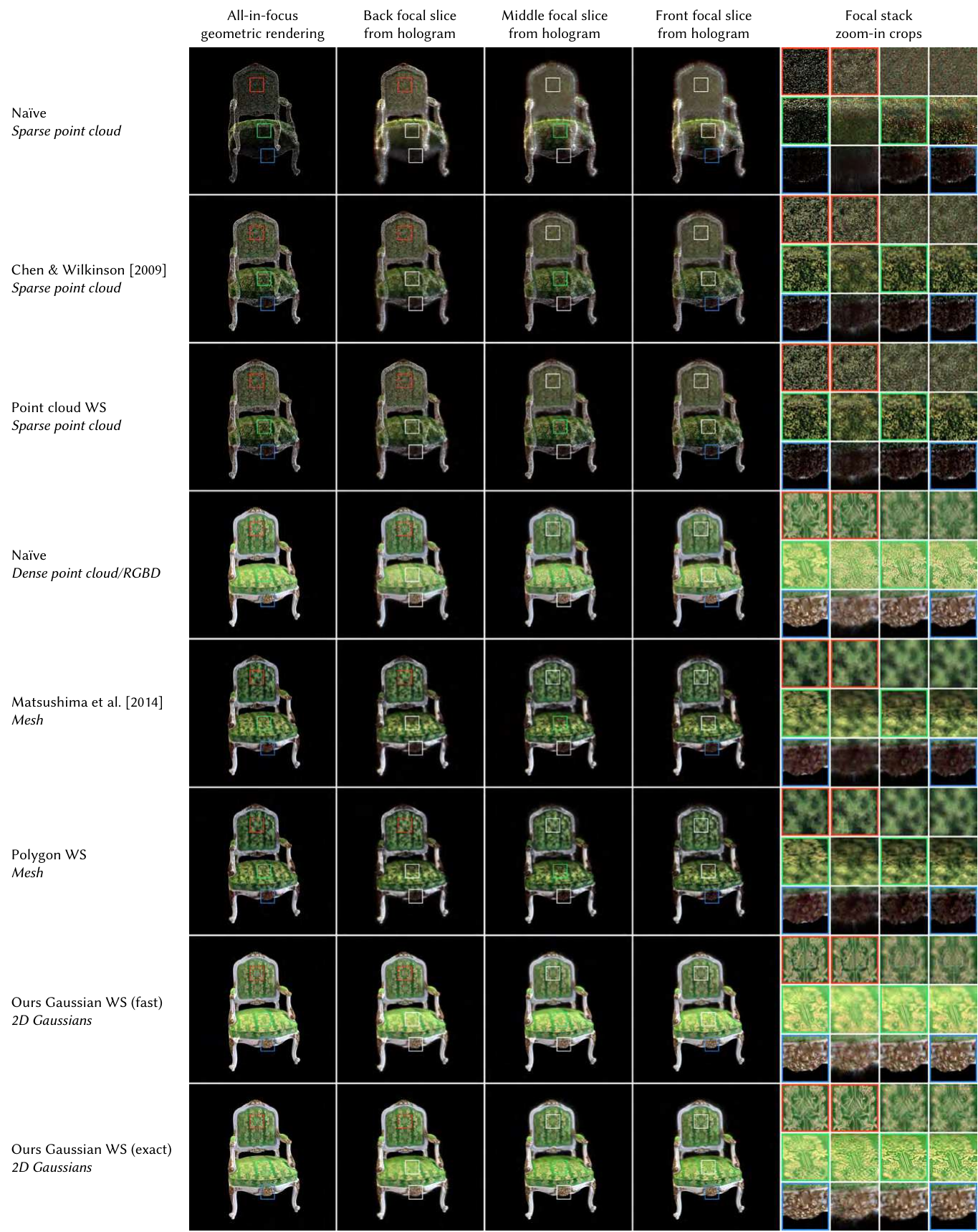}
    \caption{\textbf{Extended simulated qualitative baseline comparisons.} Point cloud CGH methods produce dim images due to sparse 3D representations. Dense point clouds like RGBD improve results but still have ringing artifacts at depth discontinuities. Polygon-based CGH lacks high-frequency details due to the per-face color constraint. In contrast, our Gaussian Wave Splatting algorithm delivers superior image quality through accurate occlusion handling and novel view synthesis enabled by alpha wave blending.}
    \label{fig:extended_baselines_2}  
\end{figure*}

\begin{figure*}
    \centering
    \includegraphics[height=0.93\textheight]{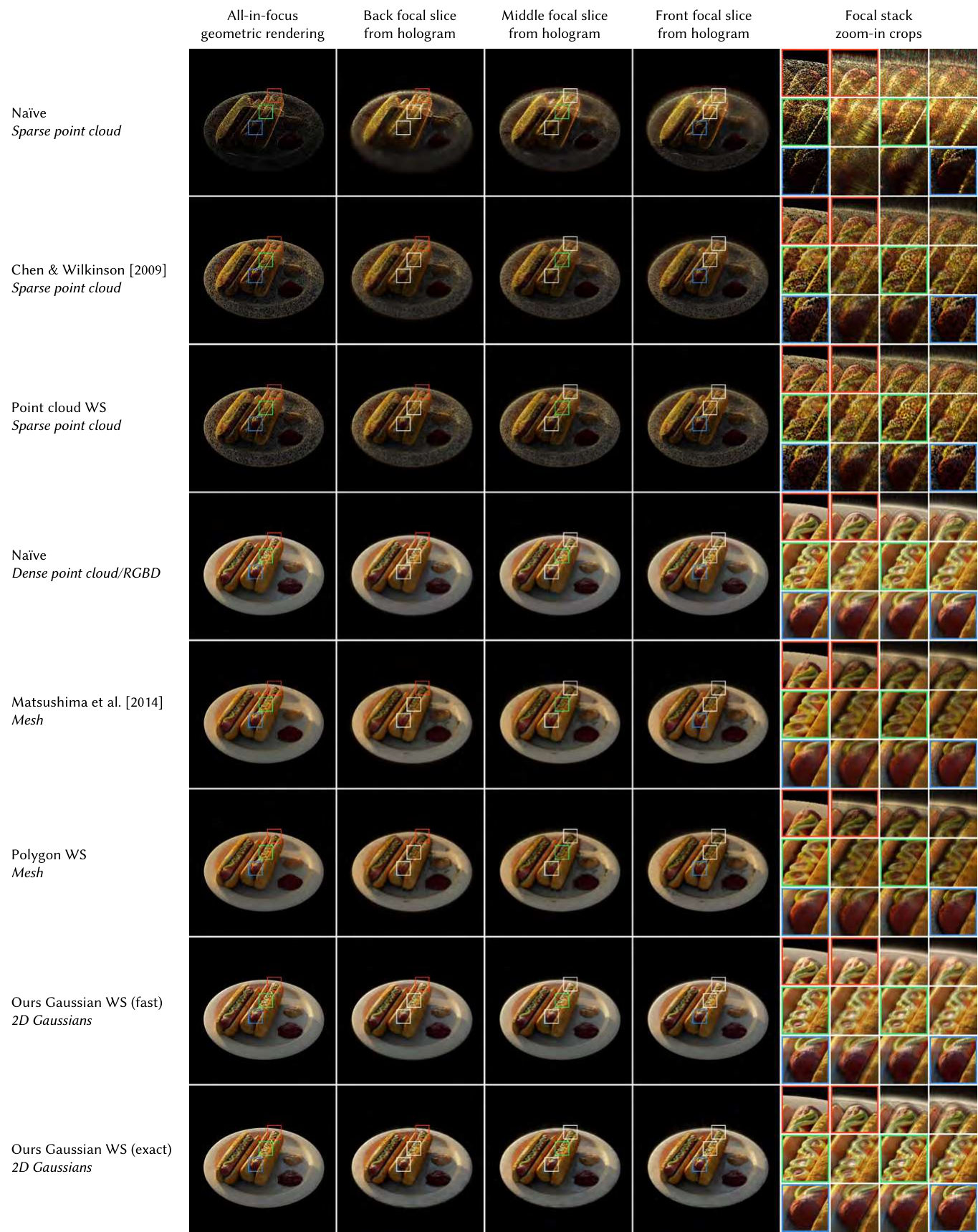}
    \caption{\textbf{Extended simulated qualitative baseline comparisons.} Point cloud CGH methods produce dim images due to sparse 3D representations. Dense point clouds like RGBD improve results but still have ringing artifacts at depth discontinuities. Polygon-based CGH lacks high-frequency details due to the per-face color constraint. In contrast, our Gaussian Wave Splatting algorithm delivers superior image quality through accurate occlusion handling and novel view synthesis enabled by alpha wave blending.}
    \label{fig:extended_baselines_3}  
\end{figure*}

\begin{figure*}
    \centering
    \includegraphics[height=0.93\textheight]{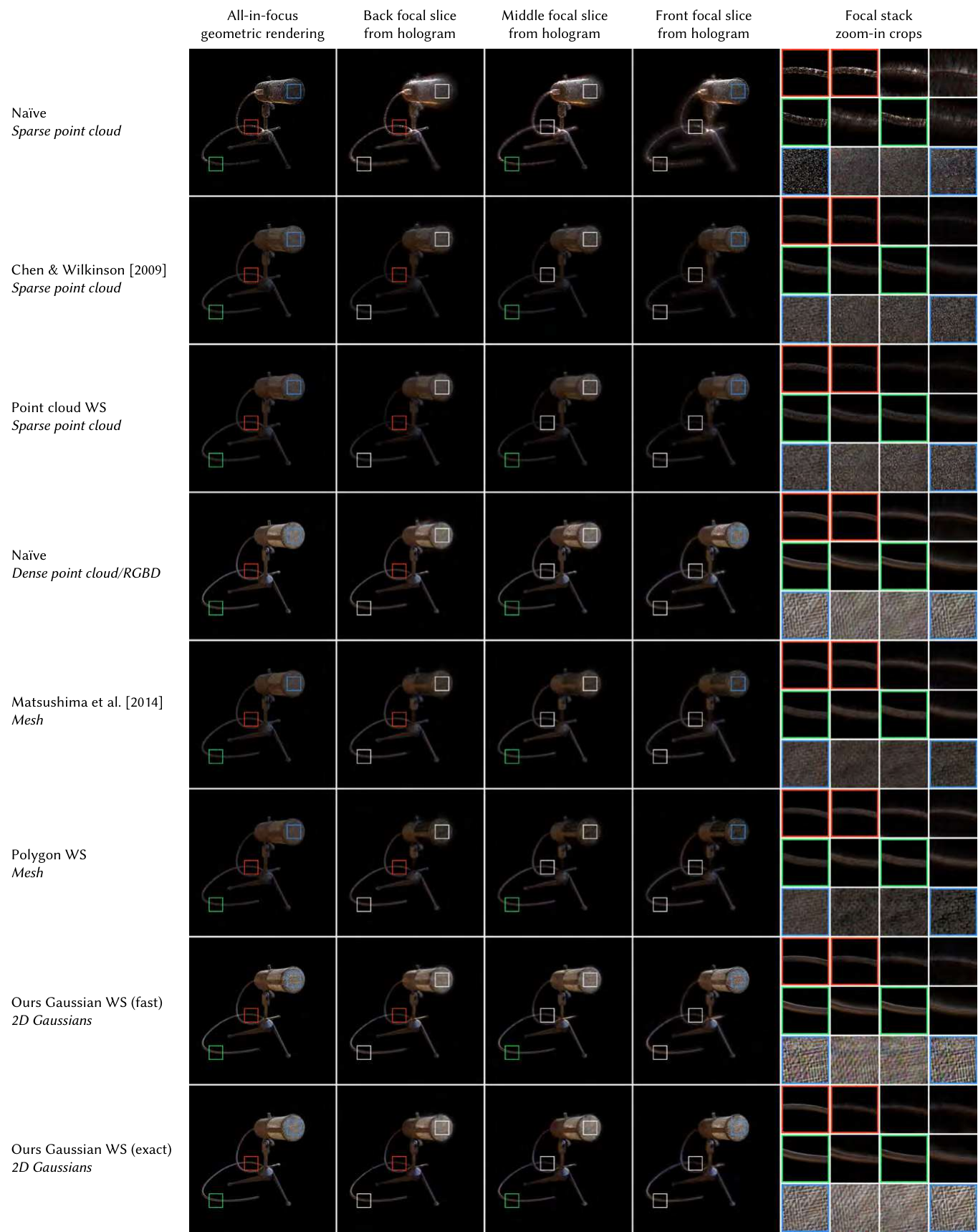}
    \caption{\textbf{Extended simulated qualitative baseline comparisons.} Point cloud CGH methods produce dim images due to sparse 3D representations. Dense point clouds like RGBD improve results but still have ringing artifacts at depth discontinuities. Polygon-based CGH lacks high-frequency details due to the per-face color constraint. In contrast, our Gaussian Wave Splatting algorithm delivers superior image quality through accurate occlusion handling and novel view synthesis enabled by alpha wave blending.}
    \label{fig:extended_baselines_4}  
\end{figure*}

\begin{figure*}
    \centering
    \includegraphics[height=0.93\textheight]{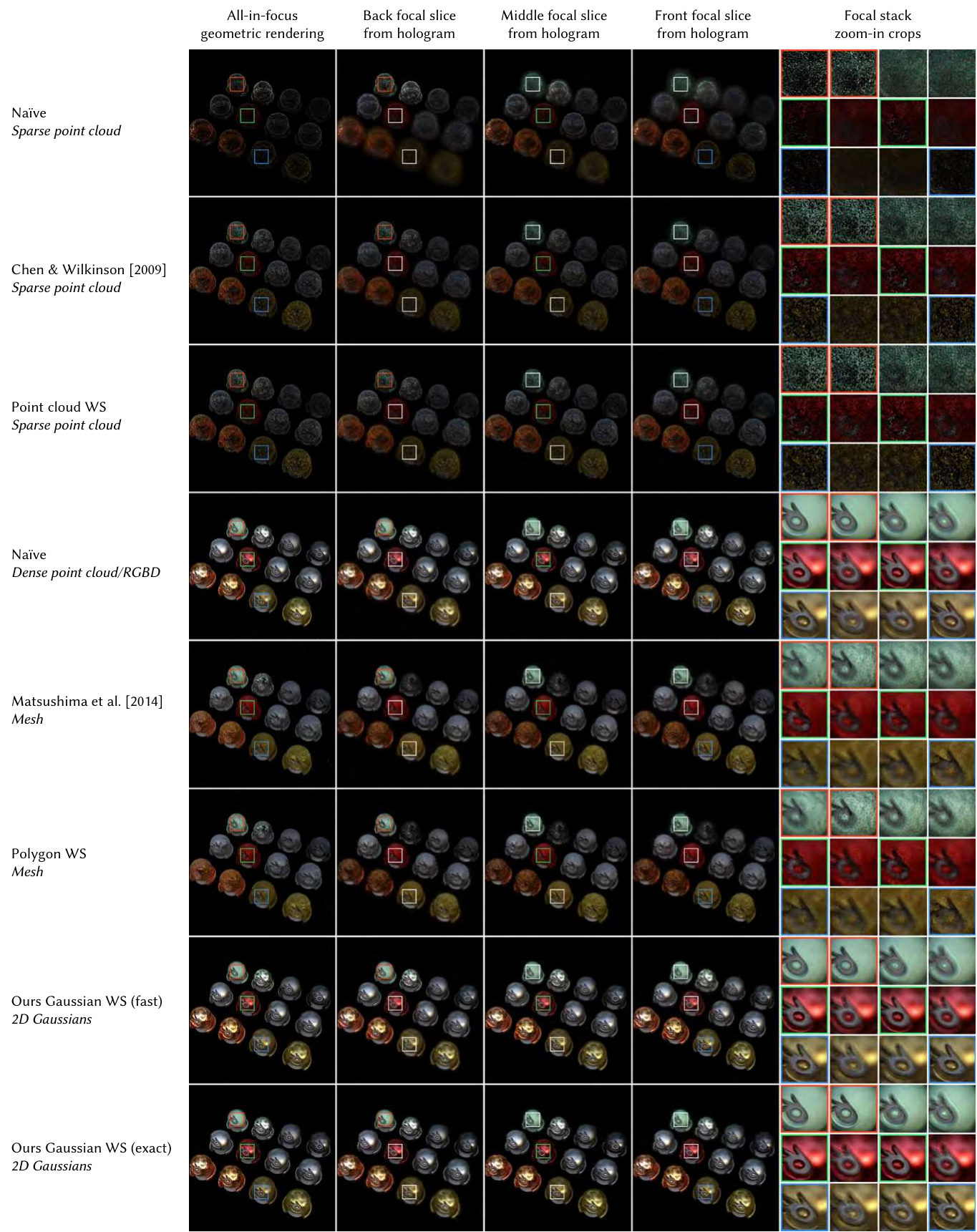}
    \caption{\textbf{Extended simulated qualitative baseline comparisons.} Point cloud CGH methods produce dim images due to sparse 3D representations. Dense point clouds like RGBD improve results but still have ringing artifacts at depth discontinuities. Polygon-based CGH lacks high-frequency details due to the per-face color constraint. In contrast, our Gaussian Wave Splatting algorithm delivers superior image quality through accurate occlusion handling and novel view synthesis enabled by alpha wave blending.}
    \label{fig:extended_baselines_5}  
\end{figure*}

\begin{figure*}
    \centering
    \includegraphics[height=0.93\textheight]{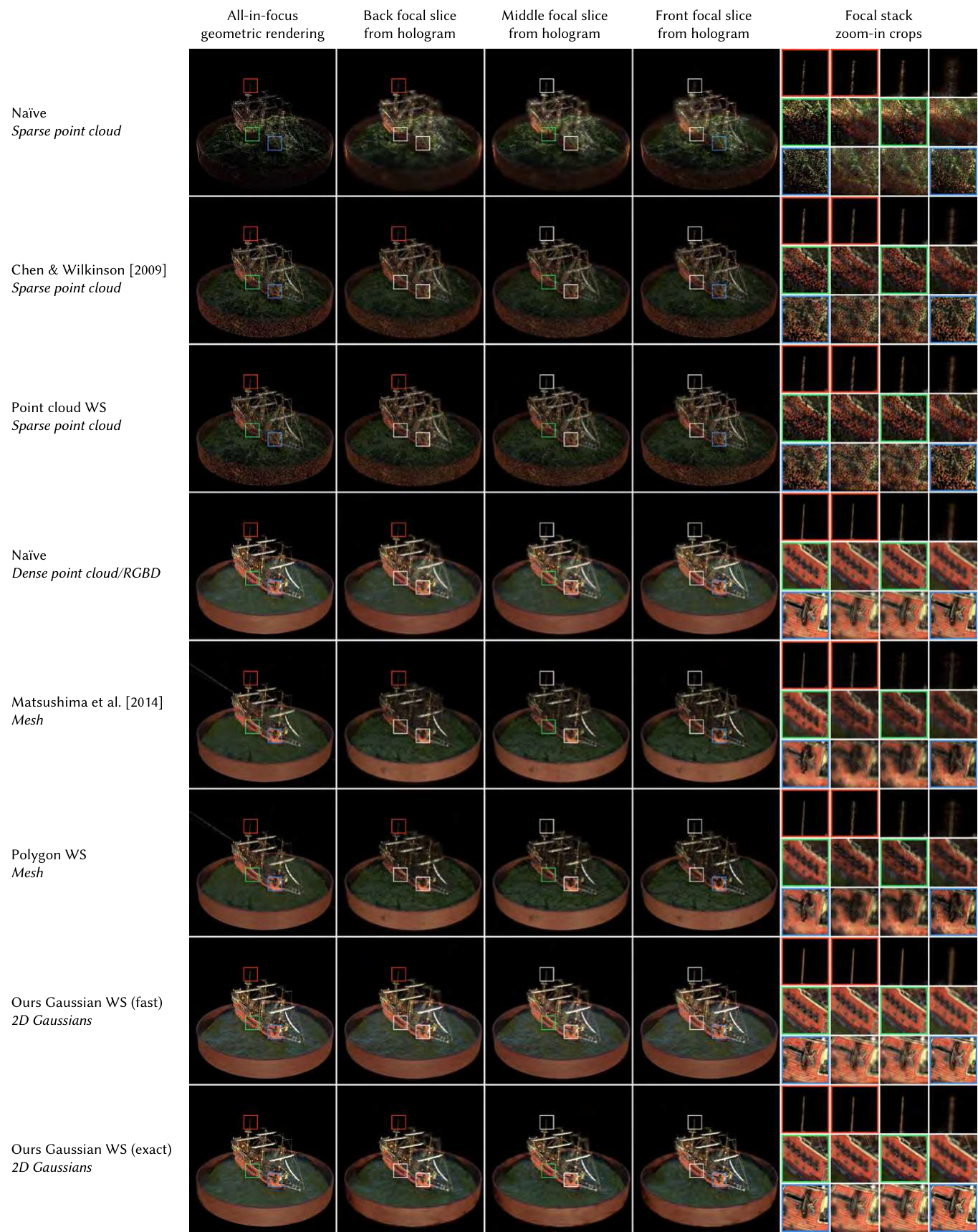}
    \caption{\textbf{Extended simulated qualitative baseline comparisons.} Point cloud CGH methods produce dim images due to sparse 3D representations. Dense point clouds like RGBD improve results but still have ringing artifacts at depth discontinuities. Polygon-based CGH lacks high-frequency details due to the per-face color constraint. In contrast, our Gaussian Wave Splatting algorithm delivers superior image quality through accurate occlusion handling and novel view synthesis enabled by alpha wave blending.}
    \label{fig:extended_baselines_6}  
\end{figure*}

\begin{figure*}
    \centering
    \includegraphics[width=\textwidth]{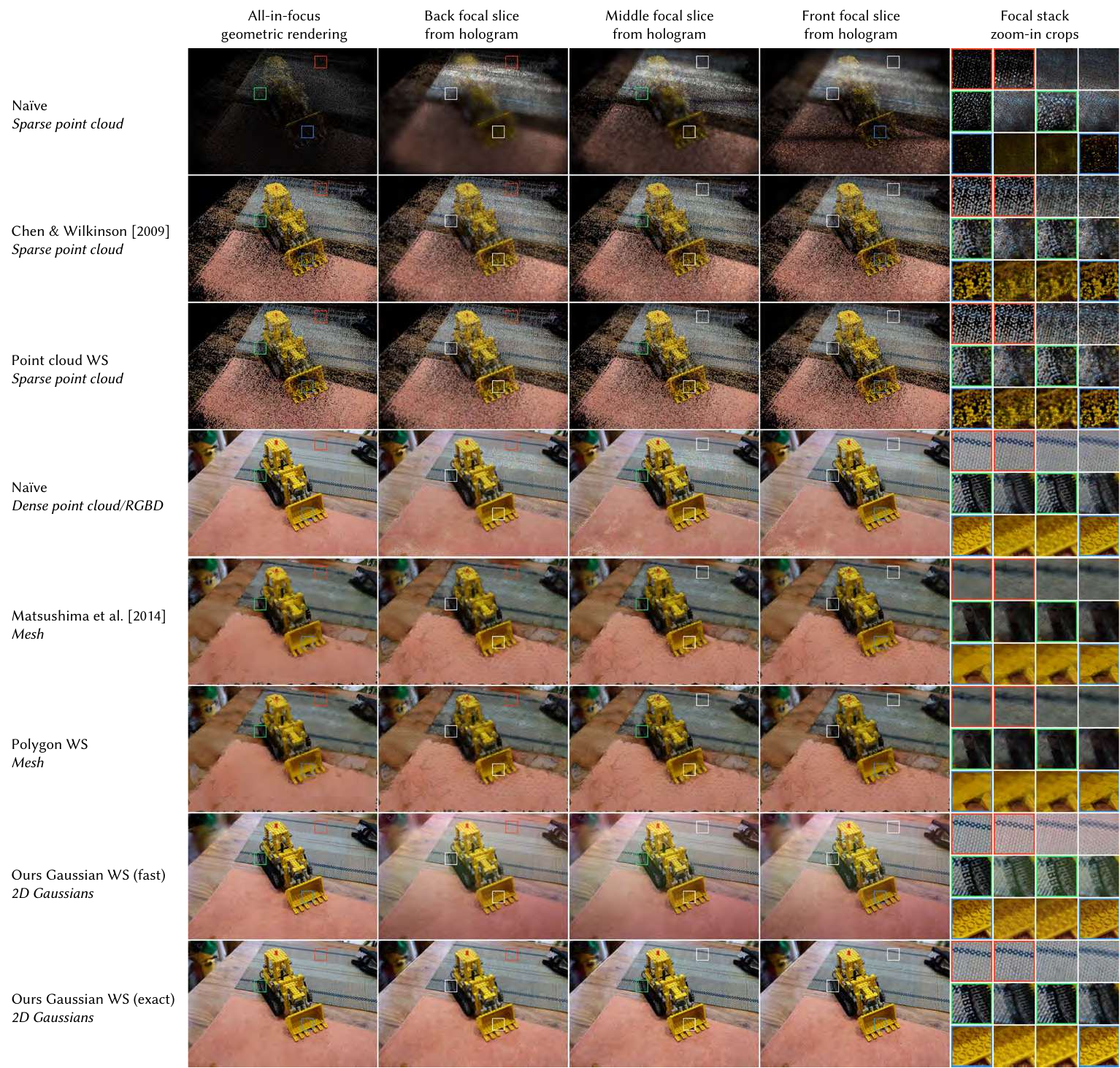}
    \caption{\textbf{Extended simulated qualitative baseline comparisons.} Point cloud CGH methods produce dim images due to sparse 3D representations. Dense point clouds like RGBD improve results but still have ringing artifacts at depth discontinuities. Polygon-based CGH lacks high-frequency details due to the per-face color constraint. In contrast, our Gaussian Wave Splatting algorithm delivers superior image quality through accurate occlusion handling and novel view synthesis enabled by alpha wave blending.}
    \label{fig:extended_baselines_7}  
\end{figure*}

\begin{figure*}
    \centering
    \includegraphics[width=\textwidth]{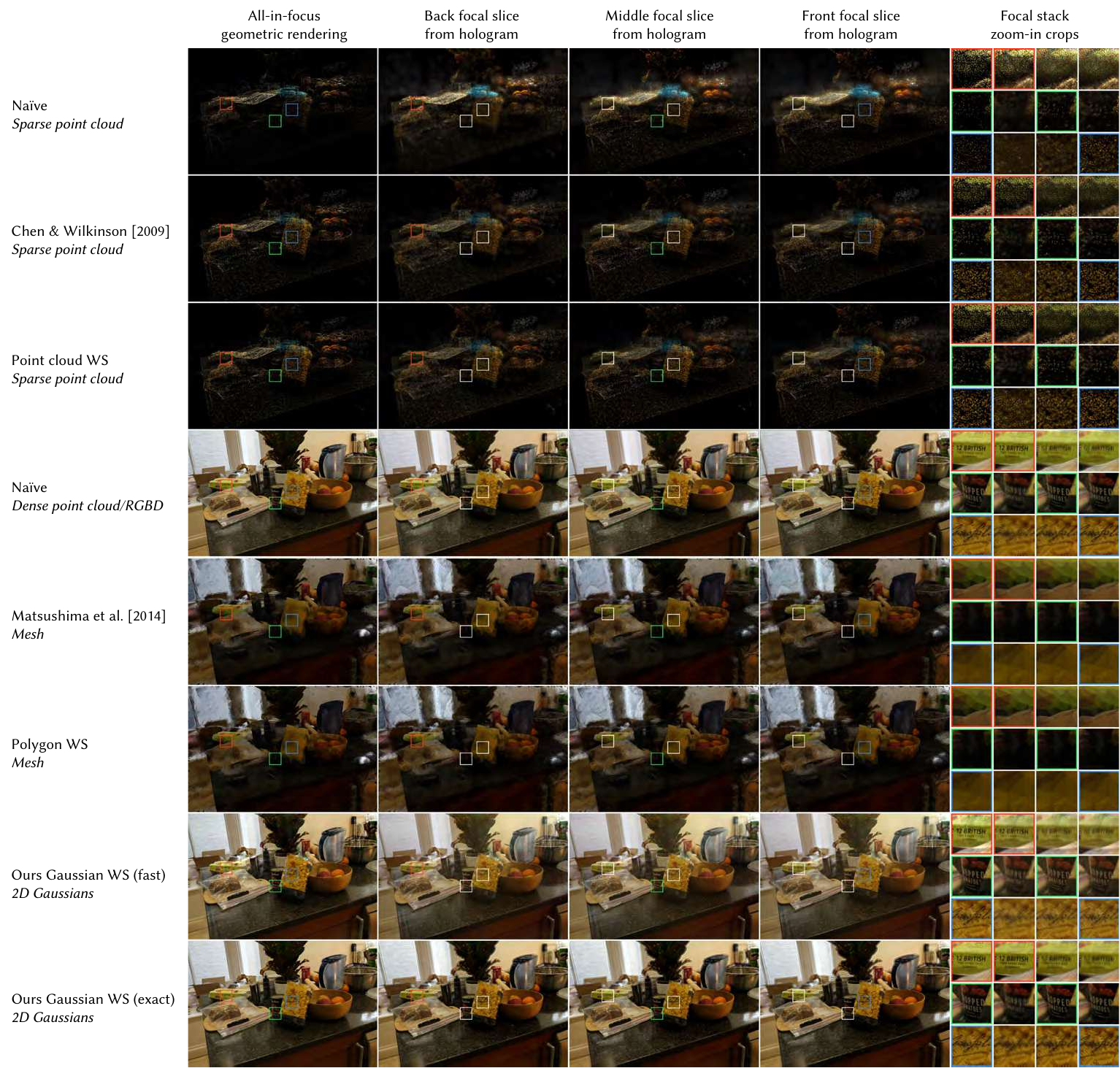}
    \caption{\textbf{Extended simulated qualitative baseline comparisons.} Point cloud CGH methods produce dim images due to sparse 3D representations. Dense point clouds like RGBD improve results but still have ringing artifacts at depth discontinuities. Polygon-based CGH lacks high-frequency details due to the per-face color constraint. In contrast, our Gaussian Wave Splatting algorithm delivers superior image quality through accurate occlusion handling and novel view synthesis enabled by alpha wave blending.}
    \label{fig:extended_baselines_8}  
\end{figure*}

\begin{figure*}
    \centering
    \includegraphics[width=\textwidth]{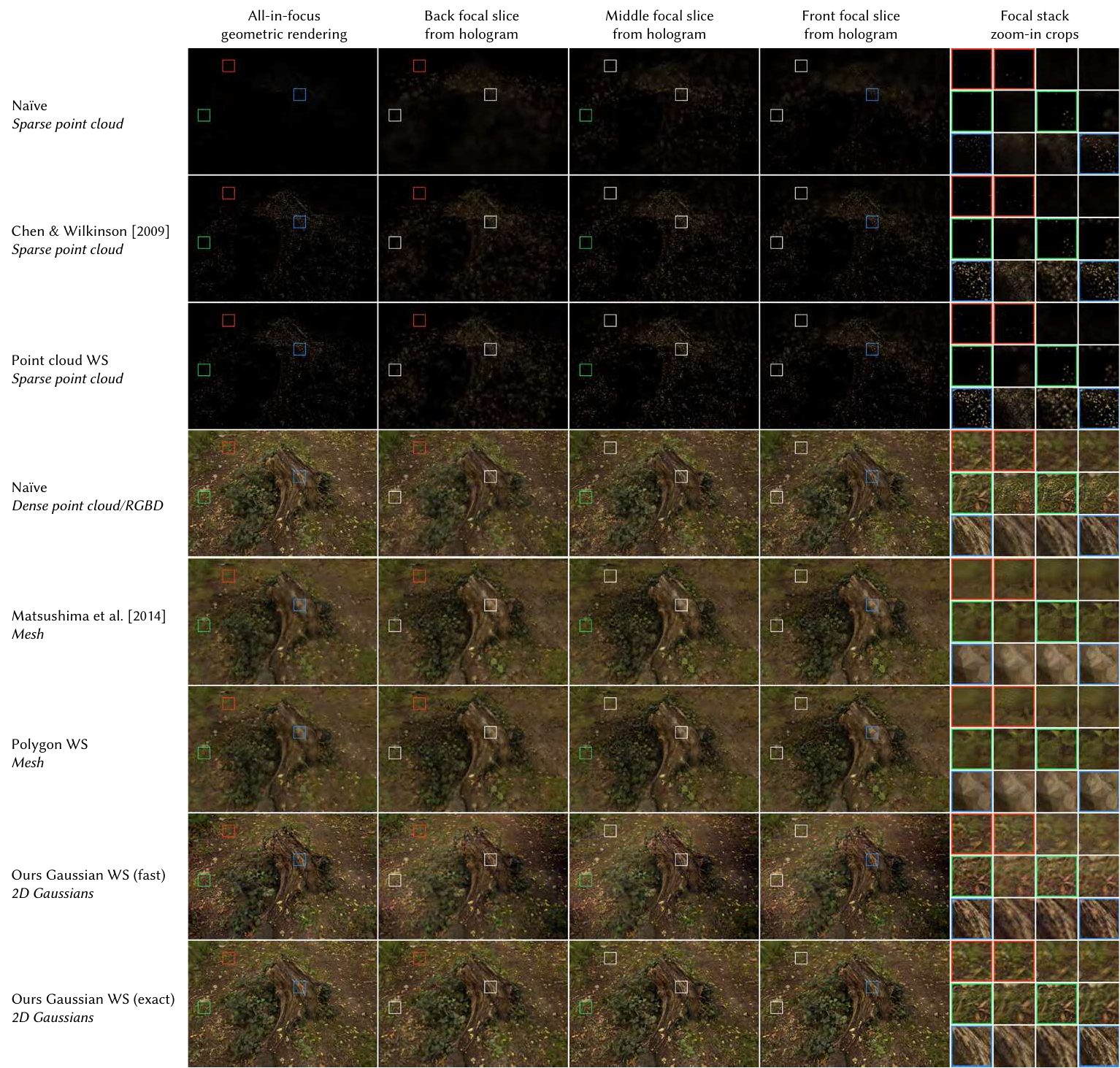}
    \caption{\textbf{Extended simulated qualitative baseline comparisons.} Point cloud CGH methods produce dim images due to sparse 3D representations. Dense point clouds like RGBD improve results but still have ringing artifacts at depth discontinuities. Polygon-based CGH lacks high-frequency details due to the per-face color constraint. In contrast, our Gaussian Wave Splatting algorithm delivers superior image quality through accurate occlusion handling and novel view synthesis enabled by alpha wave blending.}
    \label{fig:extended_baselines_9}  
\end{figure*}

\begin{figure*}
    \centering
    \includegraphics[width=\textwidth]{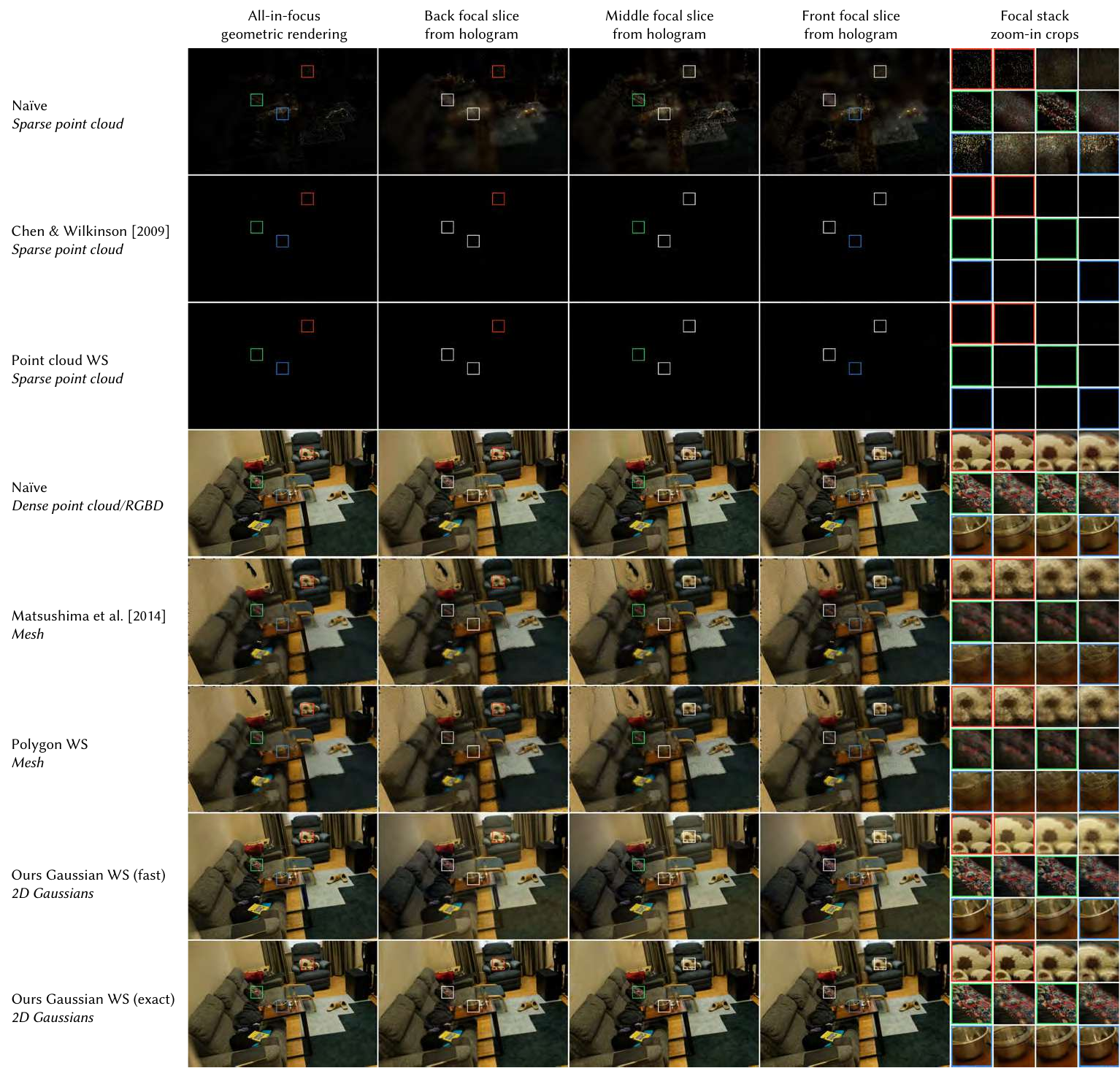}
    \caption{\textbf{Extended simulated qualitative baseline comparisons.} Point cloud CGH methods produce dim images due to sparse 3D representations. Dense point clouds like RGBD improve results but still have ringing artifacts at depth discontinuities. Polygon-based CGH lacks high-frequency details due to the per-face color constraint. In contrast, our Gaussian Wave Splatting algorithm delivers superior image quality through accurate occlusion handling and novel view synthesis enabled by alpha wave blending.}
    \label{fig:extended_baselines_10}  
\end{figure*}

\begin{figure*}
    \centering
    \includegraphics[width=\textwidth]{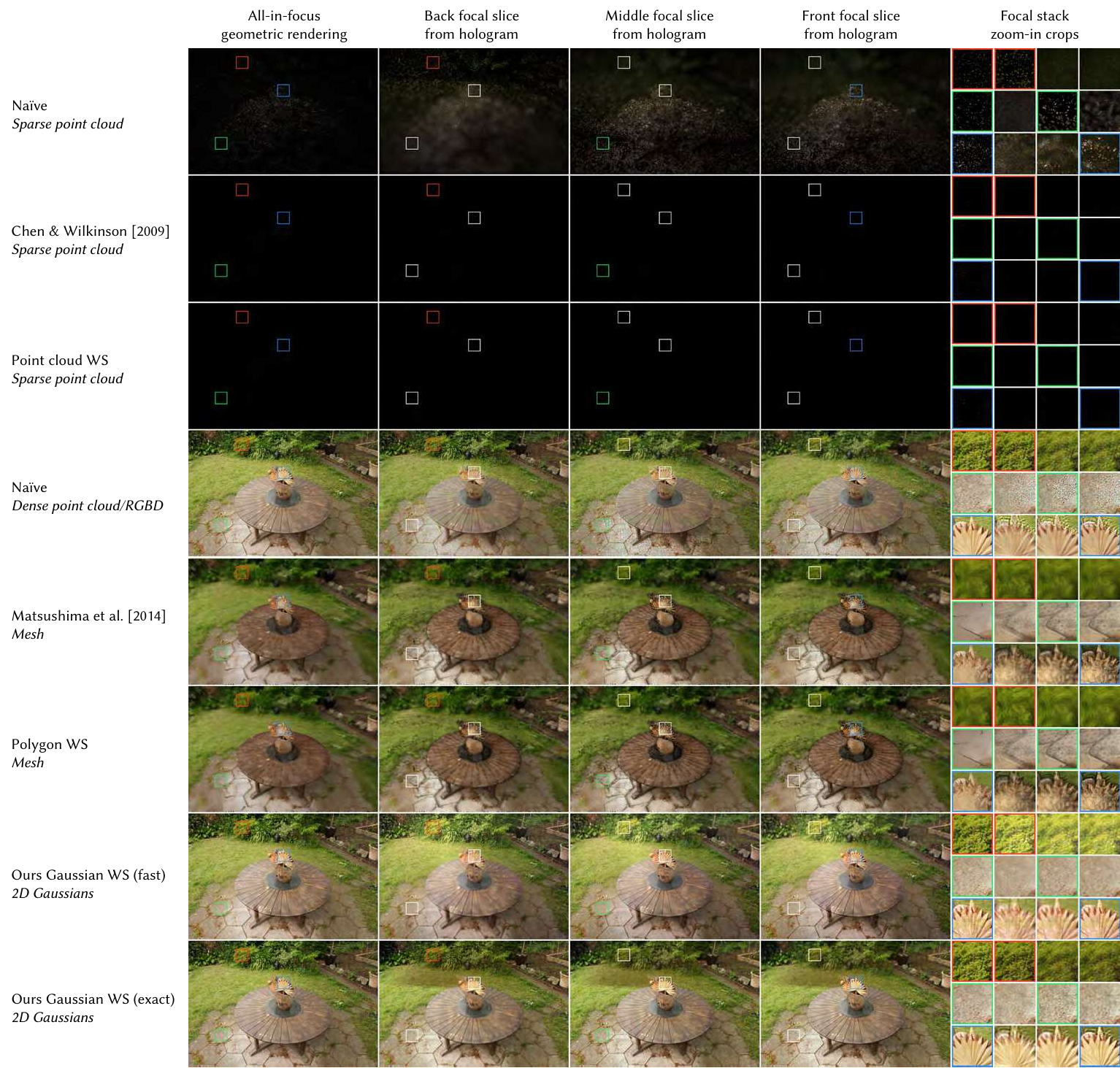}
    \caption{\textbf{Extended simulated qualitative baseline comparisons.} Point cloud CGH methods produce dim images due to sparse 3D representations. Dense point clouds like RGBD improve results but still have ringing artifacts at depth discontinuities. Polygon-based CGH lacks high-frequency details due to the per-face color constraint. In contrast, our Gaussian Wave Splatting algorithm delivers superior image quality through accurate occlusion handling and novel view synthesis enabled by alpha wave blending.}
    \label{fig:extended_baselines_11}  
\end{figure*}

\begin{figure*}
    \centering
    \includegraphics[width=\textwidth]{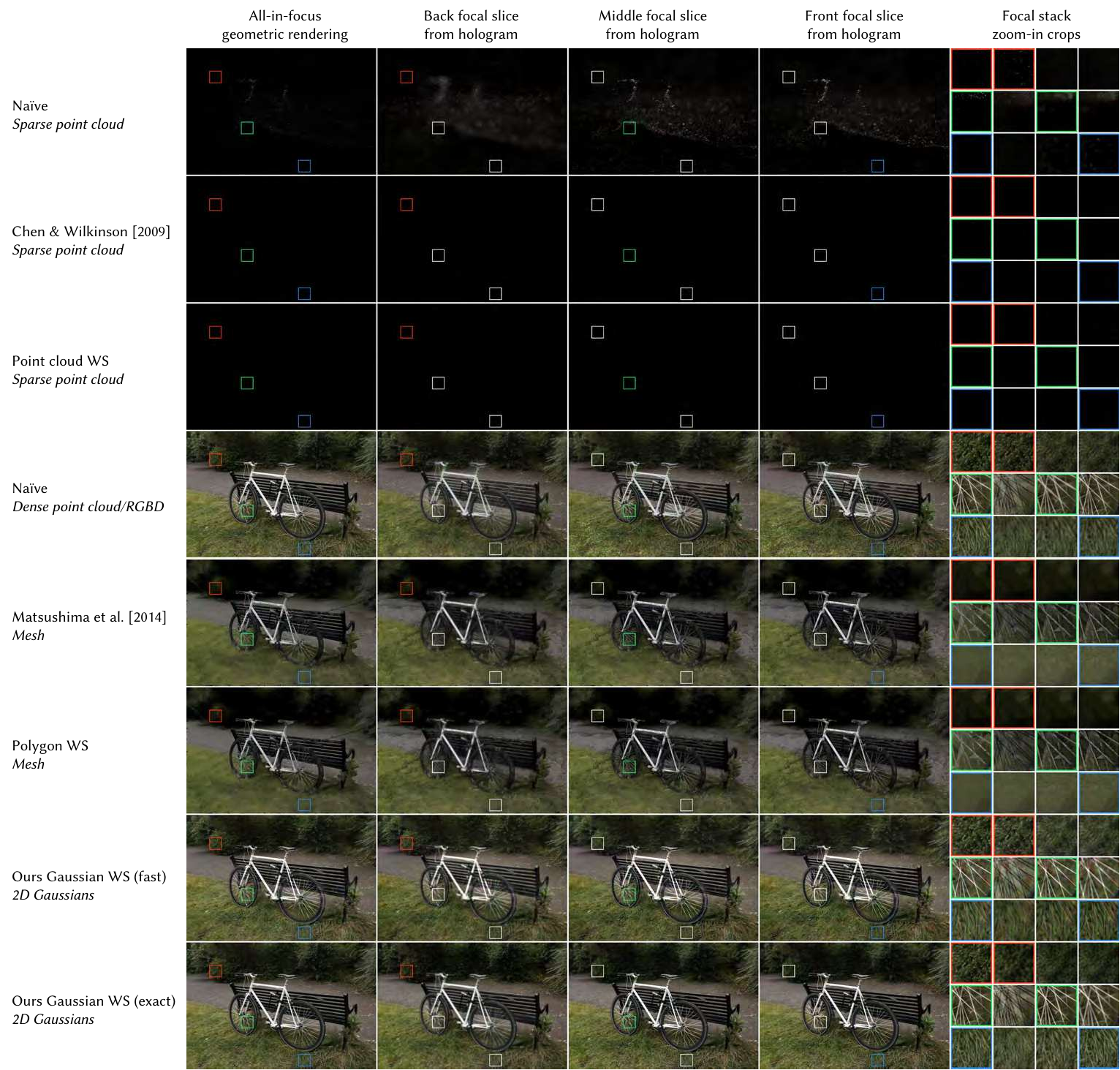}
    \caption{\textbf{Extended simulated qualitative baseline comparisons.} Point cloud CGH methods produce dim images due to sparse 3D representations. Dense point clouds like RGBD improve results but still have ringing artifacts at depth discontinuities. Polygon-based CGH lacks high-frequency details due to the per-face color constraint. In contrast, our Gaussian Wave Splatting algorithm delivers superior image quality through accurate occlusion handling and novel view synthesis enabled by alpha wave blending.}
    \label{fig:extended_baselines_12}  
\end{figure*}

\begin{figure*}
    \centering
    \includegraphics[width=\textwidth]{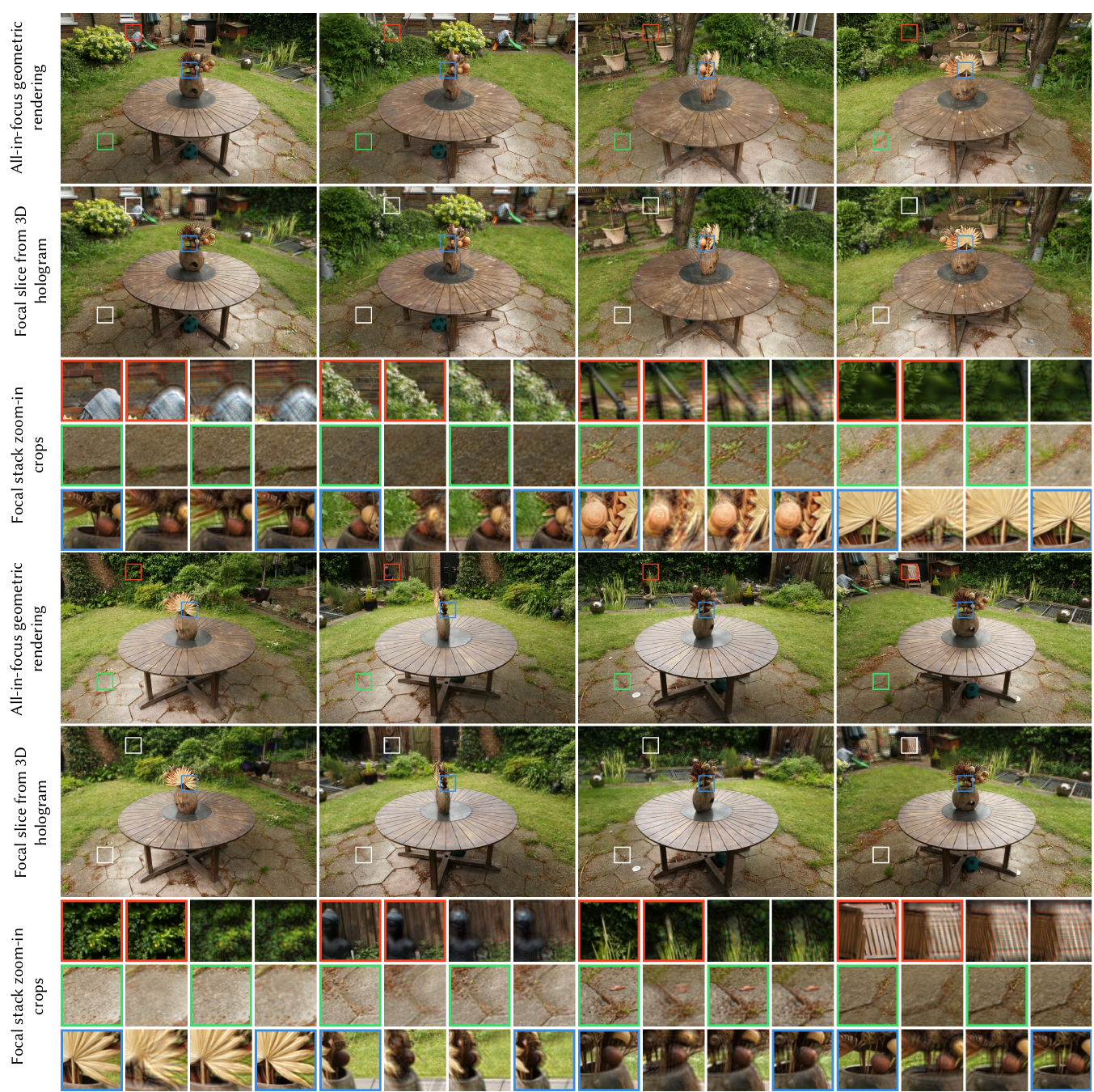}
    \caption{\textbf{Simulated 3D focal stack of holograms of novel views.} We synthesize holograms using exact Gaussian Wave Splatting from different novel viewpoints, and verify 3D focusing capabilities. Gaussian Wave Splatting accurately reconstructs sharp details in focused regions and synthesizes natural blur in defocus regions at \textit{all} viewpoints, demonstrating the robustness of our method. }
    \label{fig:nvs_fs}
\end{figure*}


\begin{figure*}
    \centering
    \includegraphics[width=\textwidth]{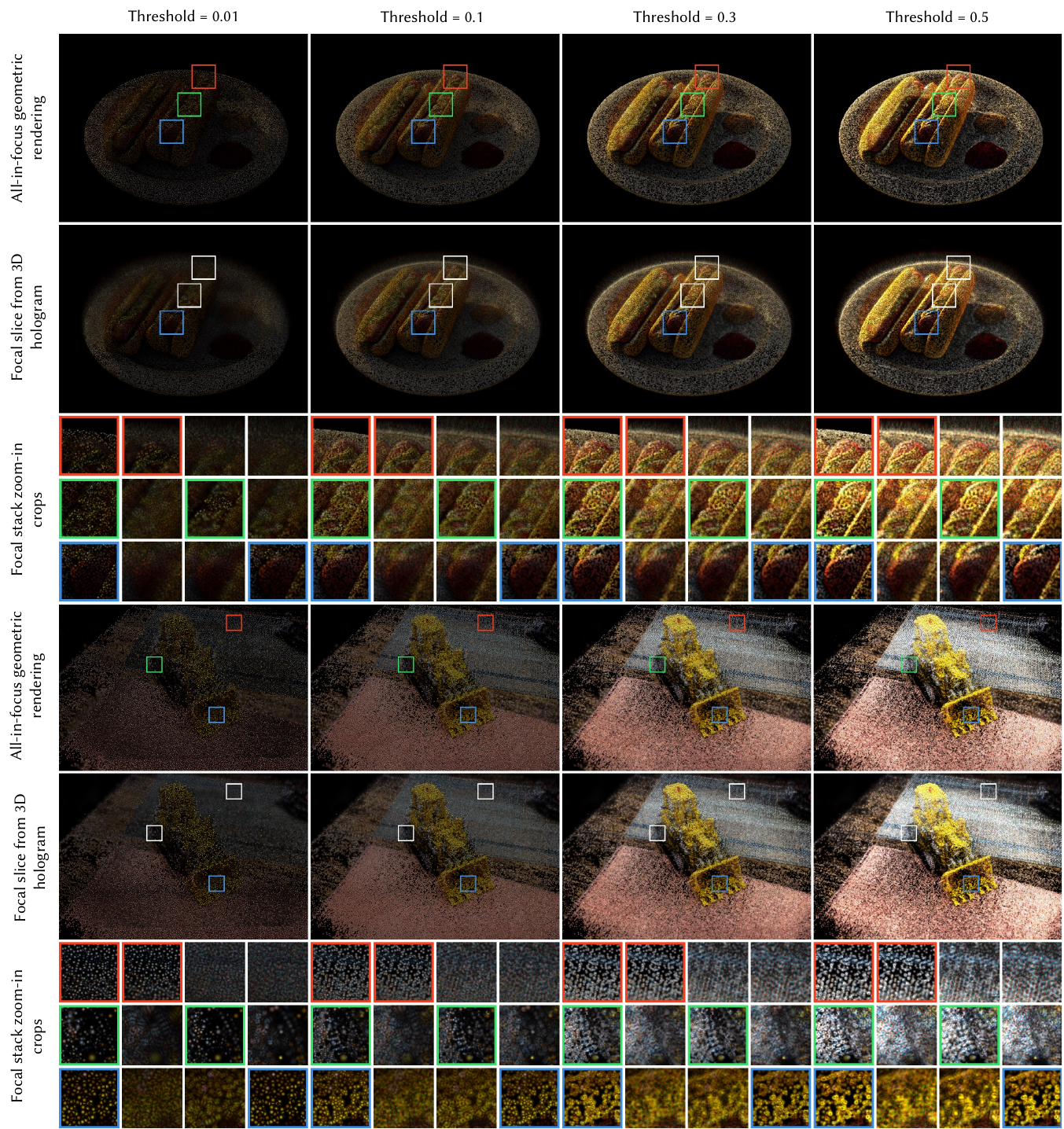}
    \caption{\textbf{Ablation studies on visibility test disk size in point cloud CGH. } We vary the binarization threshold that determines the disk size used in the visibility test described in Chen \& Wilkinson \cite{chen2009computer}. Smaller threshold leads to larger disks, resulting in dimmer images. As the threshold increases, occlusion handling becomes more lenient and might lead to oversaturated colors. However, this threshold is not robust and needs to be heuristically tuned. }
    \label{fig:pc_disk_ablation}  
\end{figure*}

\begin{figure*}
    \centering
    \includegraphics[width=\textwidth]{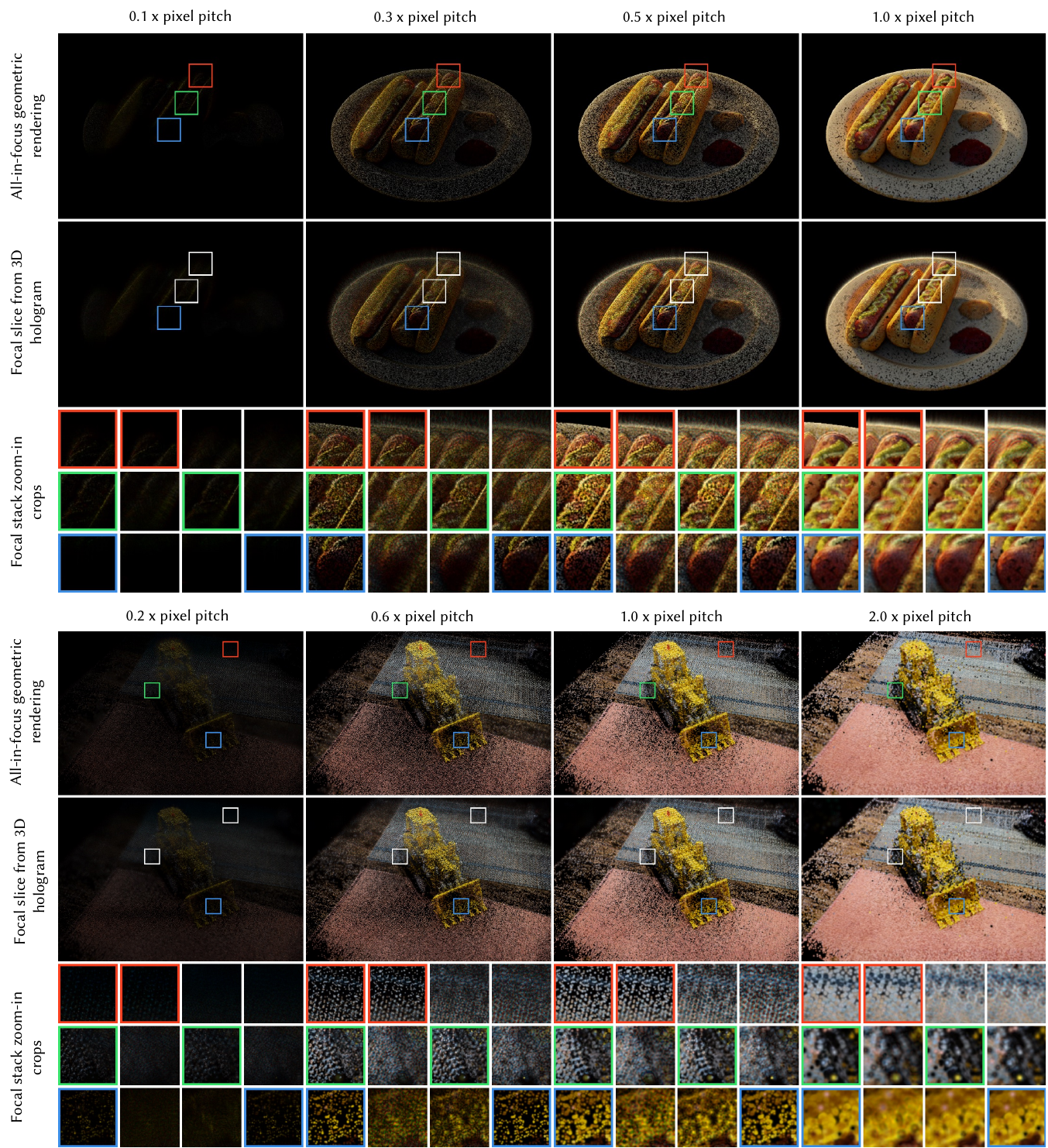}
    \caption{\textbf{Ablation studies on Gaussian scale in point cloud CGH. } We vary the Gaussian scale when treating points as isotropic Gaussians and applying Wave Splatting on the isotropic Gaussians. We define the size of the point as a multiple of the SLM pixel pitch. Smaller scale leads to smaller points, resulting in dimmer images and dark holes. As the Gaussian scale increases, the points become bigger and the image is more densely populated. However, this scale parameter is not robust and needs to be heuristically tuned.  }
    \label{fig:pc_awb_ablation}  
\end{figure*}

\begin{figure*}
    \centering
    \includegraphics[angle=90, height=0.85\textheight]{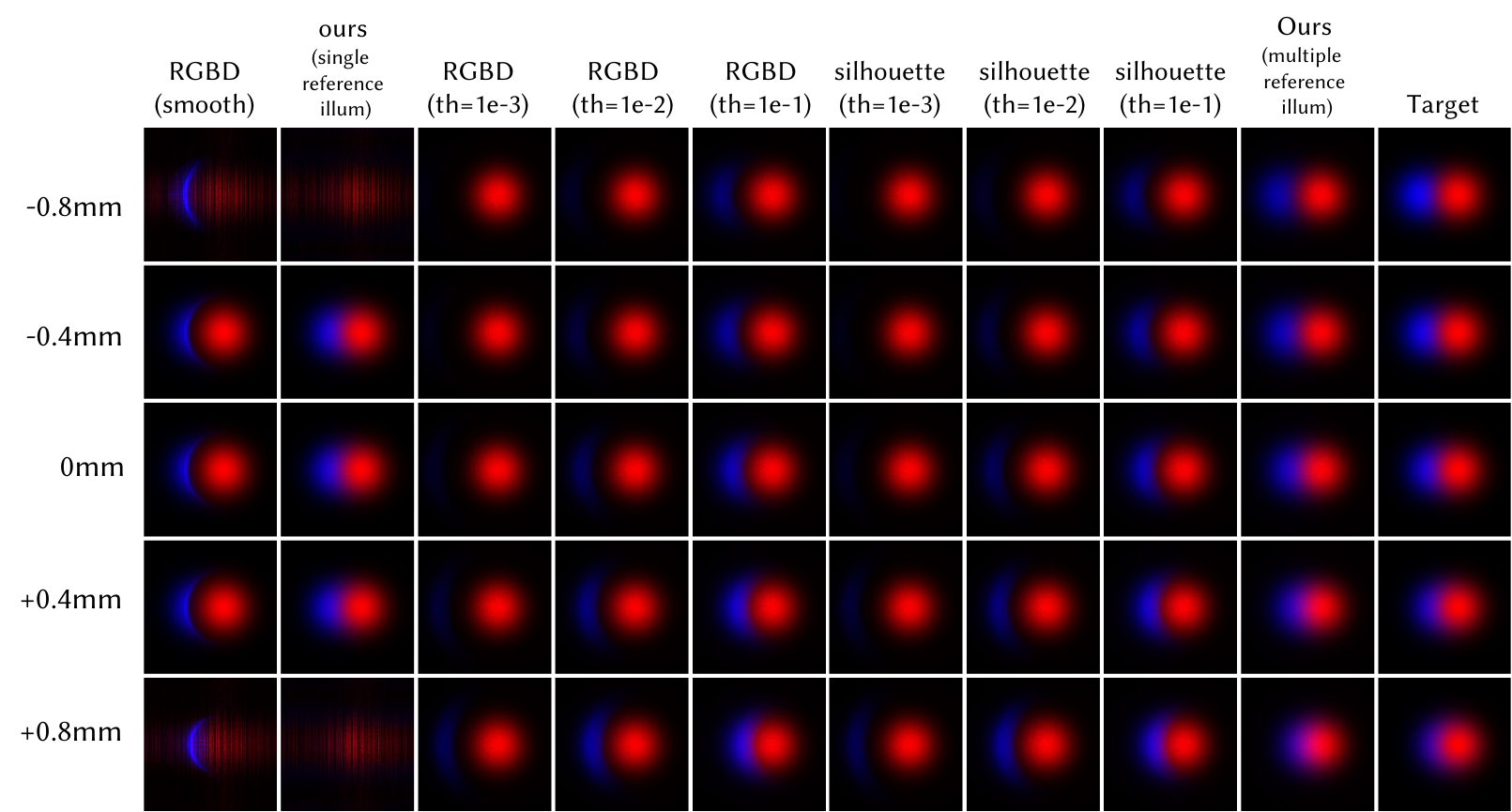}
    \caption{\textbf{Extended comparisons on occlusion handling for Gaussians.}  In this toy example, we compare occlusion handling methods for Gaussians. For all methods, we partially sample the pupil shape of a circle, translating it horizontally from -0.8 mm to 0.8 mm. This range covers the full exit pupil in our setting, with the pupil diameter being slightly smaller than the exit pupil. The first two columns show the results of the phase-matched RGBD method (\textit{first column}) and our method with a single illumination wave (\textit{second column}), which concentrates energy around the center of the exit pupil. Consequently, when pupil moves to near the edge of the eye box, the energy is filtered out. For RGBD and silhouette methods, since they require a hyperparameter (threshold) to define the depth map boundaries or the disk used to mask out waves from the rear, we test an extensive set of hyperparameters. The optimal hyperparameter would depend on the scene geometry. Notably, in all cases, the reconstruction using our method using multiple reference illuminations (sampled from a random phase distribution in the Fourier domain with zero-order spherical harmonics emission) is the closest to the target images geometrically rendered from different viewpoints.}
    \label{fig:vde_extended}  
\end{figure*}

\clearpage
\newpage
\section{Additional Experimental Results}
\label{sec:experiments}
\subsection{Experimentally Captured Baseline Comparisons}

We capture an extensive set of results of baseline methods and our Gaussian Wave Splatting algorithm applied to a wide variety of 3D scenes from the NeRF synthetic Blender \cite{mildenhall2020nerf} and Mip-NeRF 360 dataset \cite{barron2022mipnerf360}. Aside from point-based methods and polygon-based method shown in the main paper, we additionally show dense point cloud (RGBD) results.

\begin{figure*}
    \centering
    \includegraphics[width=\textwidth]{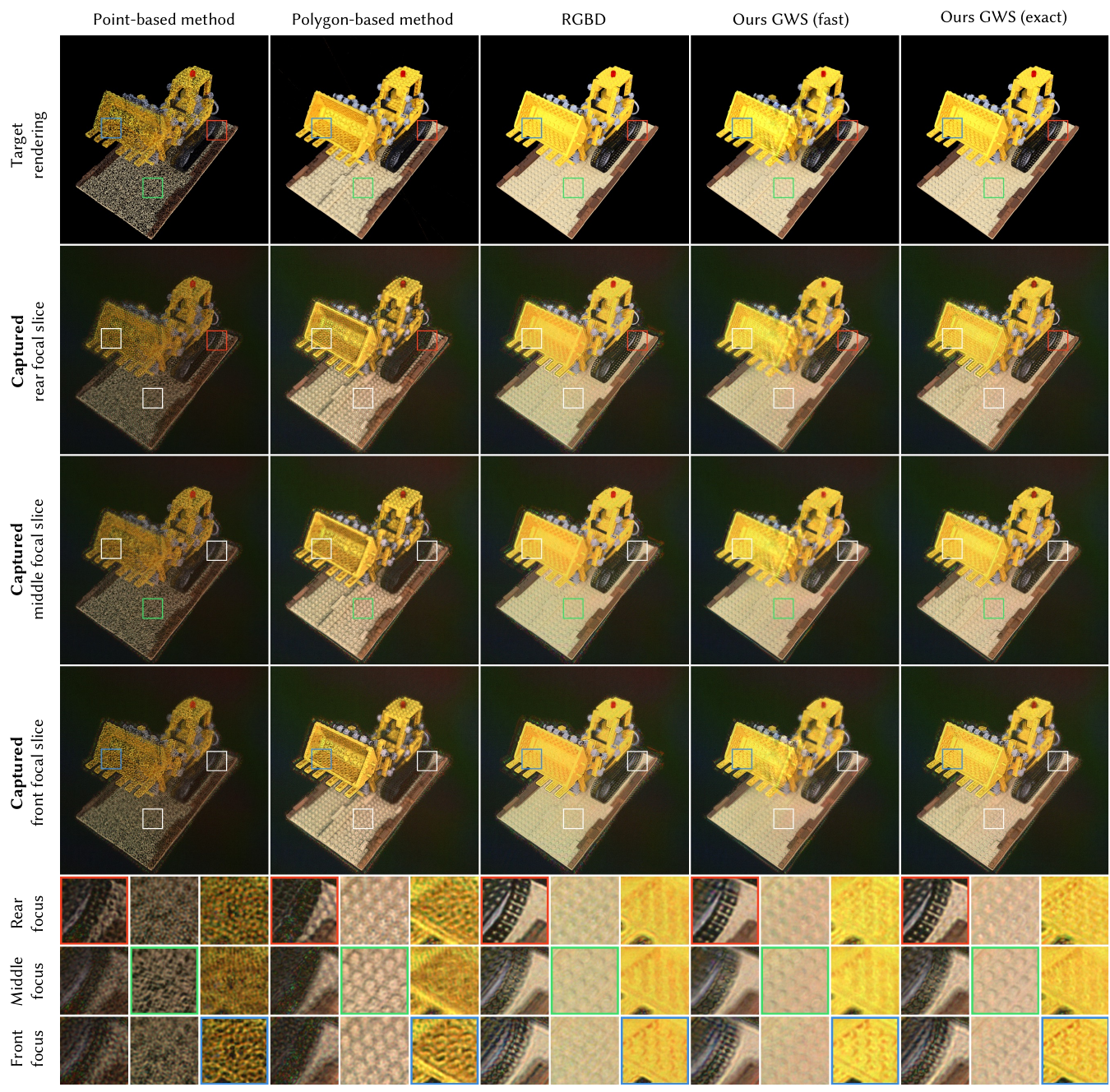}
    \caption{\textbf{Experimentally captured baseline comparisons.} We experimentally capture 3D focal stacks of holograms generated using different baseline CGH methods, and the results well match our simulations. Point-based methods result in overly dim images and poor reconstruction quality due to the sparseness of point cloud representatins. Polygon-based methods cannot reconstruct fine details. CGH from RGBD images achieve decent image quality. Our GWS achieve superior image quality over prior methods, reconstructing sharp details in focused regions.}
    \label{fig:captured_baselines_1}
\end{figure*}

\begin{figure*}
    \centering
    \includegraphics[width=\textwidth]{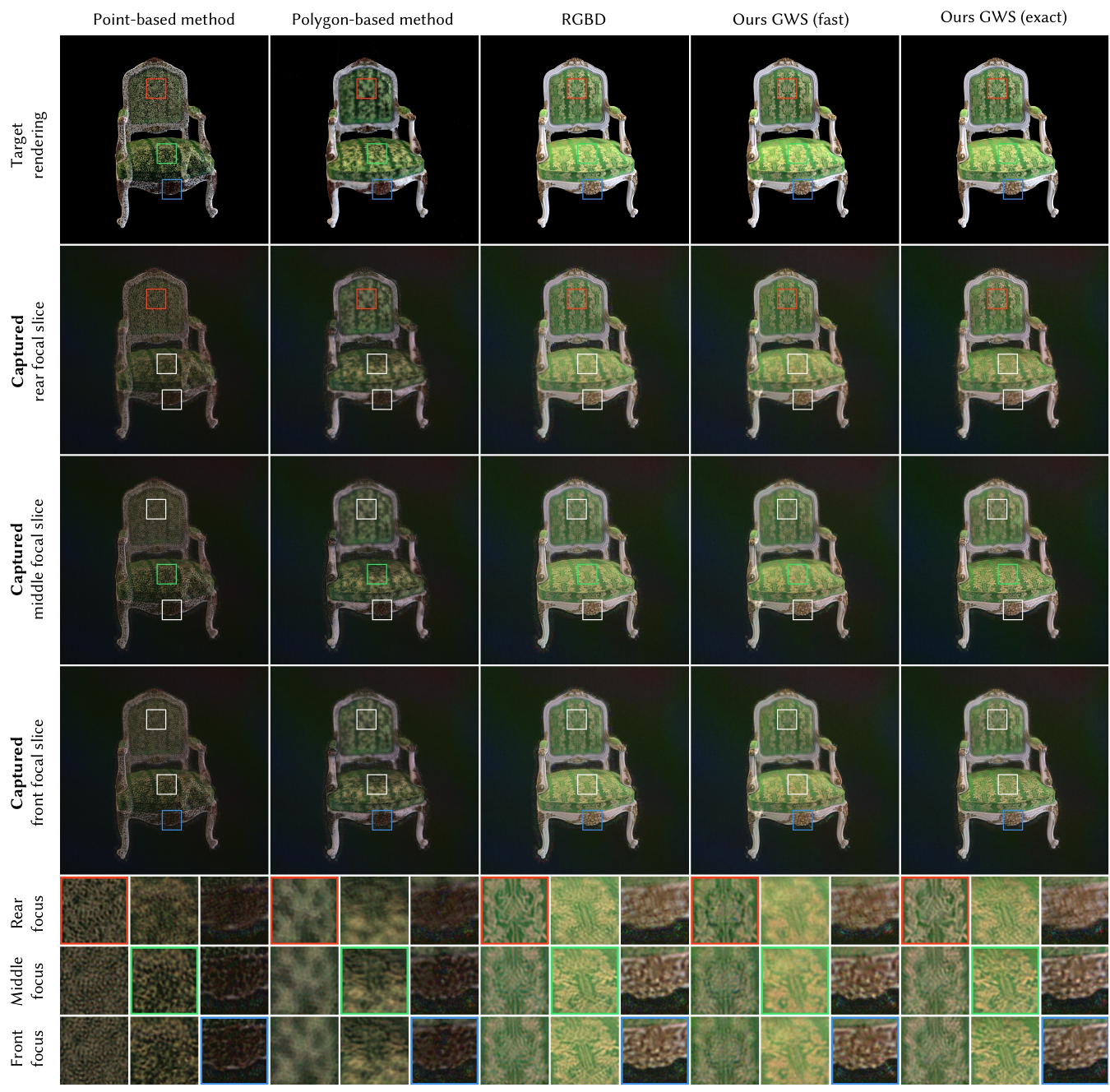}
    \caption{\textbf{Experimentally captured baseline comparisons.} We experimentally capture 3D focal stacks of holograms generated using different baseline CGH methods, and the results well match our simulations. Point-based methods result in overly dim images and poor reconstruction quality due to the sparseness of point cloud representatins. Polygon-based methods cannot reconstruct fine details. CGH from RGBD images achieve decent image quality. Our GWS achieve superior image quality over prior methods, reconstructing sharp details in focused regions.}
    \label{fig:captured_baselines_2}
\end{figure*}

\begin{figure*}
    \centering
    \includegraphics[width=\textwidth]{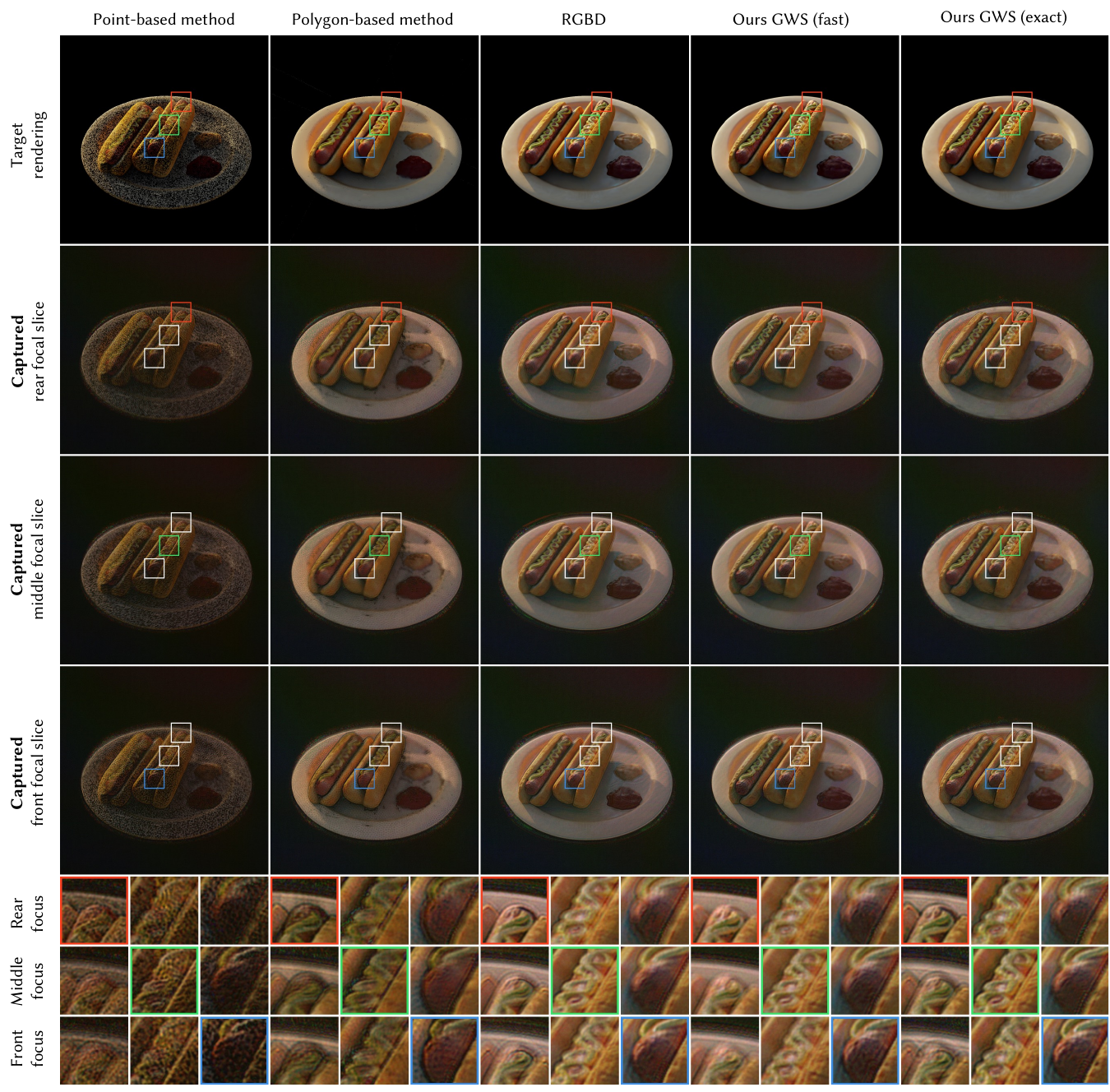}
    \caption{\textbf{Experimentally captured baseline comparisons.} We experimentally capture 3D focal stacks of holograms generated using different baseline CGH methods, and the results well match our simulations. Point-based methods result in overly dim images and poor reconstruction quality due to the sparseness of point cloud representatins. Polygon-based methods cannot reconstruct fine details. CGH from RGBD images achieve decent image quality. Our GWS achieve superior image quality over prior methods, reconstructing sharp details in focused regions.}
    \label{fig:captured_baselines_3}
\end{figure*}

\begin{figure*}
    \centering
    \includegraphics[width=\textwidth]{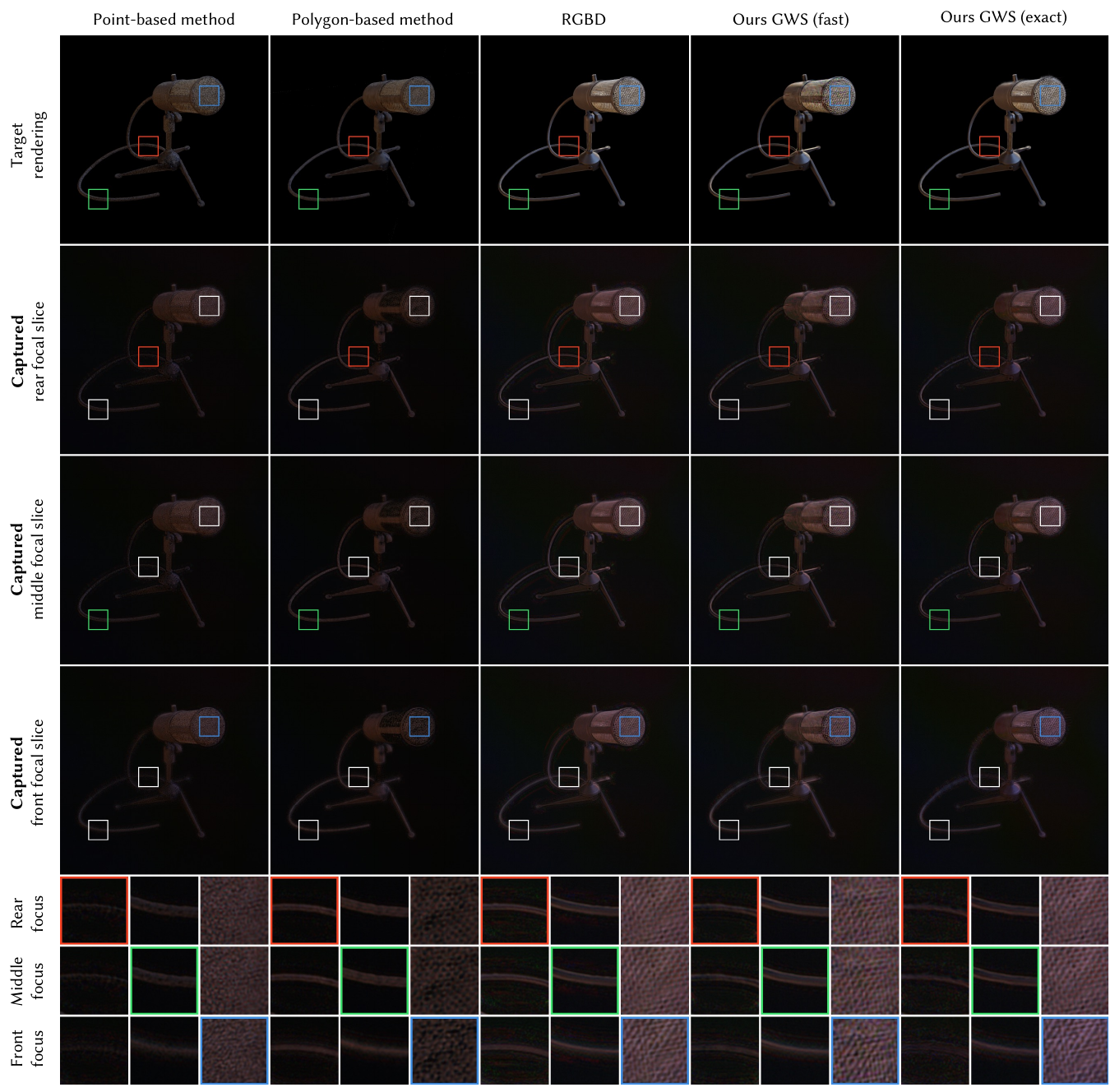}
    \caption{\textbf{Experimentally captured baseline comparisons.} We experimentally capture 3D focal stacks of holograms generated using different baseline CGH methods, and the results well match our simulations. Point-based methods result in overly dim images and poor reconstruction quality due to the sparseness of point cloud representatins. Polygon-based methods cannot reconstruct fine details. CGH from RGBD images achieve decent image quality. Our GWS achieve superior image quality over prior methods, reconstructing sharp details in focused regions.}
    \label{fig:captured_baselines_4}
\end{figure*}

\begin{figure*}
    \centering
    \includegraphics[width=\textwidth]{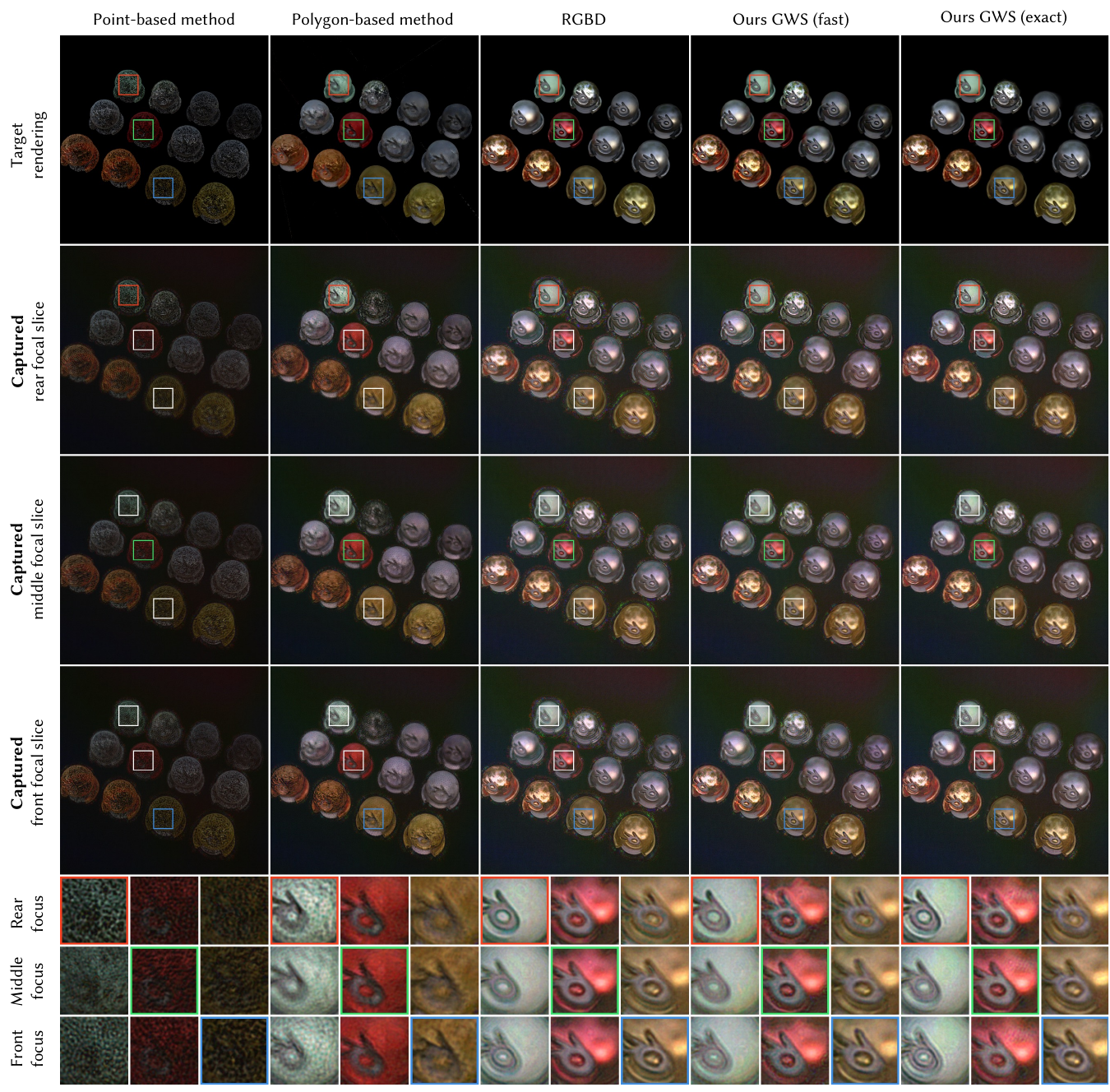}
    \caption{\textbf{Experimentally captured baseline comparisons.} We experimentally capture 3D focal stacks of holograms generated using different baseline CGH methods, and the results well match our simulations. Point-based methods result in overly dim images and poor reconstruction quality due to the sparseness of point cloud representatins. Polygon-based methods cannot reconstruct fine details. CGH from RGBD images achieve decent image quality. Our GWS achieve superior image quality over prior methods, reconstructing sharp details in focused regions.}
    \label{fig:captured_baselines_5}
\end{figure*}

\begin{figure*}
    \centering
    \includegraphics[width=\textwidth]{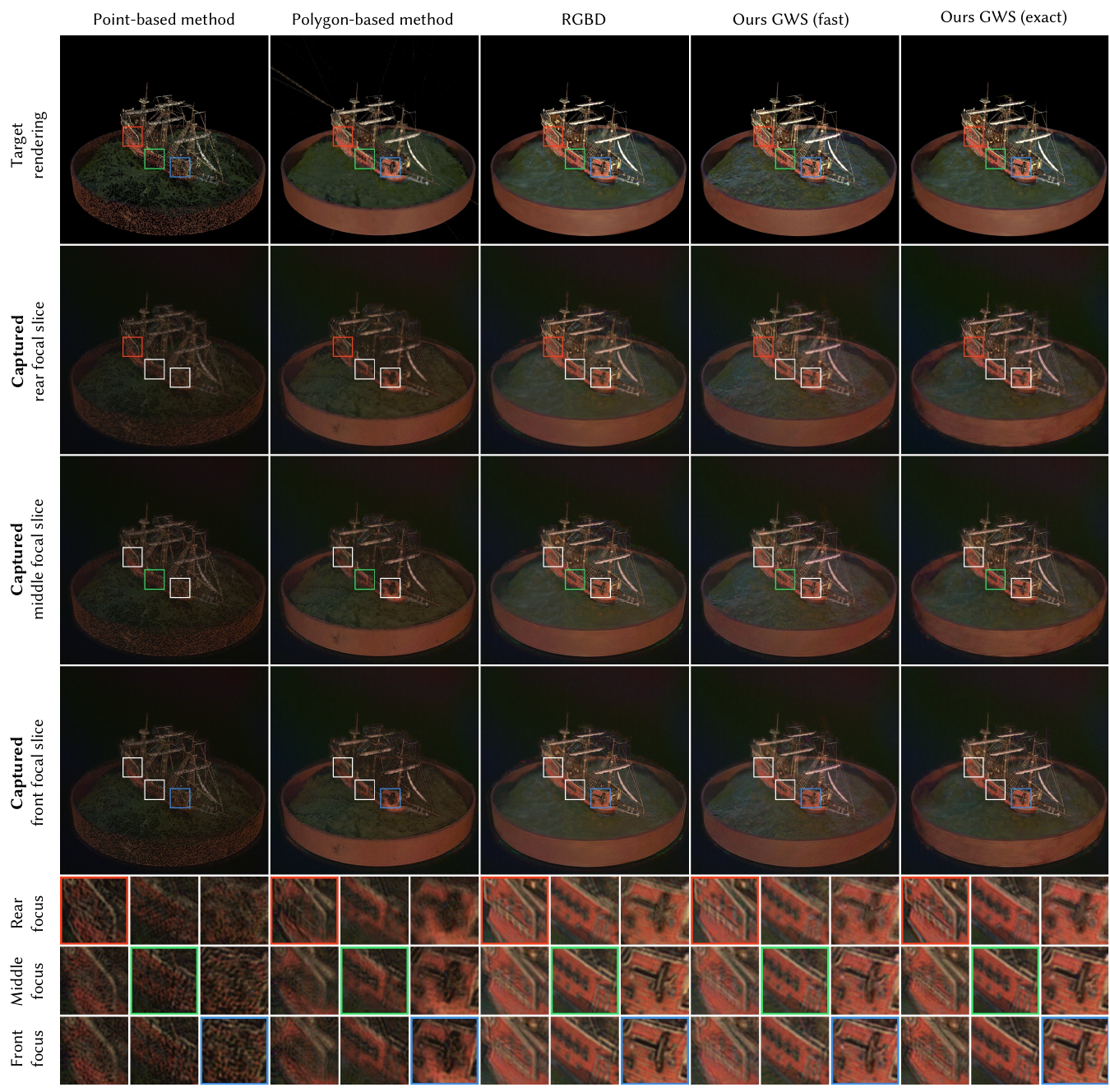}
    \caption{\textbf{Experimentally captured baseline comparisons.} We experimentally capture 3D focal stacks of holograms generated using different baseline CGH methods, and the results well match our simulations. Point-based methods result in overly dim images and poor reconstruction quality due to the sparseness of point cloud representatins. Polygon-based methods cannot reconstruct fine details. CGH from RGBD images achieve decent image quality. Our GWS achieve superior image quality over prior methods, reconstructing sharp details in focused regions.}
    \label{fig:captured_baselines_6}
\end{figure*}

\begin{figure*}
    \centering
    \includegraphics[width=\textwidth]{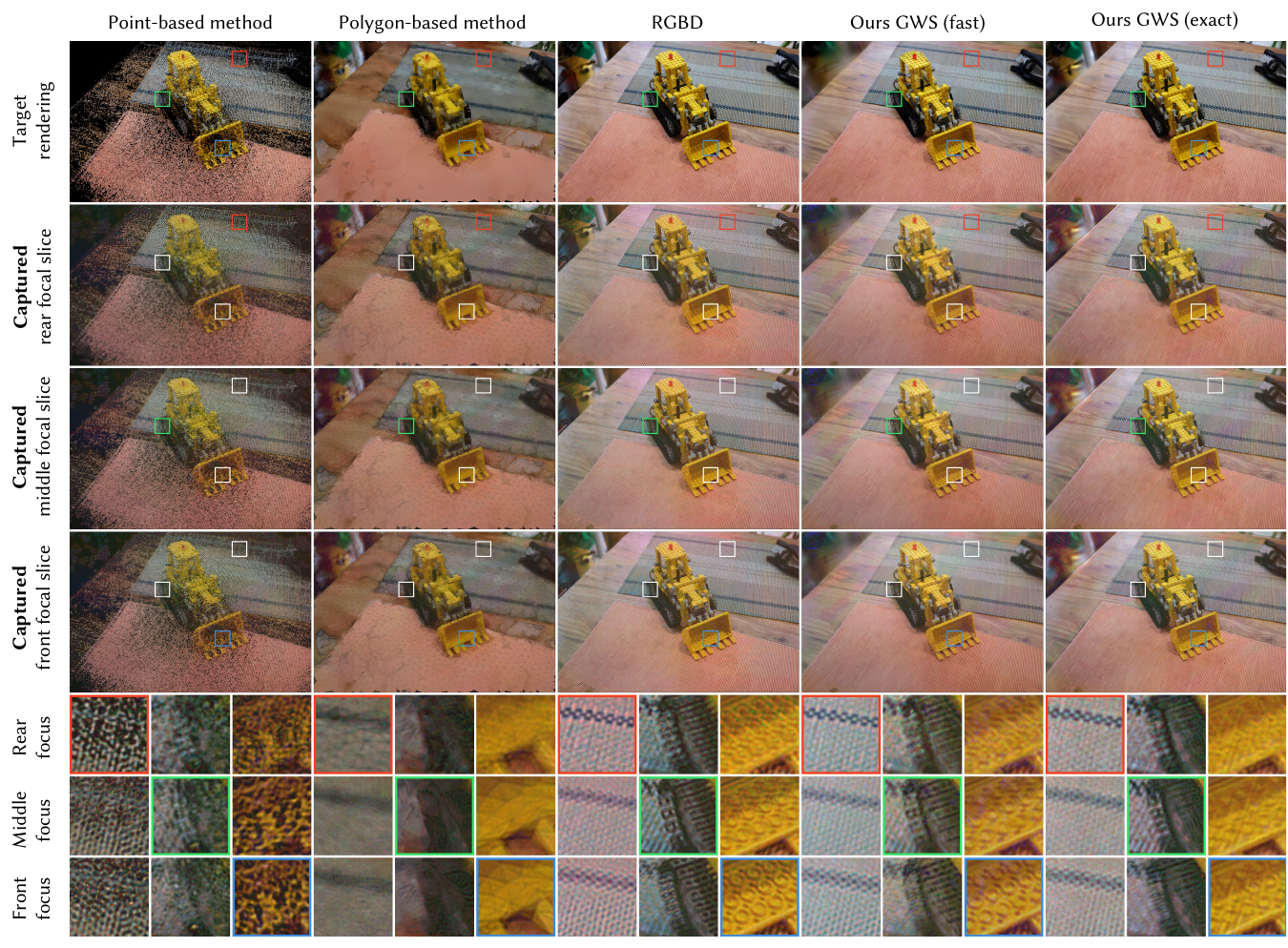}
    \caption{\textbf{Experimentally captured baseline comparisons.} We experimentally capture 3D focal stacks of holograms generated using different baseline CGH methods, and the results well match our simulations. Point-based methods result in overly dim images and poor reconstruction quality due to the sparseness of point cloud representatins. Polygon-based methods cannot reconstruct fine details. CGH from RGBD images achieve decent image quality. Our GWS achieve superior image quality over prior methods, reconstructing sharp details in focused regions.}
    \label{fig:captured_baselines_7}
\end{figure*}

\begin{figure*}
    \centering
    \includegraphics[width=\textwidth]{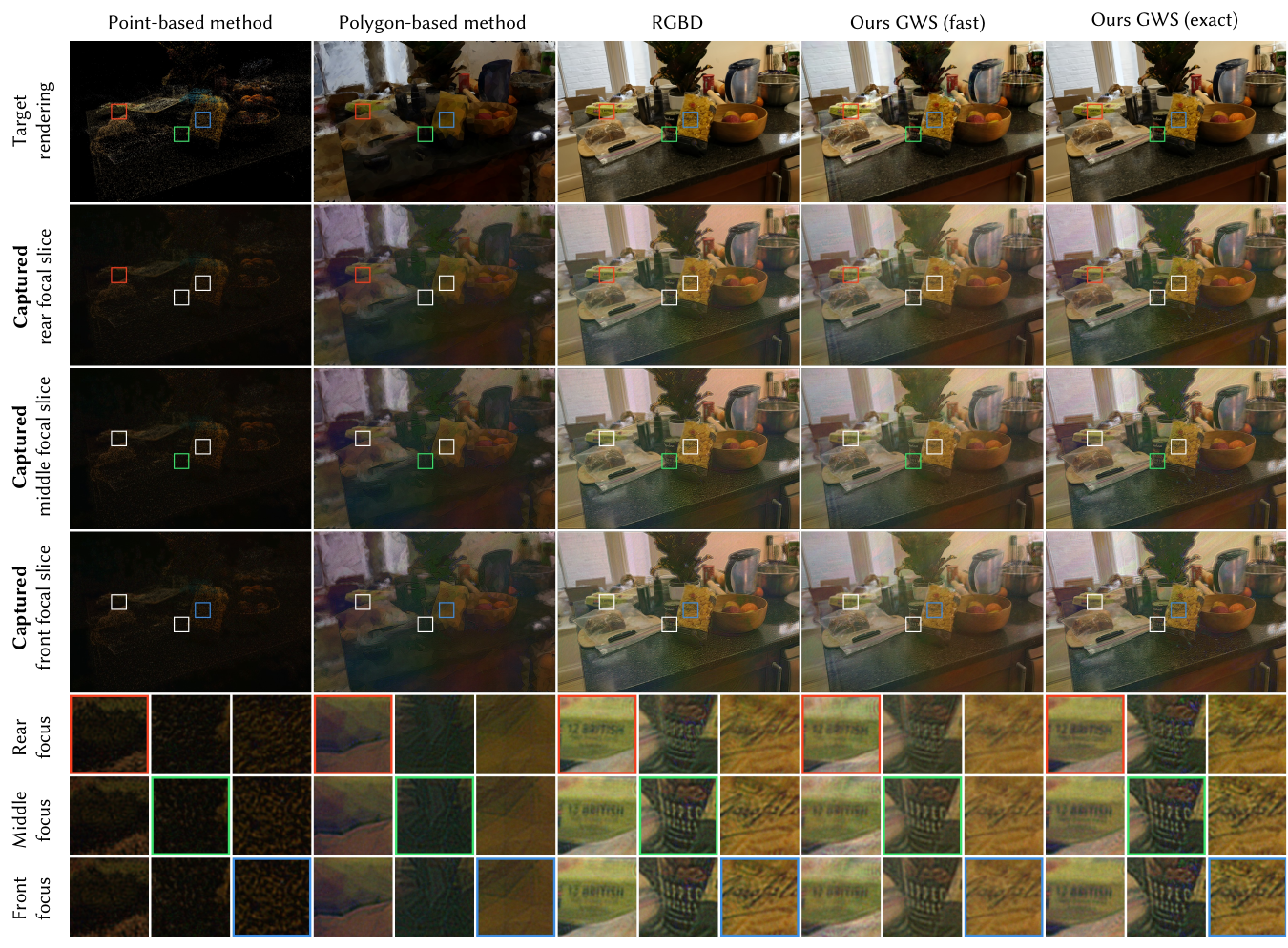}
    \caption{\textbf{Experimentally captured baseline comparisons.} We experimentally capture 3D focal stacks of holograms generated using different baseline CGH methods, and the results well match our simulations. Point-based methods result in overly dim images and poor reconstruction quality due to the sparseness of point cloud representatins. Polygon-based methods cannot reconstruct fine details. CGH from RGBD images achieve decent image quality. Our GWS achieve superior image quality over prior methods, reconstructing sharp details in focused regions.}
    \label{fig:captured_baselines_8}
\end{figure*}

\begin{figure*}
    \centering
    \includegraphics[width=\textwidth]{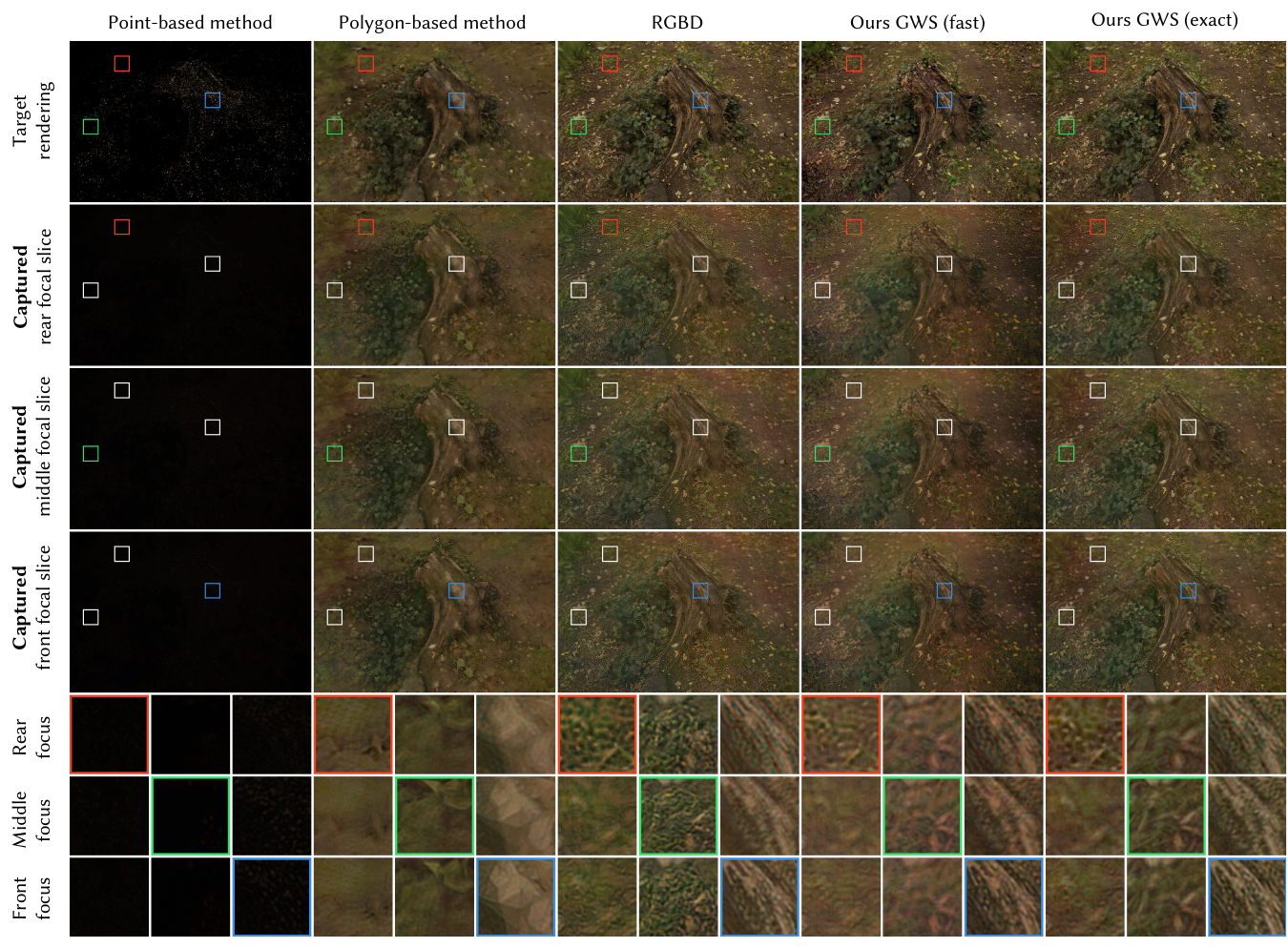}
    \caption{\textbf{Experimentally captured baseline comparisons.} We experimentally capture 3D focal stacks of holograms generated using different baseline CGH methods, and the results well match our simulations. Point-based methods result in overly dim images and poor reconstruction quality due to the sparseness of point cloud representatins. Polygon-based methods cannot reconstruct fine details. CGH from RGBD images achieve decent image quality. Our GWS achieve superior image quality over prior methods, reconstructing sharp details in focused regions.}
    \label{fig:captured_baselines_9}
\end{figure*}

\begin{figure*}
    \centering
    \includegraphics[width=\textwidth]{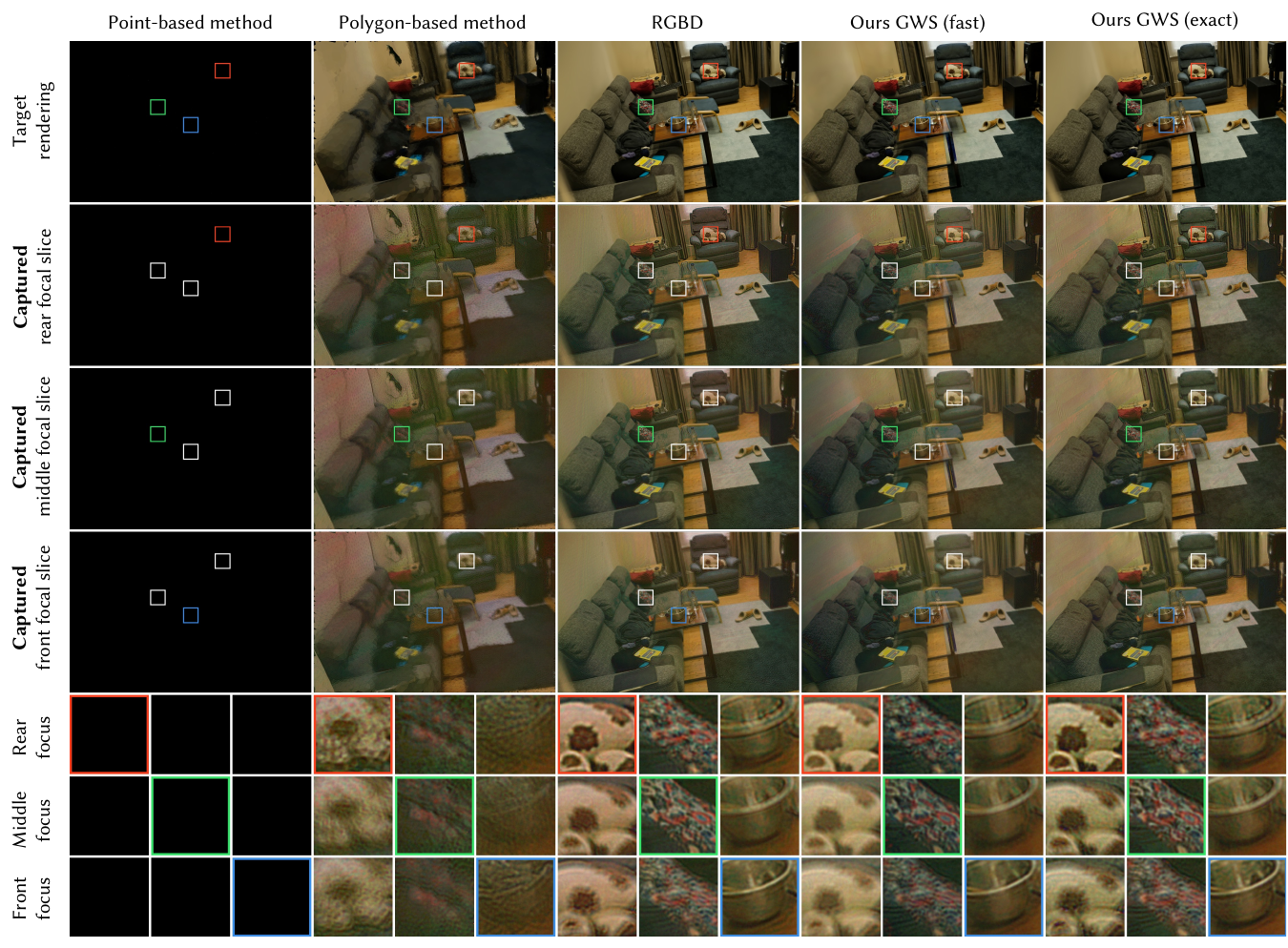}
    \caption{\textbf{Experimentally captured baseline comparisons.} We experimentally capture 3D focal stacks of holograms generated using different baseline CGH methods, and the results well match our simulations. Point-based methods result in overly dim images and poor reconstruction quality due to the sparseness of point cloud representatins. Polygon-based methods cannot reconstruct fine details. CGH from RGBD images achieve decent image quality. Our GWS achieve superior image quality over prior methods, reconstructing sharp details in focused regions.}
    \label{fig:captured_baselines_10}
\end{figure*}

\begin{figure*}
    \centering
    \includegraphics[width=\textwidth]{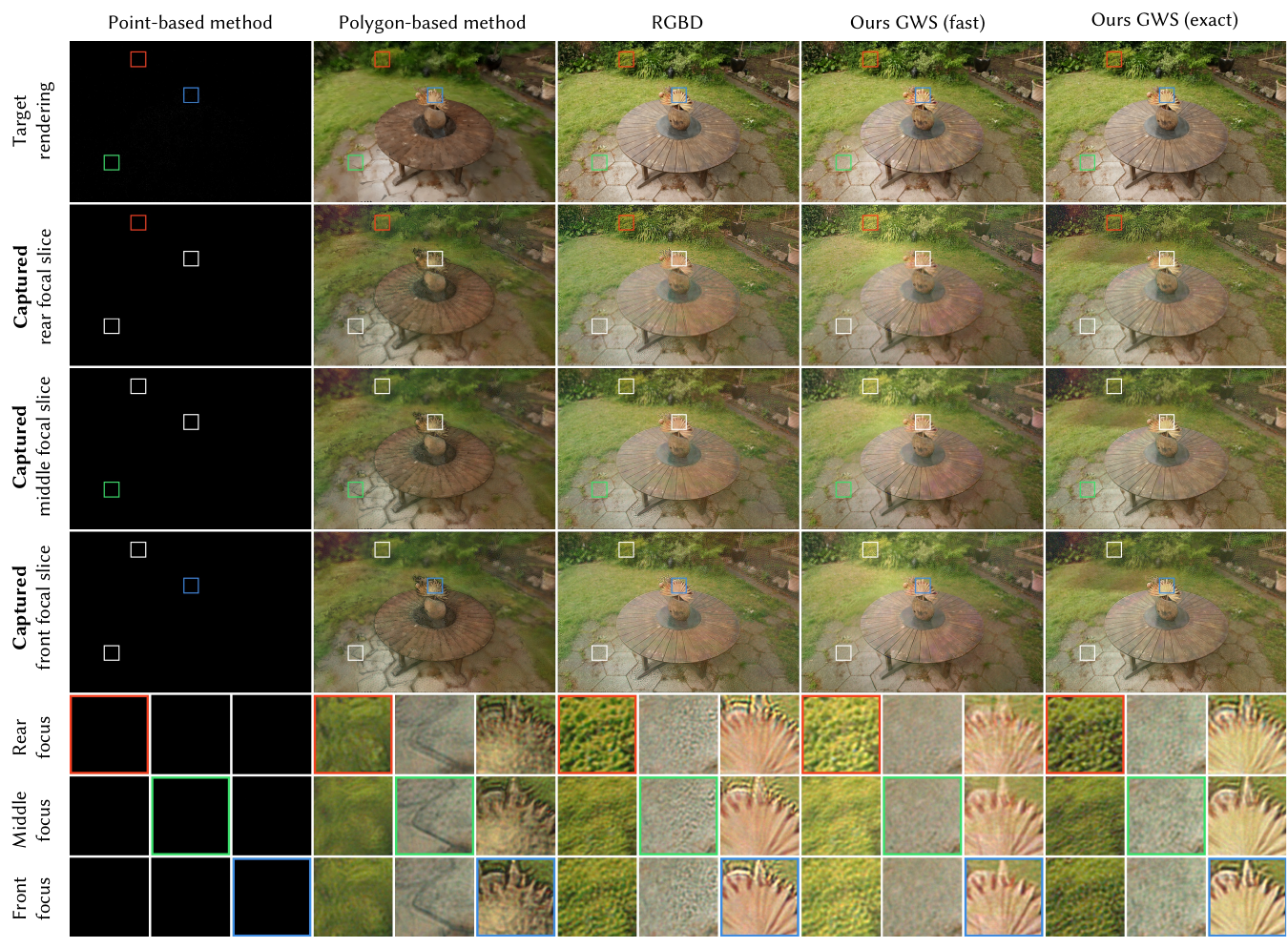}
    \caption{\textbf{Experimentally captured baseline comparisons.} We experimentally capture 3D focal stacks of holograms generated using different baseline CGH methods, and the results well match our simulations. Point-based methods result in overly dim images and poor reconstruction quality due to the sparseness of point cloud representatins. Polygon-based methods cannot reconstruct fine details. CGH from RGBD images achieve decent image quality. Our GWS achieve superior image quality over prior methods, reconstructing sharp details in focused regions.}
    \label{fig:captured_baselines_11}
\end{figure*}

\begin{figure*}
    \centering
    \includegraphics[width=\textwidth]{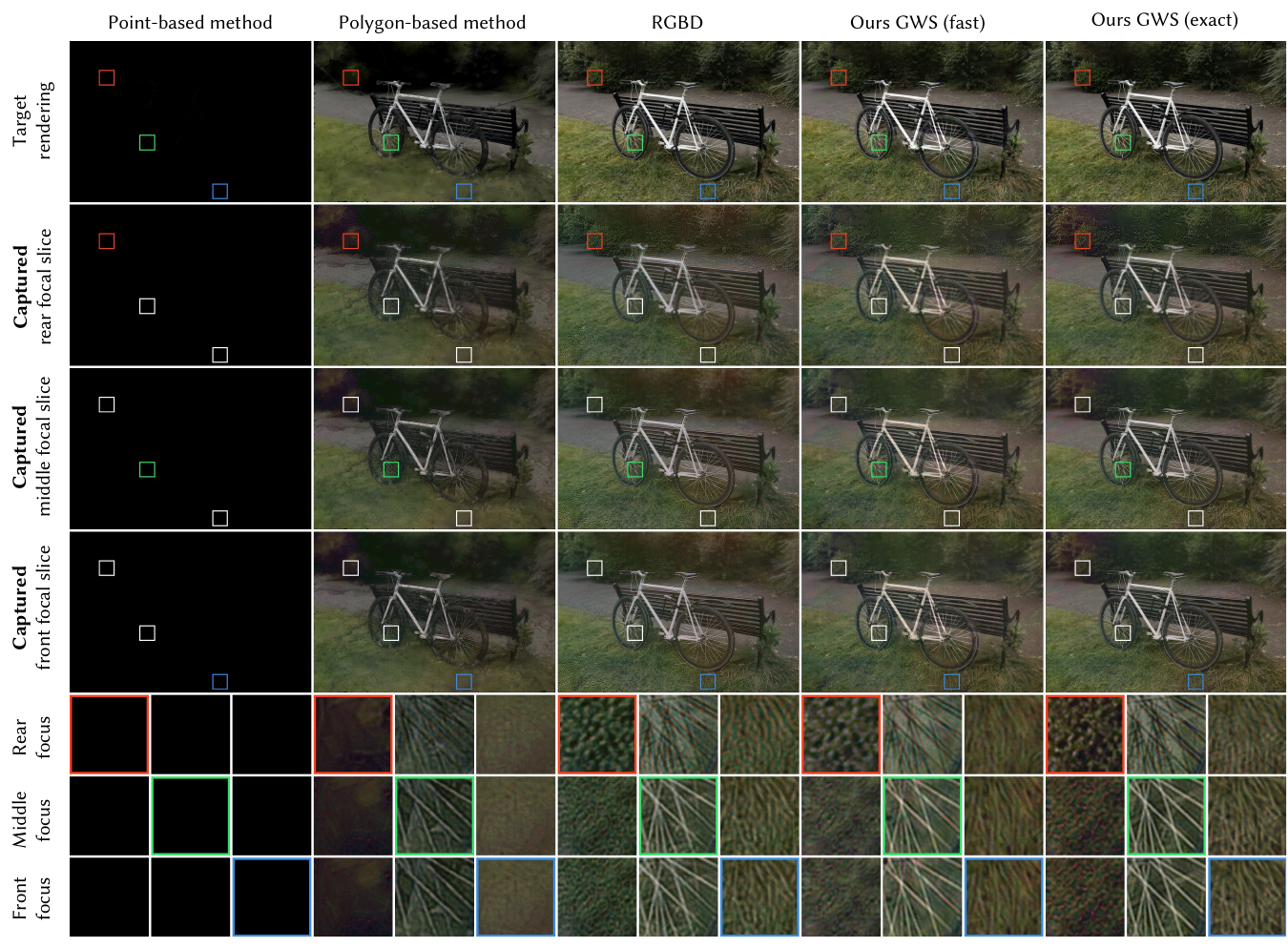}
    \caption{\textbf{Experimentally captured baseline comparisons.} We experimentally capture 3D focal stacks of holograms generated using different baseline CGH methods, and the results well match our simulations. Point-based methods result in overly dim images and poor reconstruction quality due to the sparseness of point cloud representatins. Polygon-based methods cannot reconstruct fine details. CGH from RGBD images achieve decent image quality. Our GWS achieve superior image quality over prior methods, reconstructing sharp details in focused regions.}
    \label{fig:captured_baselines_12}
\end{figure*}

Experimentally captured baseline comparisons closely follow the trend of simulated experiments, as shown in Figs. \ref{fig:captured_baselines_1} to \ref{fig:captured_baselines_12}. Point-based methods fail to accurately recosntruct the target content. Polygon-based methods cannot reconstruct sharp details due to per-face colors. RGBD holograms achieve decent image quality since the representation is generated exactly from ground truth RGB images and depth maps. However, ringing artifacts are present at depth discontinuities. 

Our Gaussian Wave Splatting method, both exact and fast, achieve significant image quality improvements over existing primitives-based CGH baselines. GWS achieves sharp reconstruction in focused regions and synthesizes more natural blur in defocused regions and does not have pronounced ringing artifacts like RGBD, well matching the simulated results. Fast GWS exhibits slight contrast reduction compared to exact GWS, but overall image quality and 3D refocusing effects are retained and achieves $30\times$ speedup as described in the main paper.

\subsection{Experimentally Captured 3D Focal Stacks of Novel Views}

\begin{figure*}
    \centering
    \includegraphics[width=\textwidth]{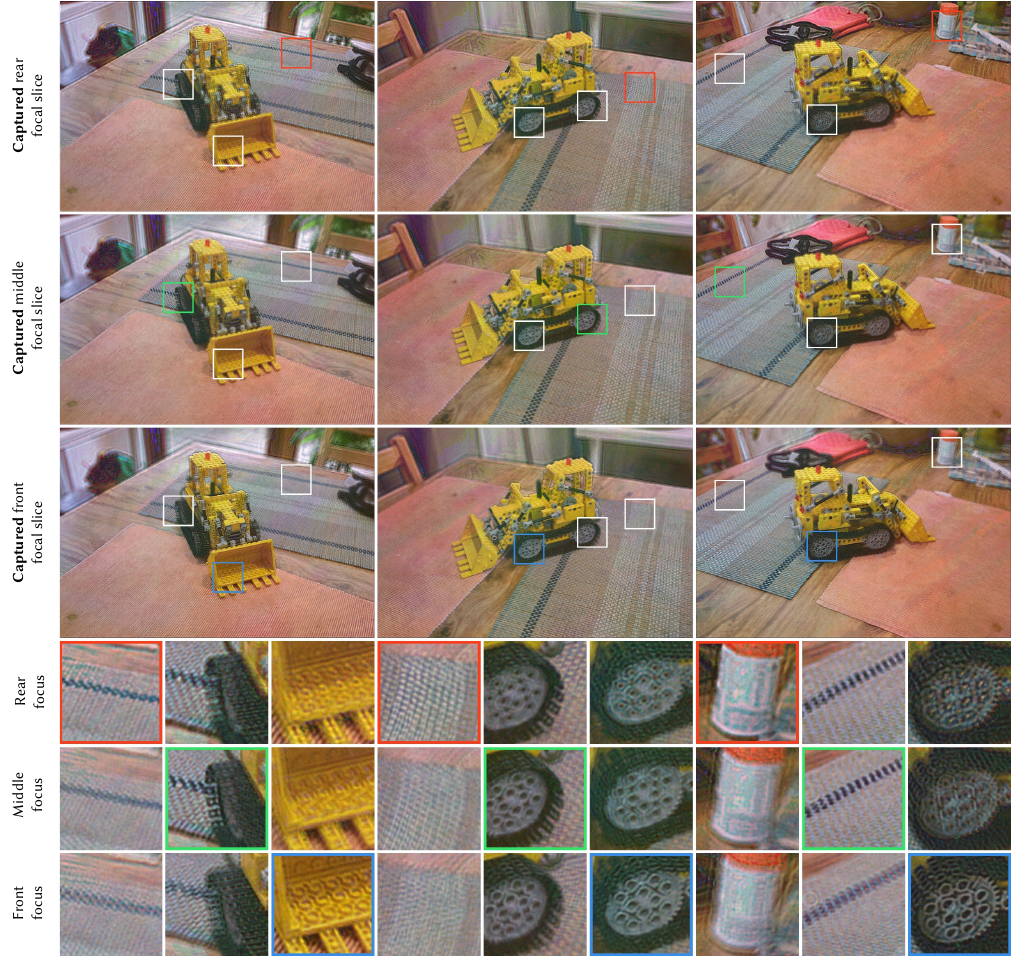}
    \caption{\textbf{Experimentally captured focal stacks of exact GWS holograms from different novel viewpoints.} We generate exact GWS holograms from various novel viewpoints and capture the corresponding 3D focal stacks at each viewpoint. GWS achieves accurate 3D refocusing effects at all viewpoints, demonstrating the robustness of our method. }
    \label{fig:nvs_fs_garden}
\end{figure*}

\begin{figure*}
    \centering
    \includegraphics[width=0.9\textwidth]{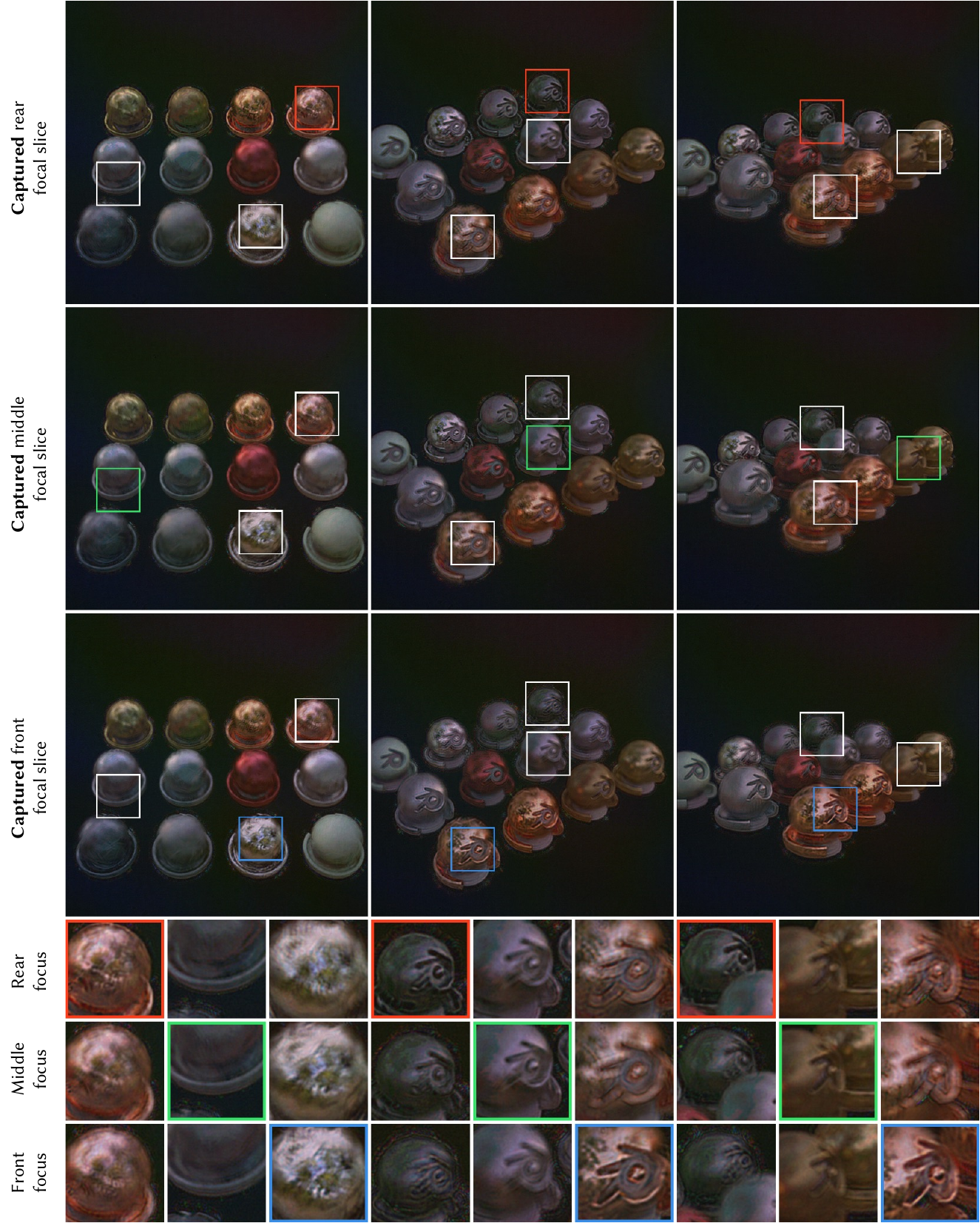}
    \caption{\textbf{Experimentally captured focal stacks of exact GWS holograms from different novel viewpoints.} We generate exact GWS holograms from various novel viewpoints and capture the corresponding 3D focal stacks at each viewpoint. GWS achieves accurate 3D refocusing effects at all viewpoints, demonstrating the robustness of our method. }
    \label{fig:nvs_fs_materials}
\end{figure*}

Aside from the ``garden'' scene shown in the main paper, we additionally capture 3D focal stacks of exact GWS holograms generated from different novel views of the ``kitchen'' scene from the Mip-NeRF 360 dataset and ``materials'' scene from the Blender dataset in Fig. \ref{fig:nvs_fs_garden} and \ref{fig:nvs_fs_materials}, respectively. GWS achieves accurate 3D refocusing effects at \textit{all} novel views, demonstrating the robustness of our method to changing perspectives. Please refer to the supplemental videos for full novel view trajectory videos of optimized GWS holograms generated from different views of the underlying 2DGS models, as well as 3D focal stacks captured at selected frames. 

\clearpage
\newpage
\section{Polygon-based methods}
\label{sec:misc}

In this section, we discuss how our Gaussian Wave Splatting method generalize well and are compatible with the pioneering works in polygon-based CGH. Here, we assume that we would transform their geometric parameters following our holographics pipeline, and discuss how their geometric parameters can be described in terms of the rotation matrix $\rot$, scaling matrix $S$, and translation $\centergaussian$ in Gaussian Wave Splatting. For more background, we refer to the review papers and the well-curated textbook on these methods~\cite{park2017recent, zhang2022polygon, matsushima2020introduction}.

\begin{figure}[ht!]
    \centering
    \includegraphics[width=1.0\columnwidth]{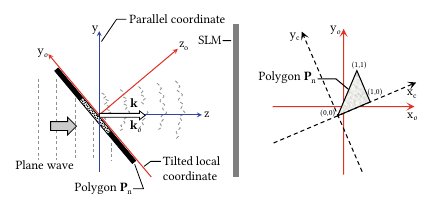}
    \caption{Illustration of three coordinate systems. The left figure is adapted from~\cite{matsushima2020introduction}. On the left, we show the canonical coordinates $\coordinate_\canonicalspace = (x_\canonicalspace, y_\canonicalspace)$ and the tilted local coordinates (or object space) $\coordinate_\objectspace = (x_\objectspace, y_\objectspace)$, which lie on the same plane and are defined by a 2D affine transformation: $\coordinate_\canonicalspace = \mathbf{A}\coordinate_\objectspace$. On the right, we show the parallel coordinate $\coordinate$ and the tilted local coordinate, which are related by a rotational transformation and translation: $\coordinate_\objectspace = \rot \coordinate + \vertice_1$. } 
    \label{fig:coord}
\end{figure}

Unlike Gaussians whose angular spectrums are also elegantly Gaussians, the angular spectrums of polygons are more complicated. In classic fully analytic polygon-based CGH, the reference triangle is assumed to be in the canonical space with vertices at $\mathbf{P}_n = \{[0,0]^\mathsf{T}, [1, 0]^\mathsf{T}, [1,1]^\mathsf{T}\}$ (see Fig.~\ref{fig:coord}). Its angular spectrum can be derived as in~\cite{kim2008mathematical, ahrenberg2008computer, zhang2022polygon}:
\begin{align}
    {\footnotesize
\angularspectrum_\canonicalspace\left(k_x, k_y, k_z\right) = \begin{cases}
    \frac{1}{2}, & \text{if } k_x = k_y = 0 \\
    \frac{1 - e^{jk_y}}{k_y^2} + \frac{j}{k_y}, & \text{if } k_x = 0, k_y \neq 0 \\
    \frac{e^{jk_x} - 1}{k_x^2} - \frac{je^{jk_x}}{k_x}, & \text{if } k_x \neq 0, k_y = 0 \\
    \frac{1 - e^{jk_y}}{k_y^2} - \frac{j}{k_y}, &  \text{if } k_x = -k_y, k_y \neq 0 \\
    \frac{e^{jk_x} - 1}{k_x k_y} + \frac{1 - e^{j(k_x + k_y)}}{k_y (k_x + k_y)}, & \text{elsewhere}, \\
\end{cases}}
    \label{eq:ref_polygon}
\end{align}
Here, we assume our polygons are already in the hologram space after appropriate transformations, with vertices at $V = [\vertice_1, \vertice_2, \vertice_3] \in \mathbb{R}^{3 \times 3}$. Note that the tilted local coordinates in \ref{fig:coord} is a jargon used in classic polygon-based CGH methods, and is the same as the \textit{object space} used in our holographics pipeline.

We can also obtain the normal vector of the triangle primitive $\mathbf{n}$ from these vertices. A common practice in polygon-based methods is to have the first vertex $\vertice_1$ be at the origin of the object space, where the $z$-axis is parallel to the normal vector $\mathbf{n}$. Thus, this is equivalent to having a translation $\bm{\mu} = \vertice_1$, and a rotation matrix $R = R_x(\theta)R_y(\phi)$, with $\theta = \tan^{-1}(\normal_x / \normal_z)$ and $\phi = \tan^{-1}(\normal_y / \sqrt{\normal_z^2 + \normal_x^2})$, where the subscript denotes the component of the vector or the axis of rotation. By applying these translation and rotation transformations, we move from hologram space to object space. There, we have three vertices at $\mathbf{V}_o = [\mathbf{0}, R(\vertice_2 - \vertice_1), R(\vertice_3 - \vertice_1)]$, whose third row is zero. Thus, we consider the upper $2\times2$ matrix as $\mathbf{V}_0$ in the following for brevity. Then, we can map these vertices to the reference vertices $\mathbf{P}_n$ with a 2D affine transform $A$ that satisfies $\begin{bmatrix} 1 & 1 \\
    0 & 1 \end{bmatrix} = A \mathbf{V}_o$. This corresponds to our 2D affine transform theorem in Eq.~8 in the manuscript. Thus, we can replace $S$ in our equation with $A=\begin{bmatrix} 1 & 1 \\
    0 & 1 \end{bmatrix} \mathbf{V}_o^{-1}$.

  In summary, we demonstrate how our formulation can encompasses classic polygon-based methods by following their well-established practices. We implement these methods by simply dialing in the parameters $R, S, \centergaussian$ as described in this section in \texttt{PyTorch} for polygon-based CGH baselines following the same implementation details provided in Sec.~\ref{subsec:details}. We also open-source the code for polygon-based CGH along with \texttt{hsplat}\footnote{\href{https://github.com/computational-imaging/hsplat}{https://github.com/computational-imaging/hsplat}}.

\newpage
\bibliographystyle{acm}
\bibliography{bibs}